\documentclass[11pt]{article}
\usepackage{graphicx}
\usepackage{epsfig}
\usepackage{amsfonts}
\usepackage{amsmath}
\usepackage{amssymb}
\usepackage{dsfont}
\usepackage{hyperref}
\usepackage{bbold}
\usepackage{color}
\usepackage{slashed}
\usepackage{cite}
\usepackage{pgfplots}
\usepackage{tikz} 
\usepackage{pgfplotstable}
\pgfplotsset{compat=1.8}
    \renewcommand{\arraystretch}{1.5}
\newcommand{\ep}{\epsilon}

\newcommand{\al}{\alpha}
\newcommand{\bt}{\beta}
\newcommand{\g}{\gamma}

\newcommand{\dt}{\delta}

\newcommand{\la}{\lambda}
\newcommand{\simu}{\sigma^{\mu\nu}}

\newcommand{\ka}{\kappa}
\newcommand{\im}{\text{Im}\,}
\newcommand{\re}{\text{Re}\,}
\newcommand{\vL}{\ensuremath{\mathcal{L}}}    
    \newcommand{\Dt}{\Delta}

\newcommand{\ga}{\gamma}
\newcommand{\GA}{\Gamma}
\newcommand{\TG}{\tilde\Gamma}

\newcommand{\Or}{\mathcal O}

\newcommand{\vp}{\varphi}
\newcommand{\sq}{^{2}}

\newcommand{\tr}{\mathrm{Tr}}
\newcommand{\dslash}[1]{#1 \llap{/\kern-0.5pt}}
\newcommand{\Dslash}[1]{#1 \llap{/\kern+1.5pt}}
\newcommand{\DDslash}[1]{#1 \llap{/\kern+2.3pt}}
\newcommand{\dslashh}[1]{#1 \llap{/\kern+1pt}}

\newcommand{\Ex}[1]{\cdot 10^{#1}}
\newcommand{\bea}{\begin{eqnarray}}
\newcommand{\eea}{\end{eqnarray}}
\newcommand{\bma}{\begin{pmatrix}}
\newcommand{\ema}{\end{pmatrix}}
\newcommand{\nn}{\nonumber}

\newcommand{\newc}{\newcommand}
\newc{\jdv}[1]{{\color{red}#1}}

\begin{document}

\begin{titlepage}

\begin{flushright}
LA-UR-21-26789\\
\end{flushright}

\vspace{2.0cm}

\begin{center}
{\LARGE  \bf 
A low-energy perspective on the \\
\vspace{3mm}
 minimal left-right symmetric model 
\vspace{3mm}
}
\vspace{1.5cm}

{\large \bf W. Dekens$^{a}$, L. Andreoli$^{b}$, J. de Vries$^{c,d}$, E. Mereghetti$^e$, and F.\ Oosterhof$^f$} 
\vspace{0.5cm}

\vspace{0.25cm}
{\large 
$^a$ 
{\it 
Department of Physics, University of California at San Diego,
La Jolla, CA 92093-0319, USA
}}

\vspace{0.25cm}
{\large 
$^b$ 
{\it 
Department of Physics, Washington University in Saint Louis,
Saint Louis, MO 63130, USA
}}

{\large 
$^c$ 
{\it Institute for Theoretical Physics Amsterdam and Delta Institute for Theoretical Physics, University of Amsterdam, Science Park 904,\\ 1098 XH Amsterdam, The Netherlands}}

{\large 
$^d$ 
{\it Nikhef, Theory Group, Science Park 105, 1098 XG, Amsterdam, The Netherlands}}

\vspace{0.25cm}
{\large 
$^e$ 
{\it Theoretical Division, Los Alamos National Laboratory,
Los Alamos, NM 87545, USA}}

\vfill
{\large 
$^f$ 
{\it Van Swinderen Institute for Particle Physics and Gravity,
University of Groningen, 9747 AG Groningen, The Netherlands}}
\end{center}

\vspace{0.2cm}

\begin{abstract}
\vspace{0.1cm}

We perform a global analysis of the low-energy phenomenology of the minimal left-right symmetric model (mLRSM) with parity symmetry. We match the mLRSM to the Standard Model Effective Field Theory Lagrangian at the left-right-symmetry breaking scale and perform a comprehensive fit to low-energy data including mesonic, neutron, and nuclear $\beta$-decay processes, $\Delta F=1$ and $\Delta F=2$ CP-even and -odd processes in the bottom and strange sectors, and electric dipole moments (EDMs) of nucleons, nuclei, and atoms. We fit the Cabibbo-Kobayashi-Maskawa and mLRSM parameters simultaneously and determine a lower bound on the mass of the right-handed $W_R$ boson. In models where a Peccei-Quinn mechanism provides a solution to the strong CP problem, we obtain $M_{W_R} \gtrsim 5.5$ TeV at $95\%$ C.L. which can be significantly improved with next-generation EDM experiments. In the $P$-symmetric mLRSM without a Peccei-Quinn mechanism we obtain a more stringent constraint $M_{W_R} \gtrsim 17$ TeV at $95\%$ C.L., which is difficult to improve with low-energy measurements alone. In all cases, the additional scalar fields of the mLRSM are required to be a few times heavier than the right-handed gauge bosons. 
We consider a recent discrepancy in tests of first-row unitarity of the CKM matrix. We find that, while TeV-scale $W_R$ bosons can alleviate some of the tension found in the $V_{ud,us}$ determinations, a solution to the discrepancy is disfavored when taking into account other low-energy observables within the mLRSM.
 

\end{abstract}

\vfill
\end{titlepage}

\tableofcontents

\section{Introduction}\label{sec:intro}

Left-right (LR) symmetric models \cite{Pati:1974yy, Mohapatra:1974hk, Senjanovic1975,Senjanovic:1978ev,Deshpande:1990ip} provide a framework for a dynamical theory of parity ($P$) violation and led to the prediction of right-handed neutrinos and the see-saw mechanism, well before neutrino oscillations were discovered \cite{Mohapatra:1979ia,Mohapatra:1980yp}. 
Apart from providing a natural explanation of parity violation and neutrino masses, LR models give rise to a rich phenomenology. For example, due to the see-saw mechanism, LR models violate lepton number, which leads to an interesting interplay of different contributions to neutrinoless double beta decay \cite{Mohapatra:1979ia,Mohapatra:1980yp,Doi:1985dx,Tello:2010am,Rodejohann:2011mu,Cirigliano:2018yza,Li:2020flq,Yang:2021ueh}. The resulting signal could very well be measurable, even in the normal hierarchy with small neutrino masses. The high-energy analogue, the so-called Keung-Senjanovi\'c process \cite{Keung:1983uu}, is a promising probe of the same source of lepton number violation at the LHC or future colliders.
In addition, the presence of right-handed charged currents mediated by $W_R$ exchange and of heavy scalar bosons with flavor-changing interactions lead
to a rich flavor phenomenology, with new contributions to a broad range of processes including CP violation in meson mixing and decays \cite{Beall:1981ze,Mohapatra:1983ae,Ecker:1983uh,Ecker:1985ei,Frere:1991db,Ball:1999mb,Bertolini:2014sua}, nuclear $\beta$-decay \cite{Gonzalez-Alonso:2018omy,Cirigliano:2013xha}, electroweak precision observables \cite{Hsieh:2010zr,Blanke:2011ry,Bernard:2020cyi}, and electric dipole moments (EDMs) of leptons, nucleons, nuclei, atoms, and molecules \cite{Ecker:1983dj,Frere:1991jt,Xu2010,Cirigliano:2016yhc}.

Direct searches for right-handed gauge bosons at colliders constrain their masses to be larger than a few TeV \cite{CMS:2018mgb,Aaboud:2017yvp,CMS:2021mux}. To accommodate the non-observation of large flavor-changing-neutral-current processes, the new scalars associated with left-right models must have even larger masses, $\gtrsim\Or(10)$ TeV. The gap between the right-handed scale, where parity is spontaneously broken, and the electroweak scale makes left-right symmetric models amenable to effective field theory (EFT) techniques. In particular, at the right-handed scale the theory can be matched onto the Standard Model EFT~\footnote{Depending on the mass scale of right-handed neutrinos, it might be appropriate to match to the SMEFT extended with right-handed neutrinos instead \cite{Li:2021tsq,Liao:2016qyd,delAguila:2008ir}. In this work, we focus on the quark sector of left-right models and do not discuss leptonic observables in great detail.} (SMEFT). Although a large number of SMEFT operators is induced,  the associated Wilson coefficients only depend on a handful of fundamental parameters. The relatively small set of parameters (compared to, for instance, supersymmetric models) allows for a global analysis of the parameter space. Several such analyses have been performed in the literature, see e.g.\ \cite{Zhang:2007da,Blanke:2011ry,Bertolini:2019out,Maiezza:2014ala}. For instance, recently Refs.~\cite{Bertolini:2019out,Maiezza:2014ala} considered the correlation between direct and indirect CP violation in kaon decays and the neutron electric dipole moment, setting lower bounds on the $W_R$ mass (for earlier work including also $\Delta F=2$ transitions in B mesons, see e.g.\ Refs.~\cite{Maiezza:2010ic,Bertolini:2014sua}). A large amount of work has also been devoted to the phenomenology of the leptonic sector of left-right models \cite{Prezeau:2003xn,Bambhaniya:2015ipg,Dev:2014xea,Tello:2010am,Nemevsek:2011aa,Nemevsek:2012iq,Barry:2013xxa}.

In this work we investigate the minimal left-right symmetric model with a generalized $P$ symmetry. In particular, we focus on the hadronic sector of the model and leave the interesting phenomenology related to the lepton sector (from neutrinoless double beta decay to lepton flavor violation) for future work. Our aim is to perform a true global analysis of the low-energy phenomenology of the $P$-symmetric minimal left-right model in order to determine the allowed parameter space of the model, focusing mainly on a potential lower bound on the $W_R$ mass. As the SMEFT operators affect many processes that are used to extract the elements of the Cabibbo-Kobayashi-Maskawa (CKM) quark mixing matrix, it is not consistent to simply use the values for the quark mixing angles and phases obtained from a SM fit. We therefore extend previous analyses and refit the CKM parameters in combination with the new parameters associated with left-right models (which we denote by LR parameters). This requires us to include a large number of observables that are discussed in detail in this work. At the same time this allows us to consider possible beyond-the-Standard-Model (BSM) solutions to recent discrepancies in some of these observables, in particular the determinations of the $V_{ud}$ and $V_{us}$ CKM elements,  in a consistent manner.
This analysis draws 
from Ref.~\cite{Alioli:2017ces} which performed a similar study for one specific dimension-six SMEFT operator that is induced in left-right symmetric models.

The hadronic observables we consider depend on perturbative and non-perturbative theoretical quantities and controlling their uncertainties is crucial to obtain strong bounds on BSM physics. 
Advances in lattice QCD have reduced the error on 
decay constants and form factors entering the theoretical expressions of leptonic and semileptonic meson decays 
to the permille level in the case of light quarks and percent level for heavy quarks \cite{Aoki:2019cca}. Similarly, 
the local matrix elements of $\Delta F=2$ operators  
required for $\varepsilon_K$ and the $B_{d,s} - \bar B_{d,s}$ mass splittings have uncertainties of a few percent.
More recently, the first complete lattice QCD calculations 
of $K \rightarrow \pi\pi$ matrix elements have appeared \cite{Abbott:2020hxn}, leading to a SM prediction for direct CP violation in kaon decays with $\sim 40\%$ error. These calculations have also helped to reduce the error on hadronic electric dipole moments \cite{Cirigliano:2016yhc,Alioli:2017ces}.  In addition to the inclusion of a large number of observables, our analysis improves upon previous literature by using state-of-the-art theoretical predictions for hadronic and nuclear matrix elements and by taking advantage of recent theoretical advances like the improved SM prediction of $\varepsilon_K$ \cite{Brod:2019rzc}. Our use of the SMEFT framework allows us to include QCD corrections, in particular those arising between $\mu=M_{W_R}$ and $\mu=m_W$, in a systematic way.
We discuss the residual theoretical uncertainties, which mostly affect the nucleon and nuclear EDMs, $\Delta F=2$ processes dominated by long-distance contributions (such as the $K-\bar K$ mass difference or $D-\bar D$ oscillations), and hadronic $B$ meson decays. 

Although the mLRSM leads to interesting signatures at high energies \cite{Keung:1983uu,Cao:2012ng, Harz:2021psp,Maiezza:2010ic}, here we focus on low-energy phenomenology and do not explicitly include LHC observables in our analysis. While such a combination is certainly interesting, an EFT analysis might not be appropriate for collider phenomenology, depending on the mass of BSM fields. The indirect bounds we find turn out to be sufficiently strong for most of the parameter space to ensure that direct production of right-handed gauge bosons is not yet accessible at the LHC. The combined analysis of low- and high-energy probes within the mLRSM is certainly very interesting and left to future work.

We start by introducing the LR model in Sect.\ \ref{introLR}. We subsequently integrate out the heavy LR fields and match onto the SMEFT in Sect.\  \ref{sec:matching}, where we also discuss the renormalization group (RG) evolution to low energies and the subsequent matching onto the $SU(3)_c\times U(1)_{\rm QED}$ invariant EFT, known as LEFT. Sect.\ \ref{sec:chiralCPodd} performs the matching onto the low-energy description of QCD, chiral perturbation theory ($\chi$PT), which is relevant for low-energy hadronic and nuclear observables. Some of the most important observables included in our analyses are described in Sect.\ \ref{sec:obs}, where we also discuss the impact of the new features of our analysis for the $\Dt F=2$ observables that have been the focus of previous works \cite{Maiezza:2014ala,Maiezza:2010ic}, while others are relegated to App.\ \ref{app:observables}. 
We finally present our results in Sect.\ \ref{sec:results} and conclude in Sect.\ \ref{sec:conclusion}, while several Appendices are dedicated to technical details.

\section{Minimal left-right models}\label{introLR}
\subsection{Particle content}
The gauge group of LR models \cite{Pati:1974yy,Mohapatra:1974hk,Senjanovic:1978ev,Senjanovic1975,Deshpande:1990ip} is given by $SU(2)_L \times SU(2)_R \times U(1)_{B-L}$. The fermions are assigned to representations of the above gauge group as follows,
\bea Q_L &=&\bma u_L\\d_L\ema \in (2,1,1/3)\,, \qquad Q_R = \bma u_R\\d_R\ema\in (1,2,1/3)\,,\nn\\
L_L &=&\bma \nu_L\\l_L\ema \in (2,1,-1)\,, \qquad L_R = \bma \nu_R\\l_R\ema\in (1,2,-1)\,.\eea
In the scalar sector, a field transforming under both $SU(2)_L$ and $SU(2)_R$, $\phi\in (2,2^*,0)$, is introduced, which allows for interactions that give rise to the mass terms of the fermions after electroweak symmetry breaking (EWSB).  
Additional scalar fields are then used to break the LR gauge group to that of the SM. We focus on the version of the LR model, called the minimal left-right symmetric model (mLRSM), in which this is done with two triplets, $\Delta_{L,R}$, assigned to $(3,1,2)$ and $(1,3,2)$, respectively. These fields can be written as 
\bea \phi = \bma \phi_1^0 & \phi_2^+\\ \phi_1^- & \phi_2^0 \ema \,,\qquad
\Delta_{L,R} = \bma \delta^+_{L,R}/\sqrt{2} & \delta^{++}_{L,R} \\ \delta^0_{L,R} & -\delta^+_{L,R}/\sqrt{2} \ema \,,
\label{scalars}\eea
and they transform as $\phi\to U_L\phi U_R^\dagger$, $\Dt_{L,R}\to U_{L,R}\Dt_{L,R} U_{L,R}^\dagger$ under $SU(2)_{L,R}$ transformations. 

Having specified the particle content we can write the complete Lagrangian as follows \bea \label{Lagrangian}
\vL &=& 
i\bar Q_L\slashed{D} Q_L+i\bar Q_R\slashed{D} Q_R+i\bar L_L\slashed{D} L_L+i\bar L_R\slashed{D} L_R \\
&&-\frac{1}{4}W_{L\, \mu\nu}^IW_{L}^{I\, \mu\nu}-\frac{1}{4}W_{R\, \mu\nu}^IW_{R}^{I\, \mu\nu}-\frac{1}{4}\mathcal B_{ \mu\nu}\mathcal B_{}^{\mu\nu}-\frac{1}{4}G_{\mu\nu}^aG_{}^{a\, \mu\nu}\nn\\
&&+{\rm Tr}\big[(D_\mu \phi)^\dagger D^\mu \phi\big]+{\rm Tr}\big[(D_\mu \Dt_L)^\dagger D^\mu \Dt_L\big]+{\rm Tr}\big[(D_\mu \Dt_R)^\dagger D^\mu \Dt_R\big]-V(\phi,\Dt_{L,R})\nn\\
&&-\bigg[\bar Q_L\big( \GA \phi + \TG \tilde \phi \big)Q_R +\bar L_L\big( \GA_l \phi + \TG_l \tilde \phi \big)L_R+\bar L_L^c i\tau_2\Dt_L Y_L L_L+\bar L_R^c i\tau_2\Dt_R Y_R L_R+\text{h.c.}\bigg]\, \nn\\
&&-\theta\frac{g_s^2}{32\pi^2}G_{\mu\nu}^a \tilde G^{a\, \mu\nu}-\theta_R\frac{g_R^2}{32\pi^2}W_{R\,\mu\nu}^I \tilde W_R^{I\, \mu\nu}-\theta_L\frac{g_L^2}{32\pi^2}W_{L\,\mu\nu}^I \tilde W_L^{I\, \mu\nu}-\theta_{B-L}\frac{g_{B-L}^2}{32\pi^2}\mathcal B_{\mu\nu} \mathcal {\tilde B}^{ \mu\nu}\,,
\nn\eea
where $I$ and $a$ are $SU(2)_{L,R}$ and $SU(3)_c$ indices,  $W_{L,R}^{\mu\nu}$, $\mathcal B^{\mu\nu}$, and $G^{\mu\nu}$ are the field strengths of the $SU(2)_{L,R}$, $U(1)_{B-L}$, and $SU(3)_c$ gauge groups, while $g_{L,R}$, $g_{B-L}$, and $g_s$ are their gauge couplings. Furthermore, $\psi^c = C \bar \psi^T$ indicates charge conjugation and  $\tilde \phi=\tau_2 \phi^*\tau_2$. Finally, $\theta_i$ denote the $\theta$ terms for each of the different gauge groups, where $\tilde X^{\mu\nu} = \frac{1}{2}\epsilon^{\al\bt\mu\nu}X_{\al\bt}$ with $\epsilon^{\mu\nu\al\bt}$ the completely asymmetric tensor  and $\epsilon^{0123} = +1$.
The first three lines give the kinetic terms of the fermions, the gauge fields, and the scalars, respectively. The fourth line gives the interactions of the fermions with the scalars. The last line describes the various $\theta$ terms.

The couplings $Y_{L,R}$ are symmetric $3\times 3$ matrices which  give rise to Majorana masses for the neutrinos, while the $\GA_{(l)}$ and $\tilde \GA_{(l)}$ matrices are general $3\times 3$ matrices which provide the Dirac masses of the fermions.  
We work in the basis where the $e_{L,R}$ and $u_{L,R}$ fields correspond to their mass eigenstates. The $d_{L,R}$  fields that reside in the quark doublets are then related to their mass eigenstates by $d_{L,R} = V_{L,R}d_{L,R}^{\rm mass}$, where $V_{L,R}$ are the left- and right-handed CKM matrices.

Finally, the covariant derivative is given by,
\bea
D_\mu = \partial_\mu -i g_s G_\mu^a t^a-i g_L T_L^I W_{L\, \mu}^I-i g_R T_R^I W_{R\, \mu}^I-i \frac{g_{B-L}}{2}(B-L) \mathcal B_\mu \,,
\eea
where $t^a$ and  $T_{L,R}^I$ are the generators of $SU(3)_c$ and $SU(2)_{L,R}$ in the representation of the field that $D_\mu$ works on.

Together with the Higgs potential, $V(\phi,\Dt_{L,R})$ (see e.g.\ Ref.\ \cite{Dekens:2014ina} for a detailed analysis),  Eq.\ \eqref{Lagrangian} specifies the complete model. However, since we will be integrating out the heavy new fields, we will need the Lagrangian in the broken phase, which requires the vacuum expectation values of the scalar fields.

\subsection{Symmetry breaking}
The breaking of the LR gauge group is realized by the vacuum expectation values (vevs) of the scalar fields
\begin{equation}
 \label{vevs}
\langle \phi \rangle =\sqrt{1/2} \bma \kappa &0\\0&\kappa' e^{i\al} \ema\, ,\quad \langle \Delta_{L} \rangle = \sqrt{1/2}\bma 0&0\\v_{L}e^{i\theta_L}&0\ema\, ,\quad \langle \Delta_{R} \rangle = \sqrt{1/2}\bma 0&0\\v_{R}&0\ema \, ,
\end{equation}
where all parameters are real after gauge transformations have been used to eliminate two of the possible phases \cite{Deshpande:1990ip}. 
The necessary conditions to obtain a symmetry-breaking pattern of this form have been discussed in Refs.\ \cite{BhupalDev:2018xya,Chauhan:2019fji,Maiezza:2016bzp,Chakrabortty:2013mha}.

We will assume that the $SU(3)_c\times SU(2)_L \times SU(2)_R \times U(1)_{B-L}$ gauge group is broken down to $SU(3)_c \times U(1)_{\rm QED}$ in two steps.
In the first step the vev of the right-handed triplet, $v_R$, breaks the $SU(2)_L \times SU(2)_R \times U(1)_{B-L}$ gauge group down to $SU(2)_L \times U(1)_{Y}$. This vev defines the high scale of the model, and gives the main contribution to the masses of the heavy fields: the right-handed gauge bosons, the right-handed neutrinos, and the heavy Higgs fields. 
At the electroweak scale the vevs of the bidoublet, $\ka$ and $\ka'e^{i\al}$, then break $SU(2)_L \times U(1)_{Y}$ to $U(1)_{\text{QED}}$, and are of the order of the EW scale, $\sqrt{\ka^2+\ka^{\prime\,2}}= v\simeq 246 \, \text{GeV}$. Finally,  $v_L$ contributes to the masses of the light neutrinos through the second to last term in Eq.\ \eqref{Lagrangian} and one would therefore expect that $v_L \lesssim \Or(1 \,\text{eV})$. 

The hierarchy between the different vevs allows us to describe the effects of the new heavy particles in an effective field theory in which the heavy fields are integrated out. This has the advantage of simplifying loop calculations and allows one to resum large logarithms. We will therefore integrate out the heavy BSM particles after the first step of symmetry breaking, i.e.\ after the right-handed triplet obtains its vev. We will work in the phase where the SM gauge group remains unbroken and match onto operators that are invariant under $SU(2)_L \times U(1)_{Y}$.

Before discussing this matching procedure we briefly describe the two possible discrete symmetries between left- and right-handed fields that can be implemented in LR models as well as the constraints they place on the model parameters.

\subsection{Left-right symmetries}\label{sec:LRsym}
One of the motivations for LR models is the possibility of having a symmetry between left- and right-handed particles at high energies. 
Here we discuss the two possible transformations that relate left- and right-handed fields,
\bea
&P:& \quad \begin{cases}
 Q_{L}\longleftrightarrow  Q_R \,, \qquad L_{L}\longleftrightarrow  L_R \,, \qquad \phi \longleftrightarrow \phi^{\dagger}\,, \qquad\Delta_{L}\longleftrightarrow \Delta_{R}\,,\\
 \tau\cdot W_{L\,\mu}^I\longleftrightarrow  \tau\cdot W_{R}^{I\,\mu}\,,  \qquad t^a  G_{\mu}^a\to  t^a G_{}^{a\,\mu}\,, \qquad\mathcal B_\mu\to \mathcal B^\mu\,,\end{cases}\nn\\
&C:&\quad 
\begin{cases}
Q_{L}\longleftrightarrow Q_R^c\, , \qquad L_{L}\longleftrightarrow  L_R^c \,, \qquad\phi\longleftrightarrow \phi^{T}\, ,\qquad \Delta_{L}\longleftrightarrow \Delta_{R}^*\,,\\ \tau\cdot  W_{L\,\mu}\longleftrightarrow (\tau\cdot W_{R\,\mu})^*\,,\qquad  t^a  G_{\mu}^a\to  \left(t^a G_{\mu}^{a}\right)^*\,,\qquad \mathcal B_\mu\to \mathcal B_\mu\,,
\end{cases}
\label{C&Ptransf}\eea 
where the first is related to parity and the second to charge conjugation \cite{Maiezza:2010ic}.

If either of these two transformations leaves an LR model invariant we will refer to it as left-right symmetric. Given our assumptions for the vevs of the scalar fields, such a symmetry will be broken by the vev of the right-handed triplet, $v_R$. Nevertheless, these symmetries still provide useful constraints on the model parameters.
For example, the $C$ and $P$ symmetries require the $SU(2)_{L,R}$ gauge couplings to be equal, $g_L=g_R$, at the LR scale and they restrict the number of parameters that appear in the Higgs potential.
In addition, they imply several relations between the couplings of the fermions to the scalars and, in the $P$-symmetric case,  set the $\theta_i$ terms to zero. This is summarized by
\bea\label{eq:PCconstraints}
P: \qquad \Gamma &=& \Gamma^\dagger\,,\qquad \TG = \TG^\dagger\,,\qquad Y_L=Y_R\,,\qquad \theta=\theta_i = 0\,,\nn\\
 C: \qquad \Gamma &=& \Gamma^T\,,\qquad \TG = \TG^T\,,\qquad Y_L=Y_R^\dagger\,. \label{implications}
\eea
For our purposes, the most important consequence of the above relations is their impact on the quark mass matrices, which can be written as
\bea
M_u = \sqrt{1/2}\ka( \Gamma +\xi e^{-i\al}\TG)\,,\qquad M_d = \sqrt{1/2}\ka (\xi e^{i\al} \Gamma +\TG)\,,\label{masses}
\eea
where $\xi\equiv \ka'/\ka$. Given our choice of basis the up-type mass matrix is already diagonal, $M_u = {\rm diag} (m_u,m_c,m_t)$, while the down-type mass matrix satisfies $V_L^\dagger M_d V_R= {\rm diag} (m_d,m_s,m_b)$.
From Eqs.\  \eqref{implications} and \eqref{masses}  one can see that the mass matrices become symmetric in the $C$-symmetric case, while the $P$-symmetric matrices are hermitian in the limit $\xi \sin\al\to 0$. 

In both cases these restrictions are enough to relate the right-handed CKM matrix to the left-handed one.  
In the $C$-symmetric case there is the simple relation \cite{Branco:1982wp}
\bea C:\qquad V_R = K_u V_L^* K_d\,, \label{CKMC}\eea
where $K_{u}=\text{diag}(e^{i\theta_u},e^{i\theta_c},e^{i\theta_t})$ and $K_{d}=\text{diag}(e^{i\theta_d},e^{i\theta_s},e^{i\theta_b})$ are diagonal matrices of phases, of which one combination can be set to zero, while the rest remains unconstrained. As a result, the mixing angles in both matrices will be equal. 

The $P$-symmetric case is somewhat more involved. Here the right-handed CKM matrix takes a simple form only in the limit where $\xi\sin\al\to 0$
\bea P :\qquad V_R = S_u V_L S_d\,,\qquad (\xi \sin \al=0)\,, \label{CKMP}\eea
where $S_{u,d}$ are diagonal matrices of signs, one combination of which is unphysical, such that there are $32$ solutions.
In the general $P$-symmetric case, the above relation is only approximately satisfied and acquires corrections $\sim \xi \sin\al$. These corrections can appear with ratios of the quark masses and so they are expected to be small as long as  $\xi \sin\al\ll m_b/m_t$ \cite{Senjanovic:2014pva}. 
The solution for $V_R$ has been derived in Refs.\ \cite{Senjanovic:2014pva,Senjanovic:2015yea} and expresses $V_R$ in terms of the quark masses, $V_L$, and $\xi \sin \al$. This implies that, although there are $32$ different solutions, $V_R$ does not introduce any additional model parameters in this case. The approximate expressions we use in this work are described in Appendix \ref{app:VR}.

Although both the $P$- and $C$-symmetric cases are phenomenologically viable, due to the more constrained and predictive nature of right-handed CKM matrix, we will focus on the scenario with a $P$ symmetry in what follows.

\subsection{Strong CP problem and $P$ symmetry}\label{strongCP}
In the case of a $P$ symmetry the QCD $ \theta$ term is explicitly forbidden, see Eq.\ \eqref{eq:PCconstraints}, and at scales where the parity symmetry is unbroken, we have $\theta =0$. However, after EWSB and the breaking of parity, the quark mass matrices generally obtain a phase which contributes to the physical combination $\bar \theta \equiv \theta + {\rm Arg\, Det}M_u M_d= {\rm Arg\, Det}V_R^\dagger$. This contribution is calculable \cite{Maiezza:2014ala} and to good approximation given by
\begin{equation}
\bar \theta \simeq \frac{m_t}{2 m_b} \sin \alpha \tan 2\beta \,,\qquad \tan 2\beta = \frac{2\xi}{1-\xi^2}\,.
\end{equation}
As the $\bar \theta$ term is a marginal operator, this source of CP violation is not suppressed by any ratio of scales. Using the current neutron EDM limit, $d_n < 1.8\cdot10^{-26}$ e cm \cite{Abel:2020gbr} and the lattice-QCD result $d_n =  -(1.5 \pm0.7) \cdot 10^{-16}\,\bar \theta$ e cm \cite{Dragos:2019oxn}, gives $|\bar \theta| < 1.2 \cdot 10^{-10}$. In the absence of another mechanism to account for the QCD $\bar \theta$ term (for instance through a Peccei-Quinn mechanism or by allowing for explicit parity violation in the mLRSM Lagrangian), this limit implies that 
\begin{equation}\label{eq:noPQlimit}
 \sin \alpha \tan 2\beta < 5.8\cdot 10^{-12}\,,
\end{equation}
which effectively forces $\sin \al =0$, for practical purposes.
Thus, the strong CP problem in the Standard Model, i.e. the smallness of $\bar \theta < 10^{-10}$, is transferred in the $P$-symmetric mLRSM to the requirement of setting $ \sin \alpha \tan 2\beta< 5.8\cdot 10^{-12}$ by hand. Of course, in both the SM and the mLRSM these are not really problems in the sense of inconsistencies. In fact, in both models these small parameters are technically natural implying that, once chosen small, there are no large radiative corrections that renormalize the parameters. It has been argued that the strong CP problem is therefore not a problem, see e.g. Ref.~\cite{Senjanovic:2020int}. 

Nevertheless, there is something uneasy about these small numbers. Why does nature prefer absence of CP violation in the strong sector? There seem to be no anthropic  arguments that motivate a small $\bar \theta$ \cite{Ubaldi:2008nf,Lee:2020tmi}. A popular way to dynamically remove the $\bar \theta$ term is through the Peccei-Quinn mechanism that leads to a new field, the axion, which can potentially be linked to Dark Matter. Of course, the Peccei-Quinn mechanism is an ad hoc addition to the mLRSM and it can be argued that it is less minimal than simply setting certain phases to be small by hand (Ref.~\cite{deVries:2021sxz} discusses how infrared and ultraviolet solutions can be separated using EDM experiments).

In this work, we do not wish to choose between these two approaches and therefore perform two analyses. In the first, we describe the EDM phenomenology in the mLRSM in presence of a Peccei-Quinn mechanism. In this case, EDMs are induced by flavor-conserving dimension-six operators and an interesting pattern of CP-violating observables appears. We will see that the Peccei-Quinn mechanism releases us from the requirement that $ \sin \alpha \tan 2\beta $ must be very small. This allows for a relatively light $M_{W_R}$ as potentially dangerous contributions to kaonic CP violation due to the CKM phase can be cancelled against contributions proportional to $\sin \alpha$.  In this case, we find a stringent lower bound on $M_{W_R}$ of order of a few TeV. These conclusions agree qualitatively with Ref.~\cite{Maiezza:2014ala, Bertolini:2019out}. In general the PQ mechanism in presence of additional sources of CP violation (beyond the $\bar \theta$ term) leads to CP-violating axion interactions with hadrons that can be limited by astrophysical constraints or searched for in dedicated experiments \cite{Moody:1984ba,Raffelt:2012sp,Bertolini:2020hjc,OHare:2020wah}. We do not specify the PQ mechanism and do not consider these couplings here. 

We also study the pure mLRSM with $P$ symmetry where no PQ mechanism is present. As this version of the mLRSM is more constrained, due to Eq.\ \eqref{eq:noPQlimit}, it leads to significantly stronger limits on the mass of right-handed gauge bosons.

\section{Matching and renormalization group equations}\label{sec:matching}
In this section we  integrate out the heavy fields and match onto gauge invariant operators in the SMEFT \cite{Grzadkowski:2010es}. In order to do so, we assume that the right-handed scalar triplet has obtained a vev, thereby breaking $SU(2)_R$, while $SU(2)_L\times U(1)_Y$ remains unbroken. At this stage there are several relevant heavy fields with masses $\Or(v_R)$:
\\\\\textbf{Gauge bosons:}
The breaking of $SU(2)_R$ leads to a charged and a neutral gauge boson, $W_R^\pm$ and $Z_R$, with $\Or(v_R)$ masses, which arise from the $W_R^I$ and $\mathcal B$ fields. The remaining linear combinations of the gauge fields make up the SM $SU(2)_L$ and hypercharge fields. The heavy charged bosons can be written as
\bea
W_{R\,\mu}^\pm = \frac{W_{R\,\mu}^1\mp iW_{R\, \mu}^2}{\sqrt{2}}\,,\qquad M_{W_R}\sq = \frac{1}{2}g_R\sq v_R\sq\, .
\eea 
The neutral $W_R^3$ and $\mathcal B$ bosons mix  and can be written in terms of mass eigenstates 
\bea 
&\bma W^3_{R\mu} \\\mathcal B_\mu \ema  = \bma c_R & s_R\\-s_R & c_R \ema \bma   Z_{R\mu} \\B_{\mu} \ema \,,\qquad s_R = \frac{g_{B-L}}{\sqrt{g_{B-L}^2+g_R\sq}}\,,\qquad  c_R = \frac{g_R}{\sqrt{g_{B-L}^{ 2}+g_R\sq}}\,,&\nn\\
&M_{B_{}}\sq =0\,,\qquad M_{Z_R}\sq = v_R\sq(g_{B-L}^{ 2}+g_R\sq)\,,&
\eea
where $B_{\mu} $ is the hypercharge field of the SM. This field then couples to hypercharge, $Y = Q-T_L^3 =\frac{B-L}{2}+T_R^3 $, with gauge coupling $g' = s_R g_R = c_R g_{B-L}$. The $W_L^I$ fields stay massless as well implying that, after integrating out the heavy gauge fields, the covariant derivative reduces to that of the SM, $D_\mu = \partial_\mu -i g_s G_\mu^a t^a-i g_L T_L^I W_{L\, \mu}^I-ig' Y B_\mu$, where $g = g_L = g_R$.
\\\\\textbf{Scalar $SU(2)_L$ doublet:}
After $\Dt_R$ acquires a vev, the bi-doublet $\phi$ can be written in terms of two $SU(2)_L$ doublets,  $\phi = (\phi_1,\, \phi_2)$, of which one linear combination obtains an $\Or(v_R)$ mass. The relation to the mass eigenstates is~\footnote{The appearance of the vevs of the bi-doublet through $\xi=s_\bt/c_\bt$  in Eq.\ \eqref{eq:HiggsRot} might be somewhat surprising as we are working in the unbroken phase of $SU(2)_L$ and $\phi$ has not acquired a vev yet. In principle, Eq.\ \eqref{eq:HiggsRot} can be written in terms of the parameters in the Higgs potential and $v_R$ alone. However, the parameters of the Higgs potential can be eliminated in favor of  $\xi$ by use of the minimum equations, see App.\ \ref{app:higgses} for details.}
\bea\bma \tilde \phi_1\\\phi_2 \ema = 
\bma -c_\bt & s_\bt e^{-i\al}\\ s_\bt e^{i\al} & c_\bt
\ema
\bma \vp\\ \vp_H\ema \,,\qquad 
M\sq_{\vp_{}}=0\,,\qquad M\sq_{{H}}=\frac{\al_3  v_R\sq}{2}\frac{1+\xi\sq}{1-\xi\sq}\,,
\label{eq:HiggsRot}
\eea
where the mixing angles are given by $s_\bt = \sin\bt$, $c_\bt=\cos\bt$, and $t_\bt = \tan\beta = \xi$, while $\vp_H$ is the heavy doublet, $\vp$ is the SM Higgs doublet, and  $\al_3$ is a parameter in the Higgs potential, in the notation of Ref.\ \cite{Dekens:2014ina}. 
\newline

In addition to the heavy states mentioned above, the right-handed neutrinos obtain an $\Or(v_R)$ Majorana mass while the right-handed triplet, $\Dt_R$, gives rise to a heavy doubly-charged and a heavy neutral scalar,  $\dt_R^{++}$ and Re$\,\dt_R^0$, respectively~\footnote{The remaining components of $\Dt_R$, namely $\dt_R^+$ and  Im$\, \dt_R^0$, are the would-be-Goldstone bosons that are eaten by the $W_R^\pm$ and $Z_R$ fields, see App.\ \ref{app:higgses} for more details.}. However, since these fields mainly couple to the leptons and scalar fields they have a limited effect on observables that probe the couplings to quarks. We therefore do not pursue the effects of the $\nu_R$, $\dt_R^{++}$, and Re$\,\dt_R^0$ fields, and focus on the matching conditions that arise from integrating out the $W_R^\pm$, $Z_R$, and $\vp_H$ fields.

\subsection{Matching conditions at $\mu=M_{W_R}$}\label{sec:SMEFTmatching}
To obtain the matching conditions, we integrate out the heavy fields and work up to dimension six in the EFT, i.e.\ we keep terms that are suppressed by up to two powers of the high scale. All the heavy fields are integrated out at a common scale which we take to be $\mu = M_{W_R}$. Since $SU(2)_R$ is explicitly broken at this stage, we now move to the mass basis for the right-handed down-type quarks. This is achieved by a rotation of the right-handed down-type quarks, $d_R\to V_R d_R$. 
The relevant interactions that receive matching contributions are a right-handed charged current, $C_{Hud}$, as well as several four-quark operators~\footnote{We have chosen a basis of dimension-six operators that is most convenient for our calculations. The comparison to the standard Warsaw basis is given in App.~\ref{app:warsaw}.}
\begin{eqnarray}\label{eq:LagSMEFT}
\mathcal L &=& 
\left(C^{ij}_{Hud}\tilde \vp^\dagger iD_\mu \vp\, \bar u^i \ga^\mu d^j+{\rm h.c.}\right)\nn\\
&&-  C^{ij\, lm}_{1\, RR}\, \bar d^i \gamma^\mu  u^j \, \bar u^l \gamma_\mu  d^m  - C^{ij\, lm}_{2\, RR}\, \bar d^i_{\alpha} \gamma^\mu  u^j_{\beta} \, \bar u^l_{\beta} \gamma_\mu  d^m_{\alpha}  \nn \\
& & + C^{ij\, lm}_{1, qd}\,  \bar q^i \gamma^{\mu} q^j\, \bar d^l \gamma_{\mu} d^m +   C^{ij\, lm}_{2, qd}\,  \bar q^i_{ \alpha} \gamma^{\mu} q^j_{ \beta}\, \bar d^l_{ \beta} \gamma_{\mu} d^m_{\, \alpha} \nn \\
&&+ C^{ij\, lm}_{1, qu}\,  \bar q^i \gamma^{\mu} q^j\, \bar u^l \gamma_{\mu} u^m +   C^{ij\, lm}_{2, qu}\,  \bar q^i_{ \alpha} \gamma^{\mu} q^j_{ \beta}\, \bar u^l_{ \beta} \gamma_{\mu} u^m_{\, \alpha} \nn \\
& & + \left(C^{ij\, lm}_{1, quqd}\, \varepsilon^{ab}\bar q^{i}_a u^{j} \, \bar q^{l}_b d^{m}   + C^{ij\, lm}_{2, quqd}\, \varepsilon^{ab}\bar q^{i}_{a \alpha} u^{j}_{\beta} \, \bar q^{l}_{b \beta} d^{m}_\alpha   +{\rm h.c.}\right)\,,
\end{eqnarray}
where $q=(u_L,\, d_L)^T$ denotes the doublet of left-handed fields, $d=d_R$ and $u=u_R$ denote right-handed fields for up- and down-type quarks, $i, \ldots, m$ are flavor indices, and  $\alpha$ and $\beta$ are color indices. 
The Wilson coefficients at the scale $\mu = M_{W_R}$ are given by
\begin{eqnarray}\label{eq:mHmatching}
C_{Hud}^{ij} &=&\frac{g_R\sq}{M_{W_R}\sq} \frac{\xi e^{i\al}}{1+\xi\sq}V_{R,\,ij}\,,\nn\\
C^{ij\, lm}_{1\, RR} &=& \frac{g^2_R}{2 M_{W_R}^2} V^*_{R, \, ji} V_{R,\, lm}\,, \qquad C^{ij\, lm}_{2\, RR}  = 0\,,\nn\\
C^{ij\, lm}_{1, qu} &=& 0\,, \qquad  C_{2, qu}^{i j\, lm} =-\frac{1}{2}Y_{uH}^{im}\left(Y_{uH}^*\right)^{jl} \left[\frac{1}{M_H\sq}+\frac{g_R^2}{32\pi^2 M_{W_R}\sq}\left(\frac{1-\xi^2}{1+\xi^2}\right)^2\left(\ln \frac{M_H^2}{M_{W_R}^2} -1\right)\right],\nn  \\
C^{ij\, lm}_{1, qd} &=& 0\,, \qquad  C_{2, qd}^{i j\, lm} =-\frac{1}{2}Y_{dH}^{im}\left(Y_{dH}^*\right)^{jl} \left[\frac{1}{M_H\sq}+\frac{g_R^2}{32\pi^2 M_{W_R}\sq}\left(\frac{1-\xi^2}{1+\xi^2}\right)^2\left(\ln \frac{M_H^2}{M_{W_R}^2} -1\right) \right], \nn  \\
C^{ij\, lm}_{1, quqd} &=& \frac{1}{M_H\sq}Y_{uH}^{ij}Y_{dH}^{lm}\, ,  \qquad C^{ij,\, lm}_{2, quqd} = 0\,,
\end{eqnarray}
where $Y_{uH,dH}$ are the Yukawa couplings of $\vp_H$,
\bea\label{eq:Hyukawa}
Y_{uH} = \frac{\sqrt{2}}{v}\frac{M_d(1+\xi\sq)-2\xi e^{i\al}M_u}{1-\xi\sq}\,,\qquad 
Y_{dH} =\frac{\sqrt{2}}{v}\frac{M_u(1+\xi\sq)-2\xi e^{-i\al}M_d}{1-\xi\sq}V_R\,.
\eea

The $C_{qd,qu}$  Wilson coefficients are important as they mediate $\Delta F = 2$ processes at low energies. They are generated by  tree-level $\vp_H$ exchange, and, at scales below $\mu = M_{W_R}$, by loop diagrams induced by $W_R$ interactions. Both types of contributions are phenomenologically relevant, as $M_H$ tends to be heavier than $M_{W_R}$. For this reason we work at tree-level for the contributions  $\sim M_H^{-2}$, while keeping loop-level contributions proportional to $(4\pi)^{-2}M_{W_R}^{-2}$. In particular, we include corrections to $C_{qu,qd}$ in Eq.\ \eqref{eq:mHmatching} scaling as $(4\pi M_{W_R})^{-2}$ that arise from self-energy graphs for $\vp_H$~\footnote{As discussed in Refs.~\cite{Chang:1984hr,Basecq:1985cr,Ecker:1985vv}, only the combination of these diagrams with box diagrams involving $W_L$ and $W_R$ bosons gives a gauge-invariant result.}, while dropping loop diagrams involving $\vp_H$ that scale $\sim (4\pi M_H)^{-2}$. 
The same approximation is used for $M_H$ in the above expressions, were we include
loop contributions due to $W_R$ interactions that are enhanced by factors of $(M_H/M_{W_R})^2$. This implies that $M_H$  corresponds to the one-loop expression for the physical Higgs mass up-to-and-including potentially large $\sim (M_H^4)/(4\pi M_{W_R})^2$ terms, but misses loop contributions without the $M_H^2/M_{W_R}^2$ enhancement, $M_H^2 = M_{H,\,\rm phys}^2\left[1+\Or\left((4\pi)^{-2}\right)\right]$.

For the loop contributions to $\Delta F=2$ operators from diagrams involving $W_L$ and $W_R$ bosons, we find that they are cancelled by those in the EFT when performing the matching at $\mu=M_{W_R}$. The finite parts of this result in principle depend on the scheme and the treatment of evanescent operators, which appear for the four-fermion interactions and impact the way Dirac structures are reduced to our basis of operators~\footnote{This scheme dependence in the matching is removed when computing physical matrix elements in the EFT.}. 
We employ $\overline{\rm MS}$ throughout our calculations, however, for the evanescent terms, we adopt a scheme in which their contributions are compensated by local counterterms \cite{Buras:1989xd,Dugan:1990df,Herrlich:1994kh}.
In particular, in the evaluation of box diagrams we use the relation $\ga_\mu\ga_\nu P_L\otimes \ga_\nu \ga_\mu P_R = d \, P_L\otimes P_R - E^{(2)}_{LR}$ to reduce the Dirac structures we encounter, where $E^{(2)}_{LR}$ is the evanescent operator that defines our scheme. We subsequently use the following Fierz identity, $(\bar\psi_1\ga_\mu P_L\psi_2) (\bar\psi_3\ga^\mu P_R\psi_4)  = -2( \bar \psi_1P_R\psi_4)(\bar \psi_3 P_L\psi_2) - E_{LR}^{(F1)}$ to further reduce the loop contributions to our basis of operators. This scheme is equivalent to that of Ref.\ \cite{Dekens:2019ept}, with $a_{\rm ev} = -1/2$.

When evolving the Lagrangian in Eq.\ \eqref{eq:LagSMEFT} from $M_{W_R}$ to the electroweak scale, the dipole operators are induced by the $C_{1,2\, quqd}$ coefficients. These dipole interactions can be written in an $SU(2)_L$-invariant form as follows
\bea\label{eq:dipLag}
\vL_{\rm dip} &=& -\frac{g'}{\sqrt{2}} \bar q\simu B_{\mu\nu}\GA_B^uu \,\tilde\vp -\frac{g'}{\sqrt{2}} \bar q\simu B_{\mu\nu}\GA_B^dd \,\vp\nn\\
&&  -\frac{g}{\sqrt{2}} \bar q\simu W^I_{L,\, \mu\nu}\tau^I \GA_W^uu \,\tilde\vp -\frac{g}{\sqrt{2}} \bar q\simu W^I_{L,\,\mu\nu}\tau^I\GA_W^dd \,\vp\nn\\
&&-\frac{g_s}{\sqrt{2}} \bar q\simu G^a_{\mu\nu}t^a\GA_g^uu \,\tilde\vp -\frac{g_s}{\sqrt{2}} \bar q\simu G^a_{\mu\nu}t^a\GA_g^dd \,\vp+{\rm h.c.}\,\,,
\eea
at low energies, the off-diagonal components of these interactions significantly contribute to $\Delta F=1$ observables, while the diagonal components give rise to EDMs.
It is useful to define the following combinations of the $\GA_{W,B,g}^{u,d}$ couplings,
\begin{align}\label{dipoles}
\frac{Q_u m_{u_j}}{v} C_{\g u}^{ij} &= -\left(\GA^u_B +\GA^u_W\right)^{ij}\,,\qquad &\frac{Q_d m_{d_j}}{v} C_{\g d}^{ij} &= -\left(V_L^\dagger \GA^d_B -V_L^\dagger \GA^d_W\right)^{ij}\,,\nn\\
\frac{m_{u_j}}{v} C_{g u}^{ij}  &= \left(\GA^u_g\right)^{ij}\,,\qquad &\frac{m_{d_j}}{v} C_{g d}^{ij} &= \left(V_L^\dagger\GA^d_g\right)^{ij}\,,
\end{align}
where $Q_u$ and $Q_d$ are the electric charges of the quarks and $C_{\g d,\, \g u}$ are the combinations that will give rise to the electromagnetic dipole moments of the quarks after electroweak symmetry breaking, while $C_{g d,\, gu}$ are the gluonic dipole moments. We introduced a CKM factor in the couplings for the down-type operators in anticipation of a later rotation to the mass basis.

\subsection{Renormalization group equations below $M_{W_R}$}\label{sec:SMEFTrge}
The evolution of the effective Lagrangian from $\mu=M_{W_R}$ to the electroweak scale requires the renormalization group equations (RGEs).
For the four-quark operators these take the form \cite{Buras:2001ra,Alonso:2013hga}
\bea\label{eq:4fermiRGE}
\frac{d}{d\ln\mu}\vec C^{ijlm} = 
\bma
\frac{\al_s}{4\pi}\g_{RR} &0&0&0\\
\frac{1}{(4\pi)^2}\g_{EW}^D& \frac{\al_s}{4\pi}\g_{LR}&0&0\\
\frac{1}{(4\pi)^2}\g_{EW}^U&0& \frac{\al_s}{4\pi}\g_{LR}&0\\
0&0&0& \frac{\al_s}{4\pi}\g_{LRLR}
\ema ^{ijlm}_{abcd} \cdot \vec C^{abcd} \,,\eea
where $\vec C^T = (C_{1RR},\, C_{2RR},\, C_{1,qd},\, C_{2,qd}, C_{1,qu},\, C_{2,qu}, C_{1,quqd}, C_{2,quqd})$.
The diagonal terms describe one-loop QCD corrections. The $\g_{RR}$ and $\g_{LR}$ terms are diagonal in generation indices 
\bea
\g_{RR} =  \dt_{ai}\dt_{bj}\dt_{cl}\dt_{dm}\bma -6/N_c &6\\ 6 & -6/N_c\ema \,,\quad \g_{LR} =  \dt_{ai}\dt_{bj}\dt_{cl}\dt_{dm}\bma 6/N_c &0\\ -6 & -6\frac{N_c\sq-1}{N_c}\ema \,,
\eea
where $N_c = 3$ is the number of colors.
For the operators with  $(\bar LR)(\bar LR)$ chiralities the anomalous dimensions are
\bea
\g_{LRLR} = \bma 2/N_c-6N_c &-4\\4 & 2/N_c+2N_c\ema \dt_{ai}\dt_{bj}\dt_{cl}\dt_{dm}+ \bma -8&8/N_c-4N_c\\8/N_c & -4\ema \dt_{ci}\dt_{bj}\dt_{al}\dt_{dm}\,.
\eea

The operators, $C_{i\, RR}$, contribute to $C_{i\,qd}$  through electroweak loops captured by $\g_{EW}^{U,D}$
\bea
\g_{EW}^D&=&\frac{2}{  v\sq}\dt_{al}\dt_{dm}(M_u^\dagger)_{bj}(M_u)_{ic}\bma 0& 1\\ 1& 0\ema \,,\qquad
\g_{EW}^U=\frac{2}{v\sq}\dt_{cl}\dt_{bm}(V_R^\dagger M_d^\dagger)_{dj}(M_d V_R)_{ia}\bma 0& 1\\ 1& 0\ema  \, .\nn\label{eq:RGE}
\eea

The dipole operators are induced through the following RGEs \cite{Degrassi:2005zd,Dekens:2013zca,Hisano1}
\bea
\frac{d}{d\ln\mu}(C_{\g u}^{ij},\, C_{g u}^{ij})^T &=&\frac{\al_s}{4\pi}\g_{\rm dip} \cdot\bma C_{\g u}^{ij}\\ C_{g u}^{ij}\ema+
\frac{1}{(4\pi)\sq}\sum_{k\in d,s,b} \frac{(V_R^\dagger M_{d}^\dagger)_{lk}}{m_{u_j}}
\g^u_{quqd}\cdot \bma C_{1\, quqd}^{kjil} \\ C_{2\, quqd}^{kjik}\ema\, ,\nn\\
\frac{d}{d\ln\mu}(C_{\g d}^{ij},\, C_{g d}^{ij})^T &=& \frac{\al_s}{4\pi}\g_{\rm dip}\cdot \bma C_{\g d}^{ij}\\ C_{g d}^{ij}\ema+
\frac{1}{(4\pi)\sq}\sum_{k\in u,c} \frac{m_{u_k}}{m_{d_j}}
\g^d_{quqd} \cdot \bma V_{L,\, il}C_{1\, quqd}^{lkkj} \\  V_{L,\, il}C_{2\, quqd}^{lkkj}\ema \,,\label{eq:dipoleRG}
\eea
where 
\bea
\g_{\rm dip} = \bma 8C_F &-8C_F\\0&16C_F-4N_c\ema\, ,\qquad \g^d_{quqd}= \bma 2\frac{Q_u}{Q_d} & 2N_c \frac{Q_u}{Q_d}\\
-2 & -4 C_F \ema \,,
\eea
where $C_F = \frac{N_c^2-1}{2N_c}$ and $\g^u_{quqd}$ can be obtained from $\g^d_{quqd}$ by $Q_u\leftrightarrow Q_d$.

Finally, the $C_{Hud}$ operator does not evolve under QCD.

\subsection{Matching at $\mu=m_{W}$}\label{sec:LEFTmatching}
After evolving the effective operators in Eq.\ \eqref{eq:LagSMEFT} to the electroweak scale we integrate out the top quark as well as the Higgs, $W$, and $Z$ bosons. Because $SU(2)_L$ has now been broken, we move to the mass basis of the left-handed down-type quarks. This can be achieved by the following flavor rotation, $d_L\to V_L d_L$, so that the left-handed quark doublet becomes, $q = (u_L,\, V_L d_L)^T$.
The relevant four-fermion operators below the electroweak scale can be written as
\begin{eqnarray}\label{eq:LowLag}
\mathcal L &=& 
-  C^{ij\, lm}_{1\, LL}\, \bar d_L^i \gamma^\mu  u_L^j \, \bar u_L^l \gamma_\mu  d_L^m  - C^{ij\, lm}_{2\, LL}\, \bar d^i_{L\,\alpha} \gamma^\mu  u^j_{L\,\beta} \, \bar u^l_{L\,\beta} \gamma_\mu  d^m_{L\,\alpha} \nn\\
&&- \left( C^{ij\, lm}_{1\, LR}\, \bar d_L^i \gamma^\mu  u_L^j \, \bar u_R^l \gamma_\mu  d_R^m  + C^{ij\, lm}_{2\, LR}\, \bar d^i_{L\,\alpha} \gamma^\mu  u^j_{L\,\beta} \, \bar u^l_{R\,\beta} \gamma_\mu  d^m_{R\,\alpha}+{\rm h.c.}\right) \nn\\
&&-  C^{ij\, lm}_{1\, RR}\, \bar d_R^i \gamma^\mu  u_R^j \, \bar u_R^l \gamma_\mu  d_R^m  - C^{ij\, lm}_{2\, RR}\, \bar d^i_{R\,\alpha} \gamma^\mu  u^j_{R\,\beta} \, \bar u^l_{R\,\beta} \gamma_\mu  d^m_{R\,\alpha} \nn\\
&&+C^{ij\, lm}_{4}  \bar d_L^{i} \gamma^\mu d^j_L\,   \bar d_R^{l} \gamma^\mu d^m_R +  C^{ij\, lm}_{5}  \bar d_{L\, \alpha}^{i} \gamma^\mu d^j_{L\,\beta}\,   
\bar d_{R \,\beta}^{l} \gamma^\mu d^m_{R\, \alpha}\nn\\
& & + \left(C^{ij\, lm}_{1, quqd}\, \varepsilon^{ab}\bar q^{i}_a u_R^{j} \, \bar q^{l}_b d^{m}_R   + C^{ij\, lm}_{2, quqd}\, \varepsilon^{ab}\bar q^{i}_{a \alpha} u^{j}_{R\, \beta} \, \bar q^{l}_{b \beta} d^{m}_{R\, \alpha}   +{\rm h.c.}\right)\nn\\
&&+C_{\tilde G}\frac{g_s}{6}f_{abc}\epsilon^{\al\bt\mu\nu}G^a_{\al\bt}G_{\mu\rho}^bG^{c\,\rho}_{\nu}\,.
\end{eqnarray}
Most of the above operators have a similar form to the $SU(2)_L$-invariant ones in Eq.\ \eqref{eq:LagSMEFT}, apart from those in the first, second, and last lines. Those in the first two lines are additional four-quark operators, generated by the SM and $C_{Hud}$, while the last line describes the so-called Weinberg operator, which is induced  through one-loop diagrams. 

The dipole operators take the following form below the electroweak scale
\bea\label{eq:Lagdip}
\vL_{\rm dip} &=& -\frac{eQ_u}{2}\sum_{i,j\in u,c}m_{u_j}C_{\g u}^{ij}\bar q_L^i\simu F_{\mu\nu}q_R^j -\frac{eQ_d}{2}\sum_{i,j\in d,s,b}m_{d_j}C_{\g d}^{ij}\bar q_L^i\simu F_{\mu\nu}q_R^j\nn\\
&& -\frac{g_s}{2}\sum_{i,j\in u,c}m_{u_j}C_{g u}^{ij}\bar q_L^i\simu G^a_{\mu\nu}t^aq_R^j -\frac{g_s}{2}\sum_{i,j\in d,s,b}m_{d_j}C_{g d}^{ij}\bar q_L^i\simu G^a_{\mu\nu}t^aq_R^j+{\rm h.c.}
\eea

The tree-level matching leads to
\bea
C_{1LL}^{ijlm} &=& 2\sqrt{2} G_F \left(V_L^\dagger\right)^{ij} \left(V_L\right)^{lm},\qquad C_{2LL}=0\,,\nn\\
C_{1LR}^{ijlm} &=&  \left(V_L^\dagger\right)^{ij} \left(C_{Hud}\right)^{lm},\qquad C_{2LR}=0\,,
\eea
while the coefficients of the remaining four-quark operators, $C_{i\,RR}$ and $C_{i\,quqd}$, are unaffected at the $W_L$ threshold.
$C_4$ and $C_5$ get a tree-level contribution from $C_{qd}$, as well as a contribution from loop diagrams involving $C^{ij\, lm}_{1,2\, RR}$ and $W_L$ exchange
\begin{eqnarray}\label{eq:matchMw}
C^{ij\, lm}_4(m_W) &=&  V^*_{L\, ai}V_{L\, bj}C^{ab\, lm}_{1, qd}(m_W)\nn\\
&& +\frac{g_L\sq}{4(4\pi)\sq}\,
 C^{lt\, tm}_{2\, RR} V^*_{L\, ti} V_{L\, tj}   x_t \left(\ln m_W\sq/\mu\sq +\frac{3}{1-x_t}+\frac{(4+(x_t-2)x_t)\ln x_t}{(1-x_t)\sq}\right)\nn\\
 &&+\frac{g_L\sq}{4(4\pi)\sq}\,
\left( C^{lc\, tm}_{2\, RR} V^*_{L\, ti} V_{L\, cj}+C^{lt\, cm}_{2\, RR} V^*_{L\, ci} V_{L\, tj}\right)\nn\\
&&\times \sqrt{x_c x_t}\frac{\left(\ln m_W\sq/\mu\sq -1\right)\left(x_t-1\right) +(x_t-4)\ln x_t}{x_t-1}\nn\\
 &&+\frac{g_L\sq}{4(4\pi)\sq}\,
 C^{lc\, cm}_{2\, RR} V^*_{L\, ci} V_{L\, cj}x_c\left(1-3\ln m_W\sq/\mu\sq\right)\,,
\end{eqnarray}
where $x_i= m_i^2/m_W^2$. The first, second and third contributions result from diagrams involving an internal $t-t$, $t-c$, and $c-c$ pair, respectively, and we dealt with the appearance of evanescent operators as described above Eq.\ \eqref{eq:dipLag}.
A similar equation holds for $C_{5}$, with the replacements, $C_{2\, RR}\to C_{1\, RR}$ and $C_{1\, qd}\to C_{2\, qd}$.

Finally, one-loop contributions to the Weinberg operator \cite{BraatenPRL,Boyd:1990bx} and dipole moments \cite{Cho:1993zb} appear
\bea\label{eq:matchMw2}
C_{\g u}^{ij}(m_{W}^-) &=&C_{\g u}^{ij}(m_{W}^+) + \frac{2}{(4\pi)^2} \sum_{k=d,s,b}\frac{m_{d_k}}{m_{u_j}Q_u}V_{L}^{ik}C_{Hud}^{*\, jk}\,,\nn\\
C_{g u}^{ij}(m_{W}^-)  &=& C_{g u}^{ij}(m_{W}^+)\,,\nn\\
C_{\g d}^{ij}(m_{W}^-)  &=& C_{\g d}^{ij}(m_{W}^+)+\frac{2}{(4\pi)^2} \sum_{k=u,c}\frac{m_{u_k}}{m_{d_j}Q_d}V_{L}^{*\, ki}C_{Hud}^{ kj}
\nn\\
&&+\frac{1}{(4\pi)^2}\frac{m_{t}}{m_{d_j}Q_d}V_{L}^{*\, ti}C_{Hud}^{ tj}\left[Q_u f_W(x_t)+g_W(x_t)\right]\,,\nn\\
C_{g d}^{ij}(m_{W}^-)  &=&C_{g d}^{ij}(m_{W}^+)    -\frac{1}{(4\pi)^2}\frac{m_{t}}{m_{d_j}}V_{L}^{*\, ti}C_{Hud}^{ tj} f_W(x_t)\,,\nn\\
C_{\tilde G}(m_{W}^-)  &=& C_{\tilde G}(m_{W}^+)  -\frac{\al_s}{8\pi}{\rm Im}\,C_{gu}^{(tt)}\,,
\eea
where $x_t = m_t^2/m_W^2$ and
\bea
f_W(x) = \frac{x^3+3x-4-6x\ln x}{2(x-1)^3}\,,\qquad g_W(x) = \frac{4+x(x-11)}{2(x-1)^2}+3\frac{x^2\ln x}{(x-1)^3}\,.
\eea

\subsection{Renormalization group equations below $\mu=m_{W}$}\label{sec:LEFTrge}

Below $\mu=m_{W}$, the QCD running for the relevant four-quark operators is equivalent to the running above the electroweak scale; the $C_{1\, LL}$ and $C_{2\, LL}$ coefficients follow the same RGEs as $C_{1\, RR}$ and $C_{2\, RR}$, while the RGEs of $C_{1\, LR}$ and $C_{2\, LR}$ (and $C_4$ and $C_5$) correspond to those of $C_{1\, qd}$ and $C_{2\, qd}$. The running of $C_{1,2 quqd}$ and $C_{1,2\, RR}$ is unchanged below  $\mu=m_{W}$.

Instead, the mixing of the $C_{i\, RR}$ operators with $C_{4,5}$ operators changes from Eq.\ \eqref{eq:4fermiRGE} to
\bea
\frac{d C_4^{abcd}}{d\ln \mu} &=& \frac{1}{4\pi\sq}m_{u_i}m_{u_j}\left[ C_{1LL}^{aij b}C_{2RR}^{c ji d}+C_{2LL}^{aij b}C_{1RR}^{c ji d}+N_c C_{2LL}^{aij b}C_{2RR}^{c ji d}\right]\,,\nn\\
\frac{d C_5^{abcd}}{d\ln \mu} &=& \frac{1}{4\pi\sq}m_{u_i}m_{u_j} C_{1LL}^{aij b}C_{1RR}^{c ji d}\,.
\eea

The RGEs for the flavor-diagonal dipole operators must be extended to include the Weinberg operator. The QCD part of the RGEs becomes 
\bea
\frac{d}{d\ln\mu}\vec C_{\rm dip} &=& \frac{\al_s}{4\pi}\g_{\rm dip}' \vec C_{\rm dip}\,,
\eea
with $\vec C_{\rm dip}=({\rm Im}\,C_{\g q}^{(qq)},\, {\rm Im}\,C_{g q}^{(qq)},C_{\tilde G})^T$, and \cite{Wilczek:1976ry,Weinberg:1989dx,BraatenPRL}
\bea
\g_{\rm dip}' =  \bma 8C_F &-8C_F&0\\0&16C_F-4N_c &2N_c\\0&0&N_c +2n_f+\bt_0\ema \,,
\eea
where $\bt_0 =(11 N_c-2 n_f)/3$, with $n_f$ the number of active flavors. The $C_{1,2\, LR}$ coefficients also contribute to dipole operators, which is captured by 
\bea
\frac{d}{d\ln\mu}(C_{\g u}^{ij},\, C_{g u}^{ij})^T &=&
\frac{\al_s}{4\pi}\sum_{k\in d,s,b} \frac{m_{d_k}}{m_{u_j}}
\g^u_{LR}\cdot \bma C_{1\,LR}^{jkik} \\ C_{2\, LR}^{jkik}\ema^* +\dots\,\,,\nn\\
\frac{d}{d\ln\mu}(C_{\g d}^{ij},\, C_{g d}^{ij})^T &=&
\frac{\al_s}{4\pi}\sum_{k\in u,c} \frac{m_{u_k}}{m_{d_j}}
\g^d_{LR} \cdot \bma C_{1\, LR}^{kjki} \\ C_{2\, LR}^{kjki}\ema +\dots \,\,,
\eea
where the dots stand for the additional terms on the right-hand side of Eq.\ \eqref{eq:dipoleRG}, and \cite{Cho:1993zb}
\bea
 \g^d_{LR}=\frac{1}{(4\pi)\sq} \bma 32\frac{Q_u}{Q_d}+\frac{64}{3} &160 \frac{Q_u}{Q_d}\\
-\frac{16}{3} & 8 \ema \,.
\eea
 $\g^u_{LR}$ can be obtained from $\g^d_{LR}$ by $Q_u\leftrightarrow Q_d$.

\subsection{Matching contributions below $\mu=m_{W}$}\label{sec:LEFTlow}
Below the electroweak scale we integrate out the bottom and charm quarks at the respective mass scales. At the bottom threshold this gives rise to matching contributions to the Weinberg operator and the dipole moments of the up-type quarks
\bea
C_{\g u}^{ij}(m_b^-)  &=& C_{\g u}^{ij}(m_b^+)+\frac{1}{8\pi^2}\frac{Q_d m_b }{Q_um_{u_j }}\left[C_{1\,LR}^{bijb}+N_c C_{2\,LR}^{bijb}\right]^*\,,\nn\\
C_{g u}^{ij}(m_b^-)  &=& C_{g u}^{ij}(m_b^+)-\frac{1}{8\pi^2}\frac{m_b }{m_{u_j }}\left[C_{1\,LR}^{bijb}\right]^*\,,\nn\\
C_{\tilde G}(m_b^-)  &=& C_{\tilde G}(m_b^+)  -\frac{\al_s}{8\pi}{\rm Im}\,C_{gd}^{(bb)}\,.
\eea
Similarly, at the charm threshold we obtain the following contributions
\bea
C_{\g d}^{ij}(m_c^-)  &=& C_{\g d}^{ij}(m_c^+)+\frac{1}{8\pi^2}\frac{Q_u m_c }{Q_dm_{d_j }}\left[C_{1\,LR}^{iccj}+N_c C_{2\,LR}^{iccj}\right]\,,\nn\\
C_{g d}^{ij}(m_c^-)  &=& C_{g d}^{ij}(m_c^+)-\frac{1}{8\pi^2}\frac{m_c }{m_{d_j }}C_{1\,LR}^{iccj}\,,\nn\\
C_{\tilde G}(m_c^-)  &=& C_{\tilde G}(m_c^+)  -\frac{\al_s}{8\pi}{\rm Im}\,C_{gu}^{(cc)}\,.
\eea
Finally, at $\mu=m_c$ we find the following matching contributions to the $C_{4,5}$ coefficients
\bea\label{eq:matchmc}
C_4^{ijlm}(m_c) &=&C_4(m_c^+)+ \frac{m_c\sq}{(4\pi)\sq}\left(1+2\ln m_W\sq/\mu\sq +2\ln x_c\right)\,\left[C_{1LL}^{i cc j}C_{2RR}^{l cc m}+C_{2LL}^{i cc j}C_{1RR}^{l cc m}+N_cC_{2LL}^{i cc j}C_{2RR}^{l cc m}\right]\,,
\nn\\
C_5^{ijlm}(m_c) &=&C_5(m_c^+)+ \frac{m_c\sq}{(4\pi)\sq}\left(1+2\ln m_W\sq/\mu\sq +2\ln x_c\right)\,C_{1LL}^{i cc j}C_{1RR}^{l cc m}\,.
\eea
At low energies the $C_{4,5}$ coefficients mediate $\Delta F=2$ processes. Working at fixed, one-loop order and collecting the matching contributions at $\mu = M_H,M_{W_R}, m_W$ and $m_c$, as well as the electroweak running contributions in between these thresholds, we reproduce the expressions in Ref.\ \cite{Bertolini:2014sua}, up to terms $\sim \Or(\frac{1}{(4\pi)^2M_H^2})$ that we neglect as explained below Eq.\ \eqref{eq:Hyukawa}.

In our analysis we include QCD corrections by solving the RGEs of the four-quark operators 
thereby evolving their Wilson coefficients from one threshold to the next.
Formally, our approach is then accurate up to leading-log precision. I.e.\ it takes into account terms of order $\frac{1}{(4\pi)^2}\ln\times \left(\frac{\al_s}{4\pi}\right)^n\ln^{n}$, but does not include all of those at order $\frac{1}{(4\pi)^2}\ln\times\left(\frac{\al_s}{4\pi}\right)^n\ln^{n-1}$. 
Some of these terms are included in our matching equations, e.g.\ through the non-log terms in Eqs.\ \eqref{eq:matchMw} and \eqref{eq:matchmc}, but we neglected contributions at the same order that would arise from  two-loop matching at the different thresholds. 
In the same way we include the leading-log contributions to the dipole operators, $C_{\g q}$ and $C_{g q}$.

This strategy is similar to the one followed in Refs.\ \cite{Maiezza:2010ic,Bertolini:2014sua} for the contributions to $C_{4,5}$ mediated by $t-t$ graphs, but differs somewhat for those with intermediate $c-t$ or $c-c$ quarks. For the latter, 
Ref.\ \cite{Bertolini:2014sua} employed the approach outlined in Refs.\ \cite{Ecker:1985vv,Vysotsky:1979tu}, which is not guaranteed to reproduce a leading-log approximation. We discuss the impact of these differences when considering $\Dt F=2$ observables in Sect.\ \ref{sec:DF2}.

\subsection{Summary}\label{sect:RGEsummary}

Using the matching conditions in Sections \ref{sec:SMEFTmatching}, \ref{sec:LEFTmatching} and \ref{sec:LEFTlow}, and the RGEs in Sections 
\ref{sec:SMEFTrge} and \ref{sec:LEFTrge}, 
we can finally give approximate expressions for the LEFT coefficients 
at the scales relevant to low-energy observables. 
Assuming the initial scale $\mu_0$ is $\mu_0 =$ 10 TeV, 
we obtain the following numerical values for the charged-current four-quark operators at $\mu_{\rm low} = 2$ GeV,
\bea\label{eq:RGEa}
\frac{v^2}{2}C_{1LL}^{ijkl}(\mu_{\rm low}) &=& 1.15 \, V_{L\,ji}^*V_{L\,kl},\qquad \qquad \quad  \qquad \, \frac{v^2}{2}C_{2LL}^{ijkl}(\mu_{\rm low}) = -0.34\,  V_{L\,ji}^*V_{L\,kl},\nn\\
\frac{v^2_R}{2} C_{1LR}^{ijkl}(\mu_{\rm low}) &=& 0.90  \frac{\xi e^{i \alpha}}{1+\xi^2}  V_{L\,ji}^* (V_R)_{kl},\qquad \frac{v^2_R}{2} C_{2LR}^{ijkl}(\mu_{\rm low}) =0.45\frac{\xi e^{i \alpha}}{1+\xi^2} V_{L\,ji}^* (V_R)_{kl},\nn\\
v_R^2C_{1RR}^{ijkl}(\mu_{\rm low}) &=&1.36\left(V_R\right)_{ji}^*\left(V_R\right)_{kl}\,,\qquad\qquad v_R^2C_{2RR}^{ijkl}(\mu_{\rm low}) =-0.65\left(V_R\right)_{ji}^*\left(V_R\right)_{kl}\,,
\eea
while, for the scalar operators, 
\bea\label{eq:rgescalar}
C_{1,\, quqd}^{ijkl}(\mu_{\rm low})&=&4.9 \frac{Y_{dH}^{kl}Y_{uH}^{ij}}{M_H^2}+2.6 \frac{ Y_{dH}^{il} Y_{uH}^{kj}}{M_H^2}\,,\nn\\
C_{2\, quqd}^{ijkl}(\mu_{\rm low})&=&-0.95 \frac{Y_{dH}^{kl}Y_{uH}^{ij}}{M_H^2}-0.82\frac{ Y_{dH}^{il} Y_{uH}^{kj}}{M_H^2}\,.
\eea
The operators $C_4$ and $C_5$, which contribute to meson-antimeson oscillations, receive a tree level contribution from the exchange of heavy Higgses, and a loop contribution from diagrams with a $W_R$ and $W_L$ exchange. At $\mu_{\rm low} = 2$ GeV, we find
\bea
C_4^{ijkl}(\mu_{\rm low}) &=& \frac{g_R^2}{M_{W_R}^2}\sum_{a,b}a_{ab}^{(4)}\frac{m_{u_a}m_{u_b}}{m_t^2}V^*_{L\,ai}V_{L\,bj}\left(V_R\right)_{bk}^*\left(V_R\right)_{al}\,,\\
C_5^{ijkl} (\mu_{\rm low})&=& -2.01\frac{1}{M_H^2}\left( Y_{dH}\right)^*_{jk}Y_{dH}^{il}+\frac{g_R^2}{M_{W_R}^2}\sum_{a,b}a_{ab}^{(5)}\frac{m_{u_a}m_{u_b}}{m_t^2}V^*_{L\,ai}V_{L\,bj}\left(V_R\right)_{bk}^*\left(V_R\right)_{al}\,,\nn
\eea
with $Y_{uH}$ and $Y_{dH}$ defined in Eq.\ \eqref{eq:Hyukawa} and evaluated at $\mu =\mu_0$.
The RG effects are captured by the $a^{(4,5)}$ coefficients, which are given by
\bea\label{eq:a45}
a^{(4)} = \bma 
0.028 & 0.028 & 0.001\\
0.028 & 0.032 & 0.001\\
0.001 & 0.001& 0.00077\\
\ema ,\qquad 
a^{(5)} =- \bma 
0.16 & 0.16 & 0.034\\
0.16 & 0.17  & 0.037\\
0.034 & 0.037& 0.030\\
\ema. 
\eea

These results depend mildly on the scale $\Lambda$, and in our analysis we set $\mu_0 = M_{W_R}$.
If we turn off the running between $M_{W_R}$ and $m_{t}$ and integrate out 
the $W_R$ and heavy Higgses at the scale $\mu_0= m_{t}$, the values 
of $C_{1\, RR}$ and $C_{2\, RR}$ are reduced (in absolute value) by about 15\% and $40\%$, respectively,
while $C_{i\, LL}$ and $C_{i\, LR}$ are not affected.
Similarly, the prefactors of the product of Yukawa couplings $Y_{dH}^{kl}Y_{uH}^{ij}\vert_{\mu = 10\,{\rm TeV}}$  and $Y_{dH}^{il} Y_{uH}^{kj}\vert_{\mu = 10\,{\rm TeV}}$  in $C_{1\, quqd}$
decrease by $\sim 10\%$ and $\sim 30\%$, respectively, while they decrease by $\sim 30\%$ for both terms in $C_{2\, quqd}$. 
These fairly mild corrections due to the RGEs are in part due to the $\mu_0$ dependence of the Yukawa couplings, $Y_{qH}$, which partially compensate for the effects of the $\ga_{LR}$ anomalous dimensions.
Finally, using $\mu_0=m_t$, the upper-left $2\times2$ block of the $a^{(4,5)}$ coefficients in Eq.\ \eqref{eq:a45} decrease by $\sim 30\%$, while the remaining components decrease by significant factors ranging from $\sim 1/5$ to $\sim1/40$.
We collect semi-analytical results for the $\mu_0$ dependence of these Wilson coefficients in App.\ \ref{app:RGE}.

\section{The CP-violating chiral Lagrangian  }\label{sec:chiralCPodd}

In this section we discuss the low-energy chiral Lagrangian induced by CP-violating operators involving light quarks. 
The construction of this Lagrangian is relevant for the study of electric dipole moments and long-distance effects in $\varepsilon_K$.
Although the effects in EDMs and $\varepsilon_K$ of certain operators can be directly evaluated using lattice-QCD or QCD sum rules, there are several operators for which it is useful to employ Chiral Perturbation Theory ($\chi$PT). In particular, the contributions of the LR operators in Eq.\ \eqref{eq:LowLag} to EDMs have not been computed directly. In this case, chiral symmetry allows us to relate their contributions to CP-odd pion-nucleon couplings to matrix elements that have been computed for $K\to \pi\pi$ processes. The obtained pion-nucleon couplings can be used to estimate the leading contributions of these operators to diamagnetic atomic EDMs. In addition, deriving the mesonic Lagrangian in $\chi$PT allows us to estimate long-distance corrections to $K-\bar K$ mixing arising from two insertions of $\Delta S=1$ operators. 

Our starting point is the following Lagrangian at the scale of a few GeV
\begin{eqnarray}\label{LagUnaligned}
\mathcal L &=& \mathcal L^{\rm QCD}_{m_q=0} -  \bar q \mathcal M q
 - \bar \theta \frac{g_s^2}{64 \pi^2} \varepsilon^{\mu \nu \alpha \beta} G^a_{\mu\nu} G^a_{\alpha \beta} 
 -\frac{g_s}{2} \bar q ( i \sigma^{\mu\nu}\gamma_5)\tilde{d}_{CE}  t^a q\,G^a_{\mu\nu}\nn\\
 && -C_{1\, LR}^{AB}\, \bar q \gamma^\mu t^A P_L  q \, \bar q \gamma_\mu t^B P_R q    -C_{2\, LR}^{AB}\, \bar q_\al \gamma^\mu t^A P_L  q_\bt \, \bar q_\al \gamma_\mu t^B P_R q_\bt   \nn\\
  && -\left[C_{1\, LL}^{suud}\, \bar s \gamma^\mu  P_L  u \, \bar u \gamma_\mu  P_L d    +C_{2\, LL}^{suud}\, \bar s_\al \gamma^\mu  P_L  u_\bt \, \bar u_\al \gamma_\mu P_L d_\bt   +(L\leftrightarrow R)+{\rm h.c.}\right]\,,
\end{eqnarray}
where $q$ denotes a vector of light quark fields $q = (u,d,s)^T$,  $t^a$ $(t^{A,B})$ are the Gellman matrices in color (flavor) space, normalized such that $\mathrm{Tr}(t_a t_b)=\delta_{ab}/2$,  and $\mathcal M$ is the real quark mass matrix, $\mathcal M = \textrm{diag}(m_u,m_d,m_s)$. The couplings are given by
\bea
C_{i\, LR}^{AB} &=& (\dt^{A1}-i\dt^{A2})(\dt^{B1}+i\dt^{B2})C_{i\, LR}^{duud}+(\dt^{A4}-i\dt^{A5})(\dt^{B4}+i\dt^{B5})C_{i\, LR}^{suus}\nn\\
&&+(\dt^{A4}-i\dt^{A5})(\dt^{B1}+i\dt^{B2})C_{i\, LR}^{suud}
+(\dt^{A1}-i\dt^{A2})(\dt^{B4}+i\dt^{B5})C_{i\, LR}^{duus}\nn\\
&&- (\dt^{A6}-i\dt^{A7})(\dt^{B6}+i\dt^{B7})C_{3+i}^{sdds}+{\rm h.c.}
\eea
We work in a basis where the overall phase of the mass matrix
has been rotated into the $\tilde GG$ term to form  the physical combination
$\bar \theta$.
The third term in the first line of Eq.\ \eqref{LagUnaligned} denotes the CP-odd quark chromo-electric dipole moment with $\tilde d_{CE} = \textrm{diag}(\tilde d_u, \tilde d_d, \tilde d_s)$, where $\tilde d_{u} = m_{u}\,\mathrm{Im}\,C_{gu}^{uu}$ and $\tilde d_{i} = m_{i}\,\mathrm{Im}\,C_{gd}^{ii}$ for $i=\{d,s\}$. The last two lines denote various CP-odd four-quark operators introduced in previous sections. To obtain the above Lagrangian we have used the relation $\left(C_{4,5}^{sd\,ds}\right)^*= C_{4,5}^{ds\,sd}$. 

Our main goal will be to estimate the CP-odd pion-nucleon couplings that are induced by the LR operators and to discuss the long-distance contributions to $\bar K-K$ mixing generated by two insertions of the $\Dt S=1$  four-fermion terms. Compared to the Lagrangians in Eqs.\ \eqref{eq:LowLag} and \eqref{eq:Lagdip}, we have omitted contributions from $C_{1,quqd}^{ij\,lm}$ and $C_{2,quqd}^{ij\,lm}$ as the operators involving light quarks are suppressed by small Yukawa couplings and $M_H^{-2}$, so that their contributions can be safely neglected. 
We also omitted the Weinberg operator and the quark EDMs here as we will use lattice QCD and QCD sum-rule calculations to directly obtain their contributions to EDMs in Sect.\ \ref{sec:EDMs}. Finally, Eq.\ \eqref{eq:LowLag} involves $\Dt S=1$ interactions  $\sim C_{4,5}^{}$ with $sddd$, $ddsd$, $dsss$, and $ssds$ flavor structures. 
Unlike the $C_{i\, LR}^{AB}$ coefficients, which transform like $\mathbf 8_L\times \mathbf 8_R$ under chiral symmetry, the $ C_{4,5}^{}$ coefficients with  $\Dt S=1$ have different chiral symmetry properties and we neglect them in the following as these
 are only generated at loop level or are suppressed by factors of small Yukawa couplings and $M_H^{-2}$.

\subsection{Vacuum alignment and the Peccei-Quinn mechanism}
For the purpose of chiral perturbation theory it is useful to perform several field redefinitions of the quark fields to remove meson tadpoles (tadpoles describe the disappearance of neutral Goldstone bosons to the vacuum). We start by applying a global anomalous axial $U(1)$ transformation of the form
\begin{equation}\label{rotA}
q \rightarrow e^{i\theta_A \gamma^5} q\,,\,\qquad \theta_A =\frac{\bar \theta}{2n_f}\,,
\end{equation}
with $n_f=3$ the number of active quark flavors, to eliminate the gluonic $\tilde GG$ term from the Lagrangian. The price to pay is that the quark mass matrix becomes complex. In a first step, we can ignore the shifts in the higher-dimensional qCEDMs and four-quark operators as the induced terms scale as $\bar \theta/\Lambda^2$, where $\Lambda^2$ collectively denotes the masses of BSM fields such as the right-handed scalar and/or gauge bosons. However, terms proportional to  $\bar \theta/\Lambda^2$ do play an important role when we discuss the Peccei-Quinn mechanism below. After the rotation, the quark mass term becomes
\begin{eqnarray}\label{quarkmass}
\mathcal L_{m} &=& -  \bar q \mathcal M q + \bar q i \gamma^5 \left[-\frac{2}{3}(2\bar m + m_s) + (4 \varepsilon \bar m) t_3  - \frac{4}{\sqrt{3}}(\bar m-m_s)t_8 \right]{\theta_A\,} q\nn\\
&\equiv&  -  \bar q \mathcal M q + \bar q i \gamma^5 \left[\theta_0 + \theta_3 t_3  + \theta_8 t_8 \right] q\,,
\end{eqnarray}
where we introduced  $\bar m = (m_u + m_d)/2$ and $2 \bar m \varepsilon = m_d - m_u$. The terms involving $\bar q i \gamma^5 t_{3,8}q$ lead to so-called tadpole operators that allow for neutral Goldstone bosons (in this case $\pi^0$ and $\eta$) to disappear in the vacuum. In the limit of no dimension-six interactions, it is straightforward to eliminate the tadpole-inducing terms (a procedure called vacuum alignment) by performing two additional non-anomalous axial $SU(3)$ rotations 
\begin{equation}
q \rightarrow e^{i (\alpha_3 t_3 + \alpha_8 t_8) \gamma^5} q\,.
\end{equation}
By setting
\begin{equation}
\alpha_3 = \frac{\varepsilon m_s }{2 m_s +\bar m(1-\varepsilon^2)} \bar \theta\,,\qquad \alpha_8 = \frac{1}{\sqrt{3}}\frac{m_s - \bar m (1-\varepsilon^2)}{2 m_s +\bar m(1-\varepsilon^2)}\bar \theta\,,
\end{equation}
the  $\bar q i \gamma^5 t_{3,8}q$ terms are removed and the dimension-four part of the Lagrangian becomes
\begin{eqnarray}\label{quarkmass2}
\mathcal L &=& -  \bar q \mathcal M q - m_* {\bar{\theta}} \bar q i \gamma^5 q\,,
\end{eqnarray}
in terms of the reduced quark mass 
\begin{equation}
m_* = \left(\frac{1}{m_u} + \frac{1}{m_d} + \frac{1}{m_s}\right)^{-1}  = \frac{\bar m (1-\varepsilon^2)}{2} \left( 1 + \frac{\bar m (1-\varepsilon^2)}{2 m_s} \right)^{-1}\,.
\end{equation}
This is the usual result that shows that the theta term decouples if one of the quarks is massless. Keeping terms to $\Or(\bar \theta ^2)$ shows that the three chiral rotations proportional to $\theta_A$, $\alpha_3$, and $\alpha_8$ generate a term 
\begin{equation}
\mathcal L_{PQ} = \frac{1}{6}\bar \theta^{\,2}\,m_* \bar q q\,,
\end{equation}
which induces a hadronic contribution to the vacuum energy. The Peccei-Quinn mechanism becomes apparent if we promote $\bar \theta$ to include a dynamical axion field~\footnote{The performed field redefinitions become field dependent and lead to derivative axion-quark interactions. Since we do not consider axions explicitly in this paper, we do not further study these terms. } $\bar \theta \rightarrow \bar \theta + a/f_a$ where $a$ is the axion field and $f_a$ the axion decay constant. Because the vacuum energy scales as $( \bar \theta + a/f_a)^2$, the axion potential is minimized for $\langle a \rangle = - f_a \bar \theta$ eliminating the CP-violating term from the Lagrangian.  

The story is similar, but somewhat more tedious to work out, in the presence of the dimension-six operators. With just the dimension-four terms, the entire argument could be made at the quark level with minimal reference to hadronic operators. Once the dimension-six operators are included, it is convenient to refer to the hadronic Lagrangian explicitly. It is useful to construct the terms in the chiral Lagrangian that can induce tadpoles after the first field transformation that eliminates the gluonic $\bar \theta$ term. The relevant terms are given by
\begin{eqnarray}\label{mesonEps'}
 \mathcal L_{\mathrm{GB}} &=&  \frac{F_0^2}{4}\left( \textrm{Tr}\, [U^{\dagger} \chi + U \chi^{\dagger}] +\textrm{Tr}\, [U^{\dagger} \tilde\chi + U \tilde\chi^{\dagger}]\right)
+ \frac{F_0^4}{4}   \textrm{Tr}\left( U^\dagger t^B Ut^A  \right) \sum_{i=1,2} \mathcal A_{i\, LR} C_{i\, LR}^{AB}\\
&&+ \frac{F_0^4}{4} \left\{\sum_{i=1,2}  \tr\left(t^A\partial_\mu U^\dagger \partial^\mu U\right){\cal A}^{(8)}_{i\,LL}\left[C^{du\,us}_{i\,LL}(\delta_{A6}+i\delta_{A7}) + {\rm h.c.}\right]+\bma L\to R\\U\leftrightarrow U^\dagger \ema\right\}\nn\,,
\end{eqnarray} 
where  $U$ is the  matrix of the pseudo-Nambu-Goldstone (pNG) boson fields
\begin{equation}\label{eq:2.3}
U = u(\pi)^2  = \exp\left(\frac{2 i \pi}{F_0}\right), \qquad
\pi =\frac{1}{\sqrt{2}} \left( \begin{array}{c c c} 
\frac{\pi_3}{\sqrt{2}} + \frac{\pi_8}{\sqrt{6}}  & \pi^+ 						& K^+ \\
\pi^-						& - \frac{\pi_3}{\sqrt{2}} + \frac{\pi_8}{\sqrt{6}} 	& K^0 \\
K^-						& \bar{K}^0					& - \frac{2}{\sqrt{6}}\pi_8 
\end{array} \right)\, ,
\end{equation}
and 
\begin{equation}\label{chi}
\chi = 2 B \left[ \mathcal M + i \left( \theta_0+ \theta_3 t_3+  \theta_8 t_8  \right)\right]\,,\qquad \tilde \chi = - 2i \tilde B \left(\tilde d_0+ \tilde d_3 t_3  +\tilde d_8 t_8  \right)\,,
\end{equation}
where we introduced the combinations $\tilde d_0 = (\tilde d_u + \tilde d_d + \tilde d_s)/3$, $\tilde d_3 = (\tilde d_u - \tilde d_d)$, and $\tilde d_8 = (\tilde d_u + \tilde d_d-2 \tilde d_s)/\sqrt{3}$. Under $SU(3)_L\times SU(3)_R$ transformations we have $U\to RUL^\dagger$ such that the quark-level Lagrangian and its chiral analogue are formally invariant if the spurions $\chi$ and $\tilde \chi$ transform in the same way as $U$. 
The LR four-quark operators   transform as $\boldsymbol 8_L\times \boldsymbol 8_R$, so that the $\mathcal A_{i\, LR}$ part of the Lagrangian is invariant if the flavor structures transform as $t^A \to Lt^AL^\dagger$ and $t^B\to Rt^BR^\dagger$. For the LL and RR operators, we only take into account the pieces transforming as $\boldsymbol 8_{L,R}\times \boldsymbol 1_{R,L}$, as the long-distance contributions of the $\boldsymbol {27}_{L,R}\times \boldsymbol 1_{R,L}$ terms are suppressed by the $\Delta I = 1/2$ rule.

The mesonic interactions are associated with 6 low-energy constants (LECs), $B$, $\tilde B$, $\mathcal A_{\{1,2\}\,LR}$, and $\mathcal A_{\{1,2\}\,LL}=\mathcal A_{\{1,2\}\,RR}$. The first is well known and relates the masses of pseudo-Goldstone bosons to the chiral condensate, while $\tilde B$ and $\mathcal A_{\{1,2\}\,LR}$ are related to the condensates of the higher-dimensional operators
\begin{align}
&B = -\frac{1}{3}\frac{\langle 0 | \bar{q} q | 0 \rangle}{F_0^2}\, ,
&  &\tilde{B} = -\frac{1}{3}\frac{\langle 0 | \bar{q} g_s \sigma_{\mu\nu} G^{\mu\nu} q | 0 \rangle}{2 F_0^2}\, .&\\
&\delta^{AB} \frac{\mathcal A_{1\,LR}}{8}=-\frac{\langle 0 | \bar q_\al \gamma^\mu t^A  P_Lq_\al \, \bar q_\bt \gamma_\mu t^BP_R q_\bt   | 0 \rangle}{F_0^4} \,,&&  \delta^{AB} \frac{\mathcal A_{2\,LR}}{8} = -\frac{\langle 0 | \bar q_\al \gamma^\mu t^A P_L q_\bt \, \bar q_\bt \gamma_\mu t^B P_Rq_\al   | 0 \rangle }{F_0^4}&\,.\nn
\end{align}
whereas the condensates of the LL and RR operators vanish at leading order. The LEC $B$ can also be expressed as $2B = m_\pi^2/\bar m$. Using the above Lagrangian, the LECs of the four-quark operators can be determined from matrix elements of the form $\langle (\pi\pi)_{I=0,2}| O_i |K^0\rangle$ which have been computed on the lattice \cite{Bai:2015nea,Blum:2012uk,Abbott:2020hxn}. Using chiral symmetry, the same LECs can be related to matrix elements that play a role in neutrinoless double beta decay \cite{Nicholson:2018mwc} or to the bag factors appearing in $ K-\bar K$ oscillations \cite{Aoki:2019cca}, up to $SU(3)$ corrections  \cite{Cirigliano:2017ymo}.
This leads to the following relations at leading order~\footnote{These relations assume that the $\boldsymbol {27}_{L,R}\times \boldsymbol 1_{R,L}$ parts of the LL and RR operators provide negligible contributions to the $\langle (\pi\pi)_{I=0}| O_i |K^0\rangle$ matrix elements. 
These contributions can be obtained by using the LEC of the $\boldsymbol {27}_{L,R}\times \boldsymbol 1_{R,L}$ representations, $\mathcal A_{(27,1)}'$, discussed in Sect.\ \ref{sec:S=1}. Such an estimate shows that the dominant contributions to $\mathcal M_{1,2}$ indeed arise from the $\boldsymbol {8}_{L,R}\times \boldsymbol 1_{R,L}$ parts of the operators.}
\bea\label{eq:LECsKpp}
\mathcal A_{1\, LR}(3\, {\rm GeV}) &=& \frac{\mathcal A'_{(8,8)}}{3\sqrt{2} F_0}\simeq 2.2(1)\, {\rm GeV}\sq\,,\qquad \mathcal A_{2\, LR}  (3\, {\rm GeV})= \frac{\mathcal A'_{(8,8)\rm mix}}{3\sqrt{2} F_0}\simeq 10.1(6)\, {\rm GeV}\sq\,,\nn\\
\mathcal A_{1\, LL}^{(8)} (4\, {\rm GeV}) &=&\mathcal A_{1\, RR}^{(8)}(4\, {\rm GeV})=- \frac{ \mathcal M_2}{\sqrt{6}F_0(m_K\sq-m_\pi\sq)}\simeq -2.8(3)\,,\nn\\ \mathcal A_{2\, LL}^{(8)}(4\, {\rm GeV})  &=&\mathcal A_{2\, RR}^{(8)}(4\, {\rm GeV})= -\frac{ \mathcal M_1}{\sqrt{6}F_0(m_K\sq-m_\pi\sq)}\simeq 1.8(3)\,,
\eea
where $\mathcal M_{1,2}$ and $\mathcal A'_{(8,8)\rm (mix)}$ are related to matrix elements $\sim\langle (\pi\pi)_{I=0,2}| O_i |K^0\rangle$, which were determined in Refs.\ \cite{Bai:2015nea,Blum:2012uk,Abbott:2020hxn}.

The Lagrangian in Eq.~\eqref{mesonEps'} leads to tadpoles as can be seen by expanding out the various terms. Introducing the ratios $\tilde r = \tilde B/B$ and $r_i = (F_0^2/B) \mathcal A_{i\, LR}$, the tadpole Lagrangian becomes
\begin{eqnarray}\label{Tad}
 \mathcal L_{\mathrm{tadpole}} &=&F_0 B \bigg\{ \pi^0 \left[ \theta_3 - \tilde r \,\tilde d_3 + \frac{1}{2}\sum_{i=1,2}r_i\left(2 \textrm{Im}\,C_{i\, LR}^{du\, ud} +\textrm{Im}\,C_{i\, LR}^{su\, us}+\textrm{Im}\,C_{3+i}^{sd\, ds}\right)  \right]
 \nn\\
 && \qquad\,\,+ \eta  \left[  \theta_8 - \tilde r  \,\tilde d_8 + \frac{\sqrt{3} }{2}\sum_{i=1,2}r_i\left( \textrm{Im}\,C_{i\, LR}^{su\, us} - \textrm{Im}\,C_{3+i}^{sd\, ds}\right)  \right] 
  \nn\\
 && \qquad\,\,-\frac{\bar K^0+K^0}{\sqrt{2}}   \frac{1}{2}\sum_{i=1,2}r_i\left( \textrm{Im}\,C_{i\, LR}^{su\, ud} + \textrm{Im}\,C_{i\, LR}^{du\, us}\right)  \nn\\
  && \qquad\,\,{-\frac{i(\bar K^0-K^0)}{\sqrt{2}}  \frac{1}{2}\sum_{i=1,2}r_i\left( \textrm{Re}\,C_{i\, LR}^{su\, ud} - \textrm{Re}\,C_{i\, LR}^{du\, us}\right) } \bigg\}\,.
 \end{eqnarray}
It is in principle possible to eliminate these leading tadpoles by a suitable redefinition of Goldstone fields at the hadronic level. Such a rotation, however, requires a corresponding complicated field redefinition of baryon fields, see Refs.~\cite{deVries:2012ab,Mereghetti:2010tp,Bsaisou:2014oka} for details. 
The baryon transformation was omitted in Ref.~\cite{Haba:2018byj} and led to erroneous conclusions as was also pointed out in Ref.~\cite{Bertolini:2019out}. In this work, we follow Ref.~\cite{Cirigliano:2016yhc} and only perform field transformations at the quark level. We reconstruct the chiral Lagrangian after each quark transformation. This leads to the same conclusions as Ref.~\cite{deVries:2012ab} (and thus in disagreement with Ref.~\cite{Haba:2018byj} and the $m_s/(m_u+m_d)$ enhancement found there). 

We begin by performing four axial chiral rotations on Eq.~\eqref{LagUnaligned}, 
  now including  $\al_6t_6$ {and $\al_7t_7$} rotations to remove the $K^0$ tadpole terms, resulting in the Lagrangian $\mathcal L'$. We then construct the hadronic Lagrangian in Eq.~\eqref{mesonEps'}, that now depends explicitly on $\alpha_{3,6,{7,}8}$, and solve for $\alpha_{3,6,{7,}8}$ by demanding that the $\pi^0$, $K^0$,  and $\eta$ tadpoles vanish. The  solutions are given by
 \begin{eqnarray}
 \alpha_3  &=&  \frac{ -\varepsilon m_s}{2m_s + \bar m (1-\varepsilon^2)}\bigg\{-\bar \theta  + \frac{\tilde r}{2 m_s}\left[ \frac{\bar m + 2 m_s}{ \varepsilon\bar m} \tilde d_3 + \sqrt{3} \tilde d_8\right]- \sum_{i=1,2}\frac{r_i}{2  \varepsilon \bar m m_s}\nn\\
 && \times\left[(\bar m+2 m_s) \textrm{Im}\,C_{i\, LR}^{du\, ud} +\frac{2 m_s + \bar m (1 + 3  \varepsilon)}{2} \textrm{Im}\,C_{i\, LR}^{su\, us} + \frac{2 m_s + \bar m (1 - 3  \varepsilon)}{2} \textrm{Im}\,C_{3+i}^{sd\, ds}\right]\bigg\}\,,\nn\\
\alpha_6 &=&-\frac{1}{2(m_d+m_s)}\sum_{i=1,2} r_i \left(\textrm{Im}\,C_{i\, LR}^{duus}+\textrm{Im}\,C_{i\, LR}^{suud}\right)\,,\nn\\ 
\alpha_7 &=&{-\frac{1}{2(m_d+m_s)}\sum_{i=1,2} r_i \left(\textrm{Re}\,C_{i\, LR}^{duus}-\textrm{Re}\,C_{i\, LR}^{suud}\right)}\,,\nn\\ 
  \alpha_8  &=&  \frac{-1}{\sqrt {3}}\frac{ 1}{2m_s + \bar m (1-\varepsilon^2)}\bigg\{ -\left[m_s - \bar m(1-\varepsilon^2)\right]\bar \theta + \frac{{3\varepsilon}\tilde r}{2}\left[\tilde d_3 + \frac{\sqrt{3}}{\varepsilon}\tilde d_8 \right]\nn\\
 && - \sum_{i=1,2}\frac{{3\varepsilon}r_i}{2}\left[ \textrm{Im}\,C_{i\, LR}^{du\, ud} +\frac{3+   \varepsilon}{2 \varepsilon} \textrm{Im}\,C_{i\, LR}^{su\, us} -  \frac{3-\varepsilon}{2 \varepsilon} \textrm{Im}\,C_{3+i}^{sd\, ds}\right]\bigg\}\,.
 \end{eqnarray}

After these rotations the Lagrangian can be written in the following form
\bea\label{Lagaligned}
\mathcal L &=& \mathcal L^{\rm QCD}_{m_q=0} -  \bar q \mathcal M q + \bar q \left[-m_*(\bar \theta - \bar \theta_{\mathrm{ind}}) + r \tilde d_0  + \theta_3' t_3 +  \theta_6' t_6 {+\theta'_7t_7}+ \theta_8' t_8 \right] i \gamma_5 q 
 \\
 &&-\frac{g_s}{2} \bar q ( i \sigma^{\mu\nu}\gamma_5)\tilde{d}_{CE}  t^a q\,G^a_{\mu\nu}\nn\\
  && -C_{1\, LR}^{AB}\, \bar q \gamma^\mu t^A P_L  q \, \bar q \gamma_\mu t^B P_R q    -C_{2\, LR}^{AB}\, \bar q_\al \gamma^\mu t^A P_L  q_\bt \, \bar q_\al \gamma_\mu t^B P_R q_\bt   \nn\\
  && -\left[C_{1\, LL}^{suud}\, \bar s \gamma^\mu  P_L  u \, \bar u \gamma_\mu  P_L d    +C_{2\, LL}^{suud}\, \bar s_\al \gamma^\mu  P_L  u_\bt \, \bar u_\al \gamma_\mu P_L d_\bt   +(L\leftrightarrow R)+{\rm h.c.}\right]+\dots\nn,
\eea
where the dots denote terms of dimension-eight or higher or terms proportional to $\bar \theta^2$ or $\bar \theta/\Lambda^2$. $ \theta_{\mathrm{ind}}$, $\theta_3' $, $\theta_6'$, $\theta'_{7}$,  and $\theta_8'$  depend on hadronic LECs
\begin{eqnarray}\label{align}
\bar\theta_{\textrm{ind}} &=& \tilde r \left( \frac{\tilde d_u}{m_u}+\frac{\tilde d_d}{m_d}+\frac{\tilde d_s}{m_s}\right)\nn\\
&&
-2\sum_{i=1,2} r_i {\rm Im}\, \left(\frac{m_d-m_u}{4m_um_d}C_{i\, LR}^{du\, ud}+\frac{m_s-m_u}{4m_um_s}C_{i\, LR}^{su\, us}-\frac{m_s-m_d}{4m_dm_s}\textrm{Im}\,C_{3+i}^{sd\, ds}\right),\nn\\
\theta'_3 &=& \tilde r\,\tilde d_3 -\sum_{i=1,2}r_i {\rm Im}\,\left( C_{i\, LR}^{du\, ud} + \frac{1}{2} C_{i\, LR}^{su\, us}+\frac{1}{2} \textrm{Im}\,C_{3+i}^{sd\, ds} \right) ,\nonumber \\
\theta'_6 &=& 
\frac{1}{2}\sum_{i=1,2}r_i {\rm Im}\,\left( C_{i\, LR}^{su\, ud}+ C_{i\, LR}^{du\, us} \right) ,\nonumber \\
\theta'_7 &=& 
{-\frac{1}{2}\sum_{i=1,2}r_i {\rm Re}\,\left( C_{i\, LR}^{su\, ud} -C_{i\, LR}^{du\, us} \right)} ,\nonumber \\
\theta'_8 &=&  \tilde r\,\tilde d_8
-\frac{\sqrt{3}}{2}\sum_{i=1,2}  r_i  \left(\textrm{Im} \, C_{i\, LR}^{su\, us}- \textrm{Im}\,C_{3+i}^{sd\, ds} \right)\ .
\end{eqnarray}

The term $ \bar \theta_{\mathrm{ind}}$ is introduced because $\bar \theta$ effectively relaxes to $ \bar \theta_{\mathrm{ind}}$ if a Peccei-Quinn mechanism is applied. The expression for $ \bar \theta_{\mathrm{ind}}$ can be obtained by calculating the induced vacuum energy of Eq.~\eqref{Lagaligned} supplemented by terms of $\Or(\bar \theta^2)$ and $\Or(\bar \theta/\Lambda^2)$. The latter depend linearly on $\bar\theta$ and ensure that, after a Peccei-Quinn mechanism is implemented through $\bar \theta\to \bar \theta_a = \bar \theta+a/f_a$, the minimum of the axion potential is shifted away from zero. This leads to a nonzero 
vev for the axion field and an effective theta angle (but suppressed by $1/\Lambda^2$), $\langle \bar \theta_a \rangle =\theta_{\rm ind}$,  even after implementation of the Peccei-Quinn mechanism. Once the Peccei-Quinn mechanism is applied the final Lagrangian becomes
\begin{eqnarray}\label{LagalignedFinal}
\mathcal L &=& \mathcal L^{\rm QCD}_{m_q=0} -  \bar q \mathcal M q + \bar q \left[ r \tilde d_0  + \theta_3' t_3+ \theta_6' t_6 {+\theta_7't_7}+ \theta_8' t_8 \right] i \gamma_5 q 
-\frac{g_s}{2} \bar q ( i \sigma^{\mu\nu}\gamma_5)\tilde{d}_{CE}  t^a q\,G^a_{\mu\nu}\nn\\
 && -C_{1\, LR}^{AB}\, \bar q \gamma^\mu t^A P_L  q \, \bar q \gamma_\mu t^B P_R q    -C_{2\, LR}^{AB}\, \bar q_\al \gamma^\mu t^A P_L  q_\bt \, \bar q_\al \gamma_\mu t^B P_R q_\bt   \nn\\
  && -\left[C_{1\, LL}^{suud}\, \bar s \gamma^\mu  P_L  u \, \bar u \gamma_\mu  P_L d    +C_{2\, LL}^{suud}\, \bar s_\al \gamma^\mu  P_L  u_\bt \, \bar u_\al \gamma_\mu P_L d_\bt   +(L\leftrightarrow R)+{\rm h.c.}\right]+\dots
\end{eqnarray}
It can be verified explicitly that with $\theta_{3}'$, $\theta'_{6}$, $\theta'_{7}$, and $\theta'_{8}$ given by Eq.~\eqref{align}, the hadronic Lagrangian in Eq.~\eqref{mesonEps'} does not induce tadpoles.

After eliminating the leading tadpoles in this way, one can use Eq.\ \eqref{mesonEps'} to derive the low-energy effects of the CP-odd operators. The first long-distance contributions to $\bar K-K$ mixing are induced by diagrams involving two insertions of $\Delta S=1$ operators, the result of which we discuss in Sect.\ \ref{sec:S=2}.
Instead, the most important flavor-conserving CPV interactions arise from the baryonic Lagrangian which we discuss below.

\subsection{CP-odd pion-nucleon interactions}\label{sec:piN}
The relevant CPV pion-nucleon interactions arise from
\begin{eqnarray}\label{eq:L2}
\mathcal L_{\pi N} &=&   b_0 \textrm{Tr} \left(\bar B_{} B\right) \textrm{Tr} \chi_+  +  b_D \textrm{Tr}\left(\bar B \{ \chi_+, B \}  \right) + b_F \textrm{Tr}\left(\bar B [ \chi_+, B ]  \right) \nn\\
&+&\tilde b_0 \textrm{Tr} \left(\bar B_{} B\right) \textrm{Tr} \tilde \chi_+  +  \tilde b_D \textrm{Tr}\left(\bar B \{ \tilde \chi_+, B \}  \right) + \tilde b_F \textrm{Tr}\left(\bar B [ \tilde \chi_+, B ]  \right)  \nn\\
&+& \mathcal L_{LR} \, ,
\end{eqnarray}
where $b_{0,D,F}$ are LECs that can be obtained from fits to the baryon masses, $\tilde b_{0,D,F}$ are LECs related to the dipole operators and currently unknown, and  $B$ denotes the octet baryon field
\begin{equation}\label{eq:3.0}
B = \left( \begin{array}{c c c}
\frac{1}{\sqrt{2}}\Sigma^0 + \frac{1}{\sqrt{6}}\Lambda & \Sigma^+ 							& p \\
\Sigma^-					       & -\frac{1}{\sqrt{2}}\Sigma^0 + \frac{1}{\sqrt{6}}\Lambda 	& n \\
\Xi^-						       & \Xi^0								& -\frac{2}{\sqrt{6}} \Lambda
\end{array} \right)\, .
\end{equation}
We have defined 
\begin{equation}
\chi_+ = u^{\dagger} \chi u^{\dagger} + u \chi^{\dagger} u\,,\qquad \tilde \chi_+ = u^{\dagger} \tilde \chi u^{\dagger} + u \tilde \chi^{\dagger} u\,,
\end{equation}
where $\chi$ is now given by $\chi = 2 B \left[ \mathcal M + i \left(\theta'_3 t_3+  \theta'_6 t_6+  \theta'_7 t_7+  \theta'_8 t_8  \right)\right]$.
Finally, $\vL_{LR}$ gives rise to so-called ``direct" contributions to CPV meson-baryon interactions,
\bea
\vL_{LR} &=& b_{\bf 1} \textrm{Tr}\left(\bar BB\right) l_1^{abba}+ \left\{\left[\left(\bar BB\right)_{ji}-\frac{\dt_{ji}}{3}\textrm{Tr}\left(\bar BB\right)\right] \left[ b_{\bf 8}^{(1)}l_1^{iaaj}+ b_{\bf 8}^{(2)}l_1^{ajia}\right]+\bma B\leftrightarrow \bar B\\ b^{(1,2)}_{\bf 8}\to b^{(3,4)}_{\bf 8}\ema\right\}\nn\\
&&+b_{\bf 10}^{\pm}l_1^{ijkl}\left[\frac{\bar B_{ji}B_{lk}\pm\bar B_{jk}B_{li}}{2}-\frac{\dt_{il}}{6}\left(\bar BB\right)_{jk}-\frac{\dt_{kj}}{6}\left( B\bar B\right)_{li} \mp (j\leftrightarrow l)\right]\nn\\
&&+b_{\bf 27}l_1^{ijkl}\left[\frac{\bar B_{ji}B_{lk}+\bar B_{jk}B_{li}+\bar B_{li}B_{jk}+\bar B_{lk}B_{ji}}{4}-\frac{\dt_{il}}{12}\left\{\bar B,\,B\right\}_{jk}-\frac{\dt_{kj}}{12}\left\{ B,\,\bar B\right\}_{li}+\frac{\dt_{jk}\dt_{il}}{18}\textrm{Tr}\left(\bar BB\right) \nn
\right]\nn\\
&&+\bma l_1^{ijkl}\leftrightarrow  l_2^{ijkl}\\ b_{\bf r} \to \bar b_{\bf r}\ema\,,
\eea
where $l_{1,2}^{ijkl} = C^{AB}_{1,2\,LR } (ut^Au^\dagger)_{ij}(u^\dagger t^Bu)_{kl}$, while $b_{\bf r}$ and $\bar b_{\bf r}$ denote currently unknown LECs.
We focus on the pion-nucleon couplings 
\bea\label{eq:g01}
\vL _{\pi N}\supset -\frac{\bar g_0}{2F_\pi}\bar N \boldsymbol \pi \cdot \boldsymbol \tau N-\frac{\bar g_1}{2F_\pi}\pi_0\bar N  N\,.
\eea
The four-quark operators enter in the above through $\chi_+$, see Eq.~\eqref{chi}, and  $\vL_{LR}$, where the latter  involves additional LECs that are currently unknown.
In this work, we focus on the ``indirect" contributions that we do control and neglect the terms $\sim b_{\bf r}$ and $\bar b_{\bf r}$. The direct pieces are expected to arise at the same order as the indirect pieces so that neglecting them leads to a sizable uncertainty. Matching Eqs.\ \eqref{eq:L2} and \eqref{eq:g01} gives,
\begin{eqnarray}\label{relations0}
\bar g_0|_{LR} &=&  2 (b_D + b_F) F_0^2 \sum_{i=1,2}\mathcal A_{i\, LR} \left({\rm Im}\, C_{i\, LR}^{{su}\, us}-{\rm Im}\, C_{3+i}^{sd\, ds}\right)\,
+\bar g_0|_{\rm direct}  \ , \nn \\
\bar g_1|_{LR} &=&  2 (2 b_0 + b_D + b_F) F_0^2 \sum_{i=1,2} \mathcal A_{i\, LR} {\rm Im}\, \big(2C_{i\, LR}^{{du}\,ud}+C_{i\, LR}^{{su}\,us}+{\rm Im}\, C_{3+i}^{sd\, ds}\big)+ \bar g_1|_{\rm direct} \, , 
\end{eqnarray}
where we indicated the contributions from $\vL_{LR}$ by $\bar g_{0,1} |_{\rm direct}$. In principle, we can insert values of $b_{\{0,D,F\}}$ from fits to the baryon spectrum to obtain estimates for the indirect pieces. We can improve these relations by resumming higher-order corrections \cite{deVries:2016jox,Seng:2016pfd} and instead write
\begin{eqnarray}\label{relationsA}
\bar g_0|_{LR} &=& - \sum_{i=1,2}  \left({\rm Im}\, C_{i\, LR}^{su\, us}-{\rm Im}\, C_{3+i}^{sd\, ds}\right)  \frac{r_i}{4} \frac{d\, \delta m_N}{d \bar m \varepsilon}  +\bar g_0|_{\rm direct}  \ , \nn \\
\bar g_1|_{LR} &=&   -\sum_{i=1,2}{\rm Im}\big(2C_{i\, LR}^{{du}\,ud}+C_{i\, LR}^{{su}\,us}+C_{3+i}^{sd\, ds}\big)  \frac{r_i}{2} \frac{d m_N}{d \bar m }+\bar g_1|_{\rm direct}\ , \label{relationsB}
\end{eqnarray}
where $\delta m_N = m_n -m_p$ and $2 m_N = m_n + m_p$. 
The tadpole-induced pieces, proportional to $r_i$, depend on known quantities such as the nucleon sigma term $\sigma_N = \bar m (d m_N/d \bar m) = 59.1\pm 3.5$ MeV \cite{Hoferichter:2015dsa} where $\bar m = (m_u+m_d)/2 = 3.37\pm0.08$ MeV \cite{Aoki:2016frl},  
and the nucleon mass induced by the quark mass difference: $(d\delta m_N/d \bar m \varepsilon) \simeq \delta m_N/( \bar m \varepsilon) = (2.49 \pm 0.17 \, \mathrm{MeV})/( \bar m \varepsilon)$ \cite{Borsanyi:2013lga,Borsanyi:2014jba}, where $\varepsilon = (m_d - m_u)/(2\bar m) = 0.37\pm0.03$ \cite{Aoki:2016frl}. The above allows for an estimate of $\bar g_{0,1}$ as the LECs $\mathcal A_{i\, LR}$ are known from lattice-QCD calculations. The additional unknown direct pieces were estimated to induce a $50\%$ uncertainty in  Ref.\ \cite{Cirigliano:2016yhc}.

The remaining sources of flavor-diagonal CPV in Eq.\ \eqref{LagUnaligned}, the quark CEDMs, enter through $\tilde \chi_+$ and the $\tilde b_{0,D,F}$ terms, which represent the indirect and direct contributions, respectively. In this case both the direct and indirect contributions involve unknown LECs. We will therefore employ estimates resulting from QCD sum-rule calculations \cite{Pospelov_piN}, leading to
\bea 					\label{eq:piNCEDM}
\bar g_0|_{CEDM} &=&-(5\pm 10)\frac{2F_\pi}{{\rm fm}}\left(\tilde d_u+\tilde d_d\right)\,,
\qquad
\bar g_1|_{CEDM} = -(20^{+40}_{-10})\frac{2F_\pi}{{\rm fm}}\left(\tilde d_u-\tilde d_d\right)\,.
\eea
which hold at a scale of $\mu=1$ GeV. The contributions from the strange-quark CEDM are proportional to the small $\eta$-$\pi$ mixing angle \cite{deVries:2016jox} and we neglect them.

\section{Observables}\label{sec:obs}
Before describing the expressions we employ in our analysis, we briefly discuss the different classes of experiments and the LR parameters they are most sensitive to.

\begin{itemize}
 \item Leptonic and semileptonic charged-current decays.

These observables are known very accurately. For example, uncertainties on the lifetimes of superallowed $\beta$ emitters, which enter the determination of $V_{ud}$, appear at the  $\mathcal O(10^{-4})$ level \cite{Hardy:2020qwl}. The branching ratios for $K \rightarrow \mu \nu$ and $K \rightarrow \pi \ell \nu$ have uncertainties at the permille level. Leptonic and semileptonic decays of $B$ and $D$ mesons are known at the percent level. In addition, the theoretical input required to convert the observables into bounds on SM and LR parameters is only affected by small theoretical uncertainties.  

Corrections to leptonic and semileptonic decays are induced at tree level, 
by the mixing between the left- and right-handed $W$ bosons,
and are proportional to $C_{Hud}\sim \xi V_{R}M_{W_R}^{-2}$. 
We must disentangle these contributions from those from the SM CKM matrix, $\sim V_{L\, ij}$, in order to constrain the LR parameters. We do so by exploiting measurements in different channels, sensitive to the axial-vector or vector component of the charged current.  
For example, purely leptonic decays of pseudoscalar mesons probe the axial-vector component of the charged current, while $0^+ \rightarrow 0^+$ superallowed nuclear transitions and semileptonic decays of pseudoscalar mesons are sensitive to the vector component. 
In this way it is possible to fit the SM CKM parameters $V_{L\,uj}$
and $V_{L\,cj}$, with $j\in \{d,s,b\}$, together with the corresponding LR contributions.

\item Purely hadronic charged-current decays.

These include $\Delta S=1$ processes such as $K \rightarrow \pi \pi$, in particular $\varepsilon^\prime$, which measures direct CP violation in  kaon decays, and $\Delta B =1$ processes such as $B \rightarrow J/\psi K_S$, $B \rightarrow \pi\pi$ and $B\rightarrow D K$, which, in the SM, contribute to the determination of the CKM parameters $\bar\rho$ and $\bar\eta$. 
In the mLRSM, these processes receive contributions from $W_L$-$W_R$ mixing, proportional to $C_{Hud}$, and from the exchange of $W_R$ between right-handed quarks, proportional to $C_{1\, RR}$ and $C_{2\, RR}$.
While the experimental measurements have uncertainties similar to the leptonic and semileptonic decays, theoretical uncertainties are usually much larger, so that these channels provide sensitive probes of LR parameters only if the SM contribution is suppressed.    
This is the case of $\varepsilon^\prime$, which in the SM receives contributions at one loop and is further suppressed by the small $V_{L\,td}$ and $V_{L\,ts}$ elements. 
In the mLRSM, $\varepsilon'$ receives a large mixing contributions at tree level
and is sensitive to the combination Im$\,C_{Hud}^{ij}V_L^{ik\, *}\sim \xi M_{W_R}^{-2}\,{\rm Im}(V_R^{ ij}V_L^{ik\, *}e^{i\al})$,
with $j, k \in \{d,s\}$ and $j\neq k$.
The CP asymmetries in $\Delta B=1$ decays, on the other hand, 
arise at tree level in the SM, and are thus less sensitive to the contribution of the LR model.

\item $\Delta S=1$ and $\Delta B =1$ flavor-changing-neutral-current (FCNC) processes.

These include several rare decays of $K$ and $B$ mesons, such as $B\to X_s\g$, 
$B \rightarrow \mu^+ \mu^-$, $K_L\to \pi^0 e^+ e^-$
and $K \rightarrow \pi \nu \bar \nu$. 
Both in the SM and in the LR model, these are generated through loop diagrams.
For those channels sensitive to dipole operators, such as $B\to X_s\g$ 
and $K_L\to \pi^0 e^+ e^-$, the presence of a right-handed current causes the mLRSM contributions induced by $W_L$-$W_R$ mixing to be enhanced by ratios of $m_t/m_{d,s,b}$, making these rare decays very sensitive to $C_{Hud}^{ti}$.
Channels such as $B \rightarrow \mu^+ \mu^-$ and $K \rightarrow \pi \nu \bar \nu$ do not get contributions from dipole operators and thus do not obtain enhanced contributions in the mLRSM. With the experimental sensitivity approaching the SM 
level \cite{Aaboud:2018mst,Ahn:2020opg,CortinaGil:2021nts}, 
in the near future these channels might be used for an extraction of the $V_{L\,td}$ and $V_{L\,ts}$ CKM elements free of LR contamination.

\item Meson-antimeson oscillations.

A different source of stringent limits arise from $K-\bar K$ and $B-\bar B$ oscillations. Important examples include the meson mass differences, $\Dt m_{K,B_d,B_s}$, and $\varepsilon_K$ which measures CP violation in kaon mixing. The experimental input is very accurate, for instance uncertainties on $\Delta m_K$ and $\varepsilon_K$ are about $0.2\%$ and $0.8\%$, respectively. 
For observables dominated by short-distance contributions, such as 
$\varepsilon_K$ and the $B$-meson mass differences, the theoretical error is also under control. $\Delta m_K$ and the $D$ meson oscillations parameters, on the other hand, receive sizable (dominant in the case of $D$ mesons) long-distance contributions, which are hard to calculate in lattice QCD.
The mLRSM gives large contributions to these observables, both at tree- and loop-level, which generally lead to strong bounds on $M_H$ and $M_{W_R}$, with less sensitivity to $\xi$. As the same observables are usually used to determine the CKM elements involving the top quark, $V_{L\, ti}$, we again need to fit CKM and LR parameters simultaneously.

\item Electric dipole moments.

Finally, the EDMs of the neutron and diamagnetic atoms probe flavor-diagonal CP violation. While CKM contributions to EDMs are negligible \cite{Seng:2014lea,Pospelov_review,Czarnecki:1997bu,Mannel:2012qk}, in the mLRSM EDMs receive large tree-level contributions from the mixing between left- and right-handed $W$ bosons and are sensitive to the combination Im$\,C_{Hud}^{ij}V_L^{ij\, *}\sim \xi M_{W_R}^{-2}\,{\rm Im}(V_R^{ ij}V_L^{ij\, *}e^{i\al})$.
 
\end{itemize}

We describe the most salient features of these observables and relegate details to  App.\ \ref{app:observables}.

\subsection{Leptonic and semileptonic decays}\label{sec:treeDecays}

Our analysis of leptonic and semileptonic decays follows closely Ref.\ \cite{Alioli:2017ces}, with updated input on the lattice QCD 
calculations of mesonic decay constants and form factors, taken from Ref.
\cite{Aoki:2019cca}, and on the radiative corrections to nuclear decays 
\cite{Seng:2018yzq,Czarnecki:2019mwq}.
For each $u_i \rightarrow d_j$ transition, with $i \in \{u, c\}$
and $j \in \{ d, s, b\}$, it is possible to find at least two independent channels, sensitive to the vector or axial-vector component of the charged-current. In the presence of $W_L$-$W_R$ mixing, these receive corrections of opposite sign.  
Schematically
\begin{equation}
  F_V \left| V_{L\,i j} + \frac{v^2}{2}C_{Hud}^{ i j} \right| = O_{V,\, ij}^{\rm exp} ,  \qquad
F_{A}\left|V_{L\,i j} -\frac{v^2}{2}C_{Hud}^{i j}\right|  = \, O_{A,\, ij}^{\rm exp}\,, \label{eq:Vij} 
\end{equation} 
where $O_{\{V,A\},\, ij}^{\rm exp}$ denotes the experimental input, while $F_V$ and $F_A$  denote theoretical input, such as meson decay constants or (axial) vector form factors. 
The values for the relevant meson decay constants and form factors are collected in Table \ref{LQCDinput}.
The extraction of $V_{L\, i j}$ and $v^2 C_{Hud}^{i j}$ is thus limited by both experimental and theoretical uncertainties. 

The most relevant changes with respect to the analysis in Ref.\ \cite{Alioli:2017ces} correspond to the $ud$ and $us$ channels.
For the $u\to d$ transitions, the strongest constraint on the vector component comes from superallowed $0^+\to0^+$ transitions, 
while the leptonic decay $\pi\to \mu\nu$ probes only the axial-vector part of the current. Using theory predictions for $0^+\to 0^+$ transitions of Refs.\ \cite{Seng:2018yzq,Seng:2018qru,Czarnecki:2019mwq} along with the experimental input of Refs.\ \cite{Hardy:2014qxa,Hardy:2016vhg,Zyla:2020zbs}, we have 
\begin{align}
0^+\to0^+:&\qquad &\left| V_{L\,ud} + \frac{v^2}{2}C_{Hud}^{ud} \right|& = 0.97370 \pm 0.00014\, ,  \nn\\
\pi\to\mu\nu :& \qquad &f_{\pi}\left|V_{L\,ud} -\frac{v^2}{2}C_{Hud}^{ud}\right|&  = (127.13 \pm  0.02 \pm 0.13)\, \textrm{MeV}\,  , \label{eq:Vud} 
\end{align} 
where $f_\pi$ is the pion decay constant.

Right-handed currents also affect the $\beta$ asymmetry in neutron decay \cite{Bhattacharya:2011qm,Bernard:2007cf}, 
described by the parameter $\tilde\lambda$. While in the SM this parameter is determined by the ratio of the nucleon axial and vector charges, $g_A$ and $g_V$, in the mLRSM one has
\begin{equation}
\tilde \lambda = \frac{g_A}{g_V} \left(1  - \frac{v^2 C_{Hud}^{ud}}{V_{L\,ud}} \right)\,.
\end{equation}
$\tilde\lambda$ is measured with error of $0.1\%$, $\tilde\lambda = 1.2754 \pm 0.0013$ \cite{Zyla:2020zbs}. The extraction of $C^{ud}_{Hud}$ is limited  by the uncertainty on the lattice QCD determination of $g_A$. Currently, the most precise  calculation quotes an error of $1\%$ \cite{Chang:2018uxx}, so that $\pi \rightarrow \mu \nu$ still provides a stronger constraint. With a further reduction of the uncertainties by a factor of two, however, the neutron $\beta$ asymmetry will become competitive.

For the $s\to u$ transitions, semileptonic kaon decays 
probe the vector current,
while the ratio of leptonic kaon and pion decays probe the axial interaction. From Refs.\  \cite{Aoki:2019cca,Antonelli:2010yf} one obtains,
\begin{align}\label{eq:Vus}
K \to\pi l\nu_l\:& \qquad &f^{K\pi}_+(0)\left|V_{L\,us} + \frac{v^2}{2}C_{Hud}^{us}\right|  &= 0.2165 \pm 0.0004\, , \nn\\
K\to\mu\nu:&\qquad& \frac{f_{K}\left|V_{L\,us} - \frac{v^2}{2}C_{Hud}^{us}\right| }{f_{\pi}\left|V_{L\, ud} -  \frac{v^2}{2}C_{Hud}^{ud} \right| } &=  0.2760 \pm 0.0004\, . 
\end{align}

Eq.~\eqref{eq:Vud} uses a re-evaluation of the universal ``inner radiative corrections'' in $0^+\to 0^+$ transitions \cite{Seng:2018yzq,Seng:2018qru,Czarnecki:2019mwq}, which led to a reduction in the uncertainty and a significant shift of the central value. This resulted in a $3\sigma$ shift of the SM determination of $\left. V_{ud}\right|_{0^+ \rightarrow 0^+}$ from $0.97420 \pm 0.00021$ \cite{ParticleDataGroup:2018ovx} to the value in Eq.\ \eqref{eq:Vud}, and a resulting tension with CKM unitarity. As we will discuss in Section \ref{sec:CKM}, this tension can in principle be solved by right-handed currents, but in the mLRSM this requires a relatively light $W_R$, which is ruled out by other observables. For kaon decays, a new  lattice QCD calculation of  $f_+^{K\pi}(0)$, with  $N_f = 2 + 1 + 1$  \cite{Bazavov:2018kjg}, reduced the error by a factor of $1.6$, and somewhat increases the tension with the SM.  Here we will use the $N_f = 2+1$ values in Table \ref{LQCDinput} which lead to a less pronounced deviation from the SM.

We follow a similar strategy for the remaining elements of $V_L$ and $C_{H ud}$,
and give the relevant expressions for the leptonic and semileptonic decays of $D$ and $B$ mesons, and for decays of the $\Lambda_b$ baryon, in Appendix \ref{app:treeDecays}.  $B \rightarrow D l \nu_l$, $B \rightarrow D^* l \nu_l$ as well as 
the inclusive decays $B \rightarrow X_c l \nu_l$ 
and $\Lambda_b \rightarrow \Lambda_c \mu \nu_\mu$ allow one to determine the CKM parameter $A$, while $B \rightarrow \pi l \nu_l$, $B \rightarrow X_u l \nu_l$,  $B^+ \rightarrow \tau^+ \nu_\tau$, and $\Lambda_b \rightarrow p \mu \nu_\mu$ determine $|V_{L\,ub}|$, which is proportional to $|\bar\rho - i\bar \eta|$.

In addition to lifetimes and branching ratios, in the case of semileptonic decays of particles with spin it is possible to measure the 
triple correlation $\langle \vec J\, \rangle\cdot ( \vec p_e \times \vec p_\nu)$,
where $\vec J$ is the polarization of the decaying particle, which is  sensitive to time-reversal violation \cite{Jackson:1957zz}. This correlation has been measured in the decays of neutrons and $\Sigma$ baryons \cite{Mumm:2011nd,Hsueh:1988ar}, and can be used to constrain the imaginary part of $C_{H ud}$.

\subsection{Hadronic $\Delta S=1$ and $\Delta B=1$ charged-current processes}\label{sec:S=1}

This class includes hadronic decays of $K$ and $B$ mesons, such as  $K \rightarrow \pi \pi$,  $B \rightarrow \pi \pi$ and $B \rightarrow J/\psi K_S$.
In the SM, these receive tree-level contributions from the operators $C_{1\, LL}$
and $C_{2\, LL}$, induced by the exchange of a $W_L$ between quarks. In addition they can receive important contributions from strong and weak penguin diagrams \cite{Buchalla:1995vs}.

The most important observable in this class is $\varepsilon'$ that measures direct CP violation in $K\to \pi\pi$ decays and can be written as \cite{Buras:2005xt}
\bea
\varepsilon'=\frac{ie^{i(\dt_2-\dt_0)}}{\sqrt{2}}\left(\frac{{\rm Im} \,A_2}{{\rm Re }\, A_0}-\frac{{\rm Re} \,A_2}{{\rm Re }\, A_0}\frac{{\rm Im} \,A_0}{{\rm Re }\, A_0}\right) \,.
\eea
Here $A_{0,2}$ represent the amplitudes $A_{0,2} = \frac{1}{\sqrt{2}}\langle (\pi\pi)_{I=0,2} | i H |K^0\rangle$, with $I$ the isospin state of the pions. We use the experimental values for the real parts of these amplitudes
\bea
{\rm Re}\,A_0 = 33.201 \cdot 10^{-8} \, {\rm GeV}\,,\qquad {\rm Re}\,  A_2 = 1.479 \cdot 10^{-8} {\rm GeV}\,.
\eea

In the SM, the amplitudes $A_2$ and $A_0$ are real at tree level. An imaginary part is generated by one-loop diagrams with virtual top quarks, and $\varepsilon^\prime$ is proportional to the imaginary part of
\begin{equation}
\tau = - \frac{V^*_{L\,ts} V_{L\,td}}{V_{L\,us}^* V_{L\,ud}}, 
\end{equation}
which, in the SM \cite{Abbott:2020hxn}~\footnote{Notice that the value of $\tau_{SM}$ in Eq. \eqref{tauSM}, given in Ref. \cite{Abbott:2020hxn}, differs by about $10\%$ from the one obtained with the latest CKM fits in Ref. \cite{Zyla:2020zbs}. Since in our framework we need to rescale the lattice QCD estimate of $\varepsilon^\prime/\varepsilon_K$ to 
allow CKM parameters to vary from their SM values, we use the same $\tau_{SM}$ as given in Ref. \cite{Abbott:2020hxn}.
},
\bea\label{tauSM}
 \tau_{\rm SM} = \left(1.558(65) - 0.663(33) i \right) \cdot 10^{-3}\,.
\eea
The loop and CKM suppression, and the additional suppression by the $I=1/2$ rule, ${\rm Re}\,A_2/{\rm Re}\,A_0 \sim 1/22 $, lead us to expect a rather small value, to be compared with the experimental value
\bea
{\rm Re}\left(\varepsilon'/\varepsilon_K\right)_{\rm exp}= 16.6(2.3) \cdot 10^{-4}\,.
\eea
In the SM, ${\rm Im} A_0$ and ${\rm Im} A_2$ are dominated by the matrix elements of strong and weak penguin operators, respectively (see, for example,  the discussion in Ref.\ \cite{Cirigliano:2011ny}).  
Recent first-principle calculations of these matrix element in lattice QCD have significantly reduced the error of the SM prediction \cite{Abbott:2020hxn}, 
which now reads
\bea
{\rm Re}\left(\varepsilon'/\varepsilon_K\right)_{\rm SM}= \frac{{\rm Im}\, \tau }{{\rm Im}\, \tau_{\rm SM}} \times 21.7(2.6)(6.2)(5.0) \cdot 10^{-4}\,,
\eea
where the errors are the statistical and systematic uncertainties, with the latter broken up into  isospin-conserving and isospin-violating pieces. 
This estimate is in good agreement with a recent reappraisal of the SM value of  $\varepsilon'/\varepsilon_K$ based on $\chi$PT and large-$N_c$, which yields \cite{Cirigliano:2019cpi}
\bea
{\rm Re}\left(\varepsilon'/\varepsilon_K\right)_{\rm SM}= 14(5) \cdot 10^{-4}\,.
\eea

The imaginary parts of $A_0$ and $A_2$ receive new contributions from the LR and RR operators appearing in Eq.\ \eqref{eq:LowLag}. Most of these contributions can be derived from the chiral Lagrangian discussed in Sect.\ \ref{sec:chiralCPodd}, the only additional terms arise from the parts of the RR operators that transform as $\boldsymbol{27}_{R}\times \boldsymbol 1_{L}$, which were omitted in the chiral discussion of Sect.\ \ref{sec:chiralCPodd}. These contributions were determined in Ref.\ \cite{Blum:2012uk} and, together with the other BSM contributions, give
\bea\label{eqs:A02}
{\rm Im}\,A_2&=&\frac{F_0}{2\sqrt{6}}\mathcal A_{i\, LR}{\rm Im}\,\left(C_{i\, LR}^{suud}-\left(C_{i\, LR}^{duus}\right)^*\right) +\frac{1}{12\sqrt{3}}\mathcal A_{(27,1)}'{\rm Im}\,(C_{1\, RR}^{duus}+C_{2\, RR}^{duus})\,,\\
{\rm Im}\,A_0
&=& -\frac{F_0}{\sqrt{3}}\mathcal A_{i\, LR}{\rm Im}\,\left(C_{i\, LR}^{suud}-\left(C_{i\, LR}^{duus}\right)^*\right)-\frac{\sqrt{3} F_0}{4}(m_K\sq -m_\pi\sq)\mathcal A_{i\, LL}^{(8)}{\rm Im}\,\left(C_{i\, RR}^{duus}\right)\,,\nn
\eea
where $\mathcal A_{i\, LL}^{(8)}=\mathcal A_{i\, RR}^{(8)}$, $\mathcal A_{i\, LR}$ are given in Eq.\ \eqref{eq:LECsKpp} and 
$\mathcal A_{(27,1)}'(3\, {\rm GeV}) = \,0.0461(14)\, {\rm GeV}^3$. Here we neglected the contributions to $A_0$ proportional to $\mathcal A_{(27,1)}'$ because, as mentioned in Sect.\ \ref{sec:chiralCPodd}, these terms can be shown to be small compared to the $\boldsymbol 8_R\times \boldsymbol 1_L$ contributions.

The other observables in this class include $B \rightarrow J/\psi K_S$, $B \rightarrow \pi \pi$, and other $\Delta B =1$ decays used to determine the CKM angles $\alpha$, $\beta$ and $\gamma$ \cite{Zyla:2020zbs}. In Appendix \ref{app:Beta} we argue that the LR contribution due to tree-level $W_R$ exchange to time-dependent CP asymmetry in $B \rightarrow J/\psi K_S$ can be neglected within current uncertainties, and thus the standard extraction of $\beta= {\rm Arg} (- V_{L\,cd} V^*_{L\, cb}/V_{L\,td} V^*_{L\, tb})$ can be used in the CKM fits. While similar considerations likely apply to other non-leptonic channels such as $B\to\pi\pi$ and $B\to DK$, used to determine $\alpha$ and $\gamma$, we do not explicitly include them in our analysis as hadronic matrix elements associated to LR contributions are not under control. 
Finally, the corrections to the $B^{0}_d$ and $ B^{0}_s$ widths also belong to this class. 
We compute the mLRSM corrections in App.~\ref{app:Bwidths}.

\subsection{$\Delta F =2$ processes}\label{sec:DF2}

\begin{table}
\center
\begin{tabular}{||c||c|c||c|c||}
\hline
$\Delta S=2$ & $\Delta M_K$ & $\left(5.293  \pm 0.009\right)   \, \textrm{ns}^{-1}$& $|\varepsilon_K|$ &$ \left(2.228 \pm 0.011\right) \cdot 10^{-3}$ \\
\hline
$\Delta B =2$ & $\Delta m_d$ & $\left(0.5064 \pm 0.0019\right)  \, \textrm{ps}^{-1}$& $\Delta m_s$ &$ 17.7656 \pm 0.0057 \, \textrm{ps}^{-1}$ \\
 & $\Delta \Gamma^{(d)}$ & $ (-1.3\pm 6.7)\Ex{-3} \,{\rm ps}^{-1}$&  $\Delta \Gamma^{(s)} $ &  $(0.086\pm 0.006)\,{\rm ps}^{-1}$\\
 & $a_{\rm fs}^d $ & $ -0.0020 \pm 0.0016$& $ a_{\rm fs}^s$ & $ -0.0006 \pm 0.0028$ \\
\hline 
$\Delta B = 1$ & $\text{BR}\,(B\to X_d\g)$ & $(14.1\pm 5.7)\cdot 10^{-6}$  & $\text{BR}\,(B\to X_s\g)$& $(3.32\pm 0.15)\times 10^{-4}$ \\
& $A_{CP}(B\to X_{d+s}\g)$ & $0.032 \pm 0.034$ & $A_{CP}(B\to s\g)$ &  $0.015\pm 0.02$  \\
& & & $S_{K^*\g}$ & $ -0.16\pm0.22$      \\
\hline
\end{tabular}
\caption{Experimental input for the processes discussed in Section \ref{sec:DF2} 
and for the $\Delta B =1$ processes discussed in Appendix \ref{app:DeltaB1}
\cite{Zyla:2020zbs,Amhis:2019ckw,Aaij:2021jky}.
The branching ratios $\text{BR}\,(B\to X_{d,s}\g)$ have a cut on the photon energy,  $E_\gamma > 1.6$ GeV.}
\label{DB1exp}
\end{table}

We move on to observables in $B - \bar B$ and $K - \bar K$ oscillations that severely constrain the mLRSM.
The experimental input on the $B - \bar B$ mass and width differences, $\Delta m_d$, $\Delta m_s$, $\Delta \Gamma^{(d)}$ and $\Delta \Gamma^{(s)}$, the $K - \bar K$
mass difference $\Delta m_K$, and $\varepsilon_K$, which measures CP violation in 
$K - \bar K$ mixing, are reported in Table \ref{DB1exp}.
We now discuss the theoretical input, and the leading uncertainties.

\subsubsection{$B - \bar B$ oscillations}\label{sec:BBbar}

For the $B_q$ mesons, with $q=\{d,s\}$, to good approximation we can use
\bea
\Delta m_{q} = 2|M_{12}^{(q)}| =\frac{\left| \langle \bar{B}_q^0 | \mathcal H_{eff} (\Dt B=2)| B_q^0 \rangle \right|}{m_{B_q}} \,.
\eea
Within the SM the $\Dt B = 2$ Hamiltonian involves operators of the form $(\bar b_L \ga_\mu q_L) (\bar b_L \ga^\mu q_L)$ that are generated through box diagrams. This leads to 
\bea
M_{12}^{(q)}\big|_{\rm SM} =\frac{G_F\sq m_W\sq m_{B_q}}{12\pi\sq}\left(V_{L\,tq}^*V_{L\, tb}\right)\sq f_{B_{q}}\sq \hat B_{B_q}\eta_B S_0(x_t,x_t)\, ,
\eea
with $x_i = m_i^2/m_W^2$ and  $x_t$ should be evaluated at $\mu=m_t$,  $\eta_B = 0.55\pm0.01$ \cite{Buras:2013ooa}. The loop function $S_0(x_i,x_j) = \frac{1}{4}(f_1(x_i,x_j)-f_1(0,x_j)-f_1(x_i,0)+f_1(0,0))$, with
\bea
 f_1(x_i,x_j) &=&  - \frac{x_j^2 (4  -8 x_j  + x^2_j)}{ (x_i - x_j) (-1 + x_j)^2  } \log(x_j)  + \frac{x_i^2 (4  -8 x_i  + x^2_i) }{ (-1 + x_i)^2 (x_i - x_j)} \log x_i \,.
\eea
Finally, the RG-invariant bag parameter, $\hat B_{B_q}$, is related to the matrix element of the left-handed operator mentioned above, for which we use  the FLAG average \cite{Aoki:2019cca} shown in Table \ref{TabBag}.

The BSM contributions arise from the $O_{4,5}$ operators in Eq.\ \eqref{eq:LowLag}, which are generated through exchange of heavy scalar bosons and loop diagrams involving $W_R$. The contributions are
\bea\label{eq:BBbarM12}
M_{12}^{(q)}\big|_{\rm LR} = \frac{m_{B_q}f_{B_q}^2}{2}\left[ \frac{1}{3}C_4^{bdbd}B_5\left(R_q(\mu)+\frac{3}{2}\right)+C_5^{bdbd}B_4\left(R_q(\mu)+\frac{1}{6}\right)\right]^*\,,
\eea
where $R_q(\mu) =  m^2_{B_q}/(m_b(\mu) + m_q(\mu))^2$ and the bag factors, related to the matrix elements of $O_{4,5}$, are shown in Table \ref{TabBag}.

We then use the above expressions with $\Delta m_{q} =2\big|M_{12}^{(q)}\big|_{\rm SM} +M_{12}^{(q)}\big|_{\rm LR} \big|$ to estimate the mass differences, which we compare with the experimental values \cite{Zyla:2020zbs} shown in Table \ref{DB1exp}.

\begin{table}
\center\small
\begin{tabular}{||c|ccccc||}
\hline
&  $f_{B_q}\sqrt{\hat B_{B_q}}\, ({\rm MeV})$   &$f_{B_q}^2 B_4$ (GeV$^2$)  & $f_{B_q}^2 B_5$ (GeV$^2$)  & $f_{B_q}^2 B_2$ (GeV$^2$)  & $f_{B_q}^2 B_3$ (GeV$^2$)  \\
\hline 
$B^0_d - \bar{B}^0_d$ &  225 (9) &  0.0390 (28)(8) &  0.0361 (35)(7) & 0.0285 (26)(6) &  0.0402 (77)(8)
\\
$B^0_s - \bar{B}^0_s$ &  274 (8) &  0.0534 (35)(7) & 0.0493 (36)(10) &  0.0421 (27)(8) &  0.0576 (77)(12)
\\
\hline
  & $\hat{B}_K$ & $B_4$ & $B_5 $ & & \\
  \hline
 $K_0 -\bar K_0$ & $0.7625(97)$  & $0.926(19)$ & $0.720(38)$ & &  \\
\hline 
\end{tabular}
\caption{Relevant bag parameters for $B_q - \bar{B}_q$ oscillations
and $K_0 - \bar K_0$ oscillations. 
For $B_q - \bar{B}_q$ oscillations we use the RG-invariant definition, $\hat B_{B_q}$ \cite{Aoki:2019cca}, for the SM operator, while the bag parameter for the LR model are given in the $\overline{\textrm{MS}}$ scheme, at the renormalization scale $\mu = m_b$  \cite{Bazavov:2016nty}. 
For $K_0 - \bar K_0$ oscillations, $\hat B_{K}$  is RG-invariant \cite{Aoki:2019cca}, while $B_4$ and $B_5$ are given in the $\overline{\textrm{MS}}$ scheme, at $\mu = 3$ GeV. We use the $N_f = 2 + 1$ averages reported in Ref.\ \cite{Aoki:2019cca}.
}\label{TabBag}
\end{table}

\subsubsection{$\Delta m_K$ and $\varepsilon_K$}\label{sec:S=2}
The mixing between $\bar K^0$ and $K^0$ is described by the off-diagonal matrix element,
\bea
2m_K M_{12}^* = \langle \bar K^0|H_{eff}(\Delta S=2)|K^0\rangle\,.
\eea
To good approximation, the real part of this amplitude determines the kaon mass difference 
\bea\label{eq:DeltaMk}
\Dt M_K=M_{K_L}-M_{K_S}=2{\rm Re}\, M_{12}\,,
\eea
while the imaginary part is connected to CP violation in $\bar K^0-K^0$ mixing, described by $\varepsilon_K$ \cite{Buchalla:1995vs},
\bea\label{eq:epsK}
\varepsilon_K = \frac{A(K_L\to (\pi\pi)_I=0)}{A(K_S\to (\pi\pi)_I=0)} \simeq \frac{e^{i\pi/4}}{\sqrt{2}\Dt M_K}\left({\rm Im}\, M_{12}+2{\rm Re}\, M_{12}\, \frac{{\rm Im}\, A_{0}}{{\rm Re}\, A_{0}}\right)\,,
\eea
where the second equality uses the approximation  $\Dt \Gamma_K  \simeq -2\Dt M_K $ \cite{Buras:2005xt}. 
\\\\\noindent{\bf The SM prediction\\}
Starting with the SM prediction, $M_{12}$ receives both short- and long-distance contributions. The former arise from local $\Dt S=2$ operators, which appear at loop level in the SM and give rise to
\bea\label{eq:M12SM}
M_{12}^{\rm SM}\big|_{SD} &=& \frac{G_F\sq m_W\sq }{12\pi\sq}m_K f_K\sq \hat{B}_K\bigg(\eta_{cc}\lambda_c\sq S_0(x_c)+2\eta_{ct}\lambda_c\lambda_t S_0(x_c,x_t)
+\eta_{tt}\lambda_t\sq S_0(x_t) \bigg)^*\, ,\nn
\eea
where $\lambda_{i} = V_{L\, is}^*V_{L\, id}$, $x_t$ should be evaluated at $\mu=m_t$ and $x_c$ at $\mu=m_c$ and $\hat B_K$ describes the non-perturbative matrix element, given in Table \ref{TabBag}.
From Refs.\ \cite{Aoki:2019cca,Buras:2013ooa} we have
\bea\label{eq:epsKsmInput}
\eta_{cc}&=&1.87\pm0.76,\qquad \eta_{ct}=0.496\pm0.047\, ,\qquad \eta_{tt}= 0.5765\pm 0.0065\, ,
\eea
while the loop function $S_0$ is given in Sect.\ \ref{sec:BBbar}. 
The short-distance contributions dominate in the CP-violating observable $\varepsilon_K$, allowing us to write
\bea
\varepsilon_K^{\rm SM} = \frac{e^{i\pi/4}\kappa_\varepsilon}{\sqrt{2}\Dt M_K^{\rm expt.}}{\rm Im}\, \left(M_{12}^{\rm SM}\big|_{\rm SD}\right)\,,
\eea
where $\kappa_\varepsilon = 0.94\pm0.02$ 
\cite{Buras:2013ooa} takes into account long-distance contributions. In the case of $\varepsilon_K$, it is advantageous to use the unitarity of the CKM matrix  to rewrite the contributions from $cc$, $ct$, and $tt$ graphs in Eq.\ \eqref{eq:M12SM} in terms of $ut$ and $tt$ diagrams. This leads to \cite{Brod:2019rzc}
\begin{equation}\label{eq:epsKSM}
 | \varepsilon_K^{\rm SM}| = \frac{G_F^2 f_K^2 m_{K^0} m_W^2}{6 \sqrt{2} \pi^2 \Delta M_K} \hat{B}_K \kappa_\varepsilon \, |V_{L\,cb}|^2 \lambda^2 \bar\eta \Big( |V_{L\,cb}|^2 (1 - \bar\rho) \eta_{tt} \mathcal{S}(x_t)
 - \eta_{ut} \mathcal S(x_c,x_t)\Big)\,,
\end{equation}
where $\lambda$, $\bar\eta$ and $\bar\rho$ determine the CKM matrix in the Wolfenstein parametrization  \cite{Buras:1998raa}. The loop functions are given by
\begin{eqnarray}
 \mathcal S(x_t) &=& S_0(x_t) + S_0(x_c) -2 S_0(x_c,x_t)\,, \nn\\
 \mathcal S(x_c,x_t) &=& S_0(x_c) - S_0(x_c,x_t)\,, 
\end{eqnarray}
and the running factors are 
\begin{eqnarray}
 \eta_{tt} &=& 0.55 (1 \pm 4.2\% + 0.1\%) = 0.55 \pm 0.02 \,,\nn\\
 \eta_{ut} &=& 0.402 (1 \pm 1.3\% \pm 0.2\% \pm 0.2\%) = 0.402 \pm 0.005\,, 
\end{eqnarray}
leading to a small uncertainty on $\eta_{ut}$ compared to large uncertainties in the $cc$ and $ct$ running factors, at the price of a slightly larger uncertainty on $\eta_{tt}$. We use Eq.\ \eqref{eq:epsKSM} for the SM prediction.

Unfortunately, unitarity cannot be used in the same way for the SM prediction for the real part of the amplitude that gives rise to  $\Dt M_K$. We therefore employ Eq.\ \eqref{eq:M12SM} to obtain the SM expression for the short-distance contribution to $\Dt M_K$.
In addition, long-distance contributions are significant in this case and lead to sizable uncertainties. We will assume no significant discrepancy between the SM and experimental measurement and simply use the experimental determination to estimate the SM prediction of  $\Dt M_K$. We thus assign a theoretical uncertainty of $\sigma^2_{\rm SM}=\sigma^2_{\rm SD, SM}+\left(\Dt M_K|_{\rm expt.} -\Dt M_K^{\rm SM}|_{\rm SD}\right)^2$, where  $\sigma_{\rm SD, SM}$ is the uncertainty due to $\Dt M_K^{\rm SM}|_{\rm SD}$.
\\\\\noindent{\bf The BSM contributions\\}
Short-distance LR contributions arise through the $O_{4,5}$ operators in Eq.\ \eqref{eq:LowLag}
\bea\label{eq:epsKSD}
 M_{12}^{\rm LR}\big|_{SD} = \frac{m_K f_K\sq}{2}
\left(\frac{m_K}{m_d+m_s}\right)\sq \left(\frac{1}{3}B_5C_4^{s{ds}d}+B_4C_5^{s{ds}d}\right)^*\,,
\eea
where $n_f=2+1$ lattice calculations of the matrix elements are given in Table \ref{TabBag}. Long-distance effects are induced by two insertions of $\Dt S=1$ operators, e.g.\ $C_{i\, LL}\times C_{i\, LR}$ and $C_{i\, LL}\times C_{i\, RR}$. We  neglect the parts of the LL, RR operators that transform as $\boldsymbol {27}_{L,R}\times \boldsymbol 1_{R,L}$, and use the $\boldsymbol 8_{L,R}\times \boldsymbol 1_{R,L}$ pieces to estimate these effects (see the discussion around Eq.\ \eqref{eq:LECsKpp}). The long-distance pieces can then be evaluated using the chiral Lagrangian in Eq.\ \eqref{mesonEps'}. This gives
\bea\label{eq:epsKLD}
2m_K M_{12}^{\rm LR}\big|_{LD}
&=&F_0^4G_8\frac{m_{K^0}^2(4m_{K^0}^2-3m_\eta^2-m_{\pi^0}^2)}{(m_{K^0}^2-m_{\pi^0}^2)(m_{K^0}^2-m_{\eta}^2)}\bigg[
-\frac{1}{2}{\cal A}_{i\,LR}\left(C^{su\,ud}_{i\,LR} + \left(C^{du\,us}_{i\,LR}\right)^*\right)\nn\\
&&+\frac{m_K^2}{3}{\cal A}_{i\, RR}^{(8)} C_{i\, RR}^{suud}\bigg]^*\,,
\eea
where $G_8 = \mathcal A_{i\, LL}^{(8)}C_{i\, LL}/4$ is the coefficient of the SM operators transforming as $\boldsymbol 8_{L}\times \boldsymbol 1_{R}$.
As in the SM \cite{Buras:2010pza}, these contributions vanish at LO in $\chi$PT after taking into account the Gell-Mann-Okubo relation. The first contributions then arise at N$^2$LO where loops and new LECs appear. As we do not control these LECs, we estimate the  contributions by using the experimental values for the meson masses in Eq.\ \eqref{eq:epsKLD} and assign a $50\%$ uncertainty to this result \cite{Cirigliano:2016yhc}.

We then estimate $\Dt M_K$ by using $\Dt M_K = \Dt M_K\big|_{\rm expt.}+ 2{\rm Re}\,M_{12}^{\rm LR}$, with $M_{12}^{\rm LR}=M_{12}^{\rm LR}\big|_{SD}+M_{12}^{\rm LR}\big|_{LD}$. To compute the CP violation in mixing we use $\varepsilon_K = \varepsilon_K^{\rm SM} + \varepsilon_K^{\rm LR}$. We rewrite Eq.\ \eqref{eq:epsK} 
\bea
\varepsilon_K^{\rm LR} = \frac{e^{i\pi/4}}{\sqrt{2}}\left(\frac{{\rm Im}\, M^{\rm LR}_{12}}{\Dt M_K^{\rm expt.}}+\frac{{\rm Im}\, A^{\rm LR}_{0}}{{\rm Re}\, A_{0}^{\rm expt.}}\right)\,,
\eea
where the mLRSM contributions to Im $A_0$ are given by Eq.\ \eqref{eqs:A02}.

To obtain constraints we finally compare the above theoretical expressions with the experimental measurements given in Table \ref{DB1exp}. 
We treat the experimental uncertainties and those due to Eqs.\ \eqref{eq:epsKsmInput}, \eqref{eq:epsKSD}, and \eqref{eq:epsKLD} as statistical.
\\\\
As mentioned in Sect.\ \ref{sec:LEFTlow} our analysis of the short-distance contributions to $\Dt F=2$ observables is similar to that of Refs.\ \cite{Maiezza:2010ic,Bertolini:2014sua}. Differences arise from our use of updated lattice QCD results and a somewhat different approach to the diagrams involving intermediate $c-c$ and $c-t$ quarks. Comparing numerically to the expressions of Ref.\ \cite{Bertolini:2014sua}, we find that the heavy Higgs contributions agree to within $20\%$ after turning off the running between $m_W$ and $M_{W_R}$. Similar agreement is found for the $W_R$ contributions that are due to $t-t$ diagrams, while we find the terms induced by the $c-c$ and $c-t$ graphs to be larger by a factor of $\sim 1.6$ and $3.9$, respectively. Note that these contributions are only potentially significant for the kaon system, while the $B_{d,s}$ systems are dominated by the $t-t$ graphs. In addition, we take into account the RGE evolution between $M_{W_R}$ and $m_W$, the effects of which are discussed in Sect.\ \ref{sect:RGEsummary}, with approximate formulae given in App.\ \ref{app:RGE}. 

Apart from these different treatments of LR contributions, there are slight differences in the fitting procedures.
Ref.\  \cite{Maiezza:2014ala}  constrained LR contributions by demanding that they are smaller than a certain fraction of the SM prediction, in the case of $\varepsilon_K$ and $\Dt M_K$, while using the results of a fit that assumes BSM physics to dominantly arise in  $\bar B-B$ oscillations \cite{Charles:2004jd} to constrain $M_{12}^{(d,s)}$ in the $B_{d,s}$-meson sector.
Instead, we fit theoretical results for observables (including up-to-date SM predictions) directly to experimental measurements, taking into account theoretical and experimental uncertainties as described above. This allows us to incorporate the LR contributions to other flavor observables in a consistent manner, without having to assume that LR effects are dominant in a certain sector.

\subsection{$\Delta F=0$ observables: Electric dipole moments}\label{sec:EDMs}
EDMs set stringent limits on the CP-violating interactions within the mLRSM. Here we focus on the contributions to the EDMs of hadronic and nuclear systems, the current experimental limits of which are collected in Table \ref{tab:edms}. In this section, we assume a Peccei-Quinn mechanism is active. In the absence of such a mechanism, all EDMs are dominated by the induced $\bar \theta$ term (see Sect.~\ref{strongCP}). 

\begin{table}[t!]
\center
\renewcommand{\arraystretch}{1.2}
$\begin{array}{||c|cccc||}
\hline
e\, {\rm cm}& d_n &d_{p,D}   & d_{\rm Hg} & d_{\rm Ra}  \\\hline
\mathrm{current}  & 1.8 \cdot 10^{-26}  & - &6.3\cdot 10^{-30}   & 1.2\cdot 10^{-23} \\
\mathrm{expected} &1.0 \cdot 10^{-28}& 1.0 \cdot 10^{-29} &1.0\cdot 10^{-30}  & 1.0 \cdot 10^{-27} \\
\hline
  \end{array}$
\caption{
The first row shows the current 90\% C.L limits on the EDMs of the neutron \cite{Afach:2015sja,Baker:2006ts,Abel:2020gbr}, $^{199}$Hg \cite{Griffith:2009zz,Graner:2016ses}, and $^{225}$Ra \cite{Bishof:2016uqx}. The second row shows the expected sensitivities of future EDM experiments, see Ref.~\cite{Chupp:2017rkp}. }\label{tab:edms}
\end{table}

\subsubsection{Nucleon EDMs}\label{sec:nEDMs}
The EDMs of the neutron and proton receive contributions from several operators. We start with the four-quark operators, discussed in Section \ref{sec:chiralCPodd}, that generate sizable pion-nucleon couplings. 
These operators give rise to direct and indirect contributions to the nucleon EDMs. The former are governed by so far unknown LECs, while the latter are due to loop diagrams involving the CP-violating pion-nucleon couplings of section \ref{sec:piN}. The EDMs resulting  from the four-quark operators can be written as follows \cite{Seng:2014pba}
\bea
d_n|_{LR} &=& \bar d_n(\mu)|_{LR} + \frac{e g_A \bar g_1|_{LR}}{(4\pi F_\pi)^2}\left(\frac{\bar g_0|_{LR}}{\bar g_1 |_{LR}}\left( \log\frac{m_\pi\sq}{\mu\sq}-\frac{\pi m_\pi}{2m_N}\right)+\frac{1}{4}(\kappa_1-\kappa_0) \frac{m_\pi^2}{m_N^2}\log \frac{m_\pi^2}{\mu^2}\right)\,,\nn\\
d_p|_{LR} &=& \bar d_p(\mu)|_{LR} - \frac{e g_A \bar g_1|_{LR}}{(4\pi F_\pi)^2}\bigg[\frac{\bar g_0|_{LR}}{\bar g_1|_{LR} }\left( \log\frac{m_\pi\sq}{\mu\sq}-\frac{2\pi m_\pi}{m_N}\right)\nn\\
&&-\frac{1}{4}\left(\frac{2\pi m_\pi}{m_N}+\left(\frac{5}{2}+\kappa_1+\kappa_0\right) \frac{m_\pi^2}{m_N^2}\log \frac{m_\pi^2}{\mu^2}\right)\bigg]\,,
\eea
where $ \bar g_{0,1}|_{LR}$ are given in Eq.\ \eqref{relations0} and $\bar d_{n,p}(\mu)|_{LR} $ are unknown LECs due to  the direct contributions of the four-quark operators. In addition, $g_A\simeq 1.27$, and $\kappa_0 = -0.12$ and $\kappa_1 = 3.7$ are related to the nucleon magnetic moments. We estimate these contributions by taking $\mu=m_N$ with $\bar d_{n,p}(m_N)=0$ as a central value. The impact of the associated theoretical uncertainty due to the unknown LECs was discussed in Ref.\ \cite{Cirigliano:2016yhc}.

In the case of the quark CEDMs both the direct and indirect contributions to the nucleon EDMs involve unknown LECs. We therefore employ QCD sum-rules estimates to estimate the total induced nucleon EDMs \cite{Pospelov_qCEDM,Pospelov_deuteron, Pospelov_review, Hisano1}, while we use recent QCD sum-rule \cite{Haisch:2019bml} and quark-model \cite{Yamanaka:2020kjo} calculations to estimate the contributions of the Weinberg operator. In addition, the nucleon EDMs receive contributions from the remaining CP-odd interactions, namely, the quark EDMs. Assuming a Peccei-Quinn mechanism, the sum of these terms then takes the form
\bea\label{eq:nEDMs}
d_n&=&d_n|_{LR}+
g_T^u\,d_u+g_T^d\,d_d+g_T^s\,d_s\,\nn\\
&&-(0.55\pm0.28)\,e\,\tilde d_u-(1.1\pm0.55)\,e\,\tilde d_d -20\,(1\pm 0.5)\,{\rm MeV}\,e\,g_sC_{\tilde G}\,,\nn\\
d_p&=&d_p|_{LR}+
g_T^d\,d_u+g_T^u\,d_d+g_T^s\,d_s\nn\\
&&+(1.30\pm0.65)\, e\,\tilde d_u+(0.60\pm0.30)\,e\,\tilde d_d  +18\,(1\pm0.5)\,{\rm MeV}\,e\,g_sC_{\tilde G}\,,
\eea
where $d_u =eQ_u m_u {\rm Im }\, C_{\gamma u}^{uu}$ and $d_q =eQ_q m_q {\rm Im }\, C_{\gamma d}^{qq}$ for $q=d,s$. The strange CEDM induces vanishing contributions if a Peccei-Quinn mechanism is active \cite{Pospelov_qCEDM}. The quark-EDM contributions have been determined by lattice QCD calculations \cite{Bhattacharya:2015esa, Bhattacharya:2015wna,Bhattacharya:2016zcn,Gupta:2018qil,Gupta:2018lvp}, which give at $\mu = 1$ GeV 
\bea
 g_T^u &=& -0.213 \pm 0.012\,,\qquad g_T^d = 0.82 \pm 0.029\,,\qquad g_T^s =- 0.0028\pm 0.0017\,.
\eea 
All couplings in Eq.\ \eqref{eq:nEDMs} should be evaluated at 1 GeV.

\subsubsection{Nuclear and atomic EDMs}\label{subsec:EDMs}
We finally consider expressions for the EDMs of light nuclei and diamagnetic atoms. 
The EDMs in the former category are theoretically attractive as they can accurately be described in terms of the nucleon EDMs and the pion-nucleon couplings \cite{Bsaisou:2014zwa,Yamanaka:2015qfa}. We will focus on the EDM of the deuteron in the following. Although no experimental limits have been set on the EDMs of light nuclei so far, there are advanced proposals to measure them in electromagnetic storage rings \cite{Eversmann:2015jnk}, with an expected sensitivity given in Table \ref{tab:edms}.  

In contrast, the EDMs of diamagnetic atoms are stringently constrained experimentally, especially that of $^{199}$Hg, but they are subject to much larger theoretical uncertainties. 
The main contributions to the EDMs of these systems are expected to arise from the nuclear Schiff moment, as there are no large enhancement factors to mitigate the Schiff screening by the electron cloud \cite{Schiff:1963zz}. The nuclear Schiff moment obtains large contributions from the pion-nucleon couplings, $\bar g_{0,1}$, which, however, require complicated many-body calculations. Currently, such calculations cannot be performed with good theoretical
control \cite{Dmitriev:2003sc,deJesus:2005nb,Ban:2010ea,Dzuba:2009kn,Engel:2013lsa}, leading to large nuclear uncertainties, while the contributions from the nucleon EDMs are under better control. Here we will focus on the EDMs of mercury, currently the most stringently constrained system experimentally, and radium. The experimental limit on the latter is significantly weaker than the former, but future measurements aim at improvements of several orders of magnitude. 

Collecting all the above information, we use
\bea 
d_{D} &=&
(0.94\pm0.01)(d_n + d_p) - \left [ (0.18 \pm 0.02) \,\frac{\bar g_1}{2F_\pi}\right] \,e \,{\rm fm} \, ,\nn\\
d_{\rm Hg}&=& -(2.1\pm0.5)
\Ex{-4}\bigg[(1.9\pm0.1)d_n +(0.20\pm 0.06)d_p\nn\\
&&\qquad\qquad\qquad\qquad-\bigg(0.13^{+0.5}_{-0.07}\,\frac{\bar g_0}{2F_\pi} +
0.25^{+0.89}_{-0.63}\,\frac{\bar g_1}{2F_\pi}\bigg)e\, {\rm fm}\bigg]\,,\nn\\
d_{\mathrm{Ra}} &=& (7.7\pm 0.8)\Ex{-4}\cdot\left[\left(-2.5\pm 7.6\right) \,\frac{\bar g_0}{2F_\pi} + \left(63\pm 38\right)\,\frac{\bar g_1}{2F_\pi}\right]e\, {\rm
	fm}\,,
\eea
where $\bar g_{0,1}  = \bar g_{0,1}\big|_{LR}+\bar g_{0,1}\big|_{CEDM}$ can be read from Eqs.\ \eqref{relationsA} and \eqref{eq:piNCEDM}, $d_{n,p}$ are given by Eq.\ \eqref{eq:nEDMs}, and the experimental constraints are shown in Table \ref{tab:edms}.
Within our analysis we estimate the EDMs by using the central values for the relevant hadronic and nuclear matrix elements and refer to Refs.\cite{Chien:2015xha,Cirigliano:2016yhc} for a discussion on the impact of the associated uncertainties.

\section{Results}\label{sec:results}
After computing the observables described in the previous section we construct a $\chi^2$ 
\bea
\chi^2 = \sum_{i=\{\rm obs \}}\left(\frac{O^{\rm th}_i-O^{\rm expt}_i}{\sigma_i}\right)^2\,,
\eea
where $O^{\rm th}_i$ and $O^{\rm expt}_i$ are the theoretical and experimental determinations of a particular observable and  $\sigma_i$ is determined by summing the corresponding experimental and theoretical uncertainties described in the previous section in quadrature. The $\chi^2$ function thus depends on the parameters appearing in the LR model, $M_{W_R}$, $M_H$, $\al$, and $\xi$, as well as the SM CKM elements. 

Some of the LR parameters are subject to theoretical constraints. As discussed in Ref.~\cite{Basecq:1985cr}, the masses $M_{W_R}$ and $M_H$ are both related to the vev $v_R$, so that $M_H/M_{W_R}$ is given by the ratio of parameters in the Higgs potential and the $SU(2)$ gauge coupling.
As the latter is fixed from experiment, a significant hierarchy $M_H\gg M_{W_R}$ would force
the parameters in the Higgs potential to become non-perturbatively large. Because our description breaks down in this part of parameter space, we focus on the region $M_H<8 M_{W_R}$. Note that if one wants to keep these parameters in the perturbative regime up to the Grand Unification scale, $\mu\sim 10^{16}$ GeV, stringent limits on the LR scale of $v_R\gtrsim 10$ TeV can be set as well \cite{Chauhan:2018uuy}. 

Similarly, for tuned values of  $\ka'/\ka=\xi\simeq 1$ certain parameters in the Higgs potential would have to become non-perturbatively large, see App.\ \ref{app:validity}. To avoid this region we take $|\xi|\leq 0.8$. The CP-violating combination of parameters, $ t_{2\beta}s_\alpha=  \tan 2\beta \sin \alpha$, is constrained to be  $\vert t_{2\beta}s_\alpha\vert\lesssim 2m_b/m_t$ in order to reproduce the quark masses \cite{Maiezza:2010ic,Senjanovic:2014pva}, see App.\ \ref{app:validity} for more details.
Finally, for the CKM elements we use the Wolfenstein parametrization, which parametrizes the CKM matrix in terms of $\la$, $A$, $\bar \rho$, and $\bar \eta$, and we expand the expressions up to $\Or(\lambda^6)$ \cite{Buras:1998raa}.  We then simultaneously fit the four CKM parameters along with the LR parameters. 

Obtaining constraints, e.g.\ in the $M_{W_R}-M_H$ plane, involves marginalizing over the remaining SM and LR parameters.
This minimization of the $\chi^2$ is performed using NLopt~\cite{Johnson}, a free/open-source library for nonlinear optimization which includes various global and local optimization algorithms. In particular, an Improved Stochastic Ranking Evolution Strategy~\cite{Runarsson:2005} is used. 
To obtain fits as those depicted in Fig.\ \ref{fig:noPQ}, we divide the $M_{W_R}-M_H$ plane into $40\times 40$ squares within which we marginalize over all LR and CKM parameters. For each square, $M_{W_R}$ and $M_H$ are then constrained to lie within the considered square, while the remaining parameters are varied within the ranges described above.

Before discussing the resulting constraints on the mLRSM we check our expressions by performing an analysis of the CKM parameters in the decoupling limit, $M_{H,W_R}\to \infty$. We find
\bea \label{eq:CKMfit}
\lambda \in [0.2254,\,0.2267 ]\,,\qquad A \in [0.78,\,0.82 ]\,,\qquad \bar \rho\in [0.07,\,0.16 ]\,\qquad \bar \eta\in [0.35,\,0.39 ]\,,
\eea
at $90\%$ C.L. These values are similar to the results of Ref.\ \cite{Alioli:2017ces} and are consistent with the values advocated by the PDG \cite{Zyla:2020zbs}. The ranges found here are wider than those of Ref.\ \cite{Zyla:2020zbs}, especially in the case of $\bar \rho$ and $\bar \eta$. The reason for the weaker constraints in the SM limit is that we do not include non-leptonic $B$ decays like $B\to\pi\pi$. The evaluation of these decays in the mLRSM would require additional non-perturbative matrix elements that are not currently available.

\subsection{Analysis without  a Peccei-Quinn mechanism}

\begin{figure}[t!]\center
\includegraphics[width=0.49\textwidth]{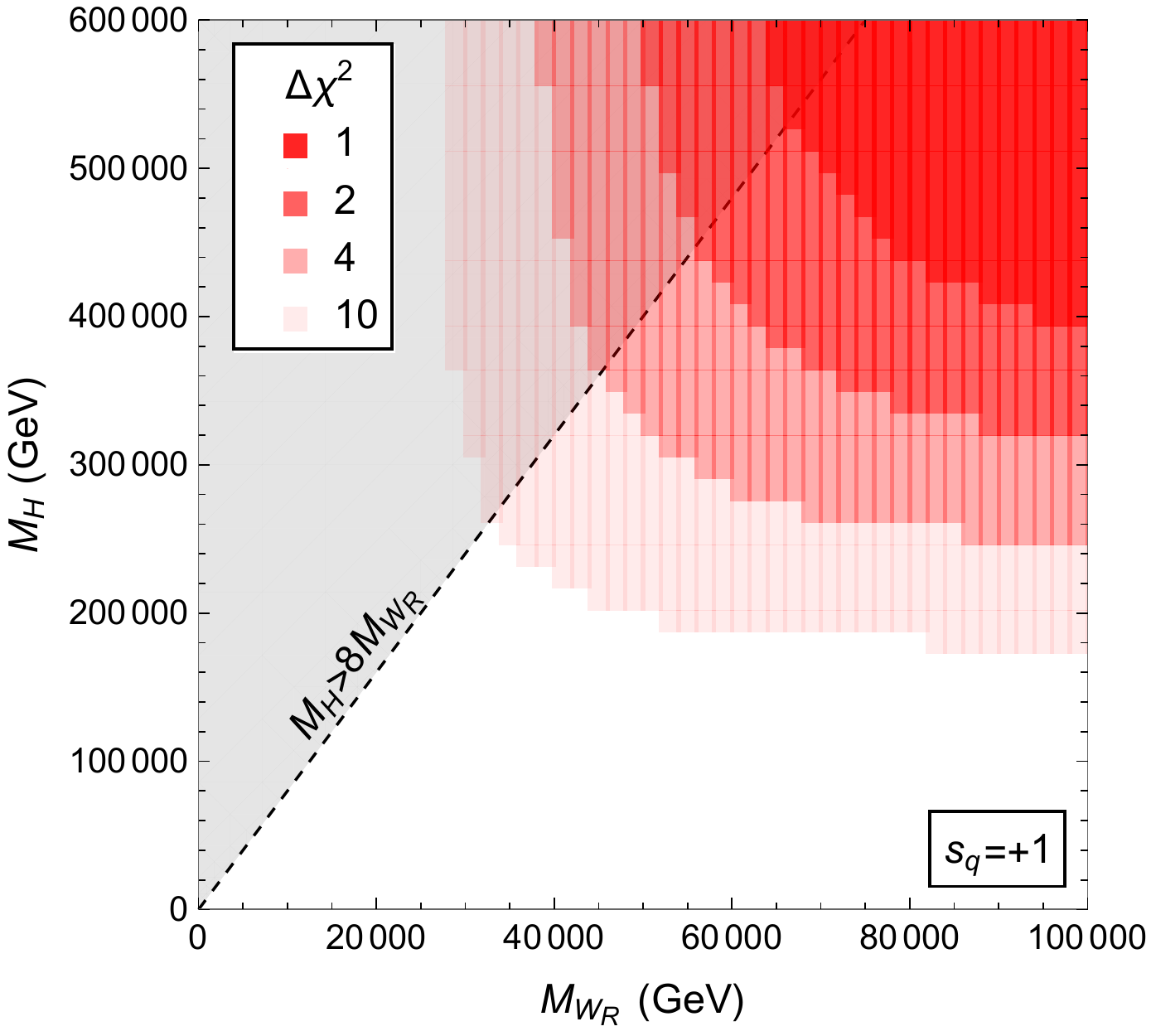}
\includegraphics[width=0.49\textwidth]{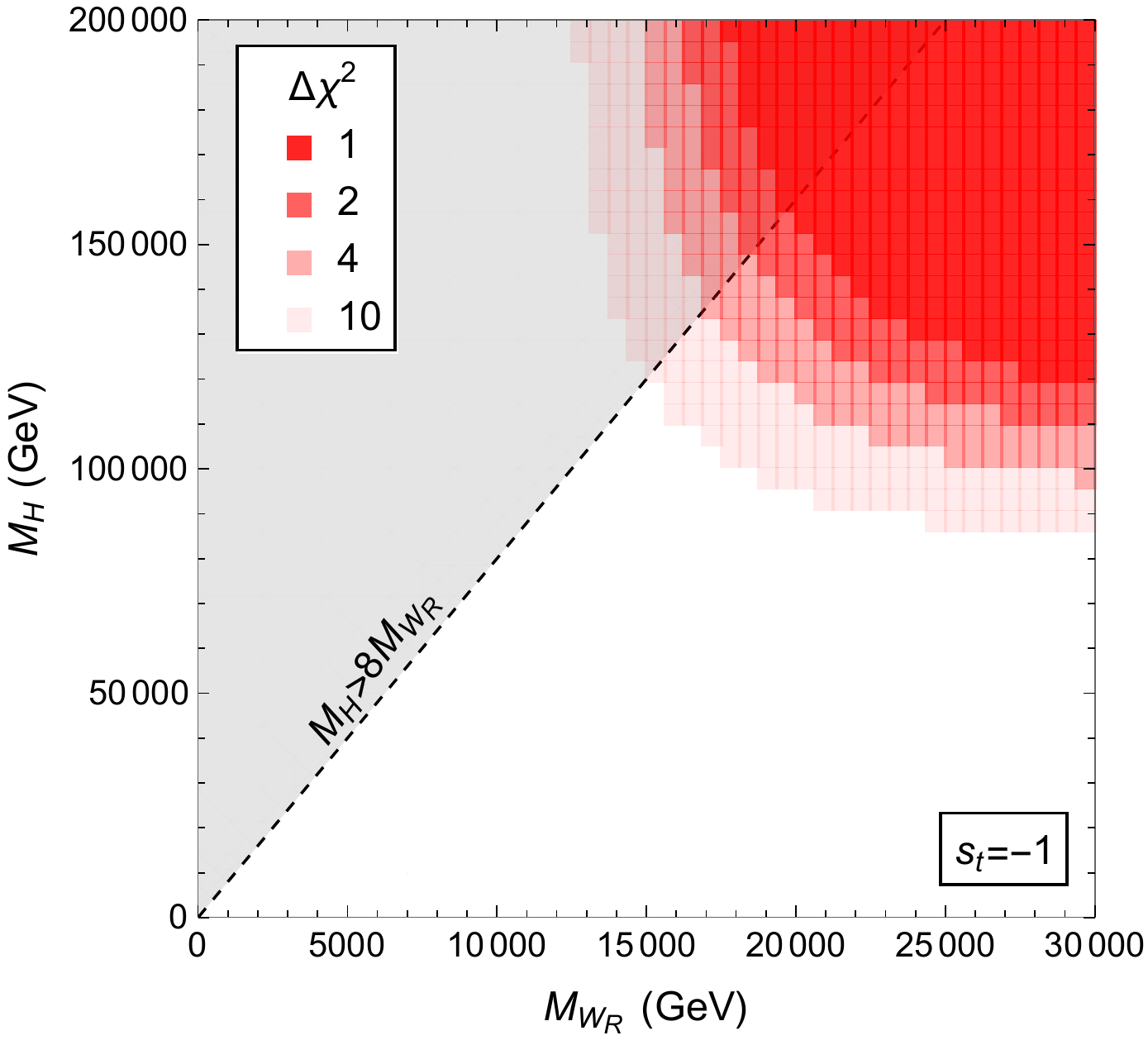}
\caption{The left panel depicts the $\Delta \chi^2 = \{1,2,4,10\}$ constraints in the  $M_{W_R}$-$M_H$ plane, after marginalizing over the other LR and CKM parameters. No Peccei-Quinn mechanism is applied. The gray line shows the $M_H>8 M_{W_R}$ region where couplings in the Higgs potential become non-perturbatively large \cite{Basecq:1985cr}. The left and right panels depict the sign configurations $s_t =+1$ and $s_t=-1$, respectively, with $s_{q\neq t} = +1$ for the remaining signs. } 
\label{fig:noPQ}
\end{figure}

We begin the analysis in the parity-conserving mLRSM without a PQ mechanism where the model itself accounts for the smallness of the CP-violating QCD vacuum angle. As discussed in Sect.~\ref{strongCP}, $\bar \theta$ now becomes a calculable function in terms of the LR parameters. Current EDM measurements then require that the spontaneous phase $t_{2\beta}s_\alpha  \simeq 0$ to very good approximation and in essence transfer the strong CP problem from $\bar \theta$ to $\alpha$. 
This effectively sets $\al=0$~\footnote{Note that $t_{ 2\beta} \rightarrow 0$ does not give rise to a different solution to the constraint $t_{2\beta}s_\alpha  \simeq 0$. The reason is that 
$\al$ always appears in the combination $t_\bt  e^{i\al}$.},
that is, the EDM constraints are so strong that they effectively remove one parameter from the analysis and, after this removal, they no longer constrain the remaining parameters. We are then left with three LR parameters ($M_{W_R}$, $M_H$, and $\xi$) and the CKM parameters that can be varied. We remind the reader that the right-handed quark-mixing matrix is expressed in terms of CKM parameters and quark masses and a set of discrete phases and reduces to $V_R = S_u V_L S_d$ in this limit, see App.~\ref{app:VR}.  We begin our analysis by setting all discrete phases to $\theta_q = 0$, and later discuss the impact of alternative sign combinations. 

The main result is shown in Fig.~\ref{fig:noPQ} which depicts $\Delta \chi^2 = \{1,2,4,10\}$ contours in the  $M_{W_R}$-$M_H$ plane, 
where each point has been minimized with respect to the remaining LR and CKM parameters. The left plot illustrates a clear lower bound on $M_{W_R} \gtrsim 38$ TeV at $95\%$ C.L. ($\Delta \chi^2=4$) in the limit of a decoupled $M_H \gtrsim 400$ TeV. Part of this parameter space however covers a range where the Higgs sector contains non-perturbatively large parameters. Constraining the parameter space to $M_H < 8 M_{W_R}$ implies a stronger bound $M_{W_R} \gtrsim 45$ TeV at $95\%$ C.L. and $M_H>240$ TeV at $95\%$ C.L. for the scalar mass. The bound on $M_{W_R}$ is very stringent in light of the current limit on the $M_{W_R} \geq 4$  TeV from direct production at the LHC \cite{Aad:2019hjw}.  

We still need to address the role of the sign choices, which in principle lead to 32 distinct variants of  the $P$-symmetric model. It turns out that choosing $s_i = +1$ for all the signs leads to significantly more stringent constraints than some other assignments. For instance, setting $s_t = -1$ while keeping the other signs the same, leads to the right panel of Fig.~\ref{fig:noPQ}. In this case, we obtain roughly $M_{W_R} \gtrsim 17$ TeV $95\%$ C.L. in the perturbative regime.
We find that each of the 32 sign combinations essentially fall in either of the two scenarios shown in Fig.~\ref{fig:noPQ}. 
While the more stringently constrained scenarios give rise to a similar value for $\chi^2\vert _{\rm min}$ as the SM, the less constrained sign combinations allow for a smaller value by about $\sim 5$. We discuss this slight improvement of the fit compared to the SM in more detail in the next subsection, in which we consider the LRM with a PQ mechanism, where a similar improvement of the fit can be achieved.

In both cases, the strong bounds are essentially driven by $\varepsilon_K$.
This observable obtains contributions due to $\sin\al$ as well as mLRSM contributions proportional to the CP-odd phase in the CKM matrix that survive even when $\al\to0$.
A low-mass $W_R$ then requires cancellations to occur between these two different LR contributions to CP-violation in $
 K^0-\bar K^0$ mixing. This only becomes possible in case of a sizable spontaneous phase $\alpha$ \cite{Zhang:2007da,Maiezza:2010ic,Bertolini:2014sua,Maiezza:2014ala} which is excluded in the absence of a PQ mechanism, leading to stringent limits. The $\varepsilon_K$ constraint is easier to satisfy for the choice $s_t s_c = -1$ and $s_d s_s = +1$ in agreement with Ref.~\cite{Maiezza:2014ala}. This leads to the least stringent limits and defines the class of signs depicted in the right panel of Fig.\ \ref{fig:noPQ}. As other observables are not as constraining, it will be difficult to further tighten the limits from low-energy constraints barring further theoretical refinements of the SM prediction of $\varepsilon_K$. 
The result $M_{W_R} \gtrsim 17$ TeV is still very strong compared to direct limits and is in good agreement with Ref.~\cite{Bertolini:2019out} that obtained $M_{W_R} \gtrsim13$ TeV. The main differences with respect to our analysis is that we applied an updated SM prediction for $\varepsilon_K$, an improved RGE analysis, and performed a fit involving both the CKM and LR parameters.

\begin{figure}[t!]\center
\includegraphics[width=0.49\textwidth]{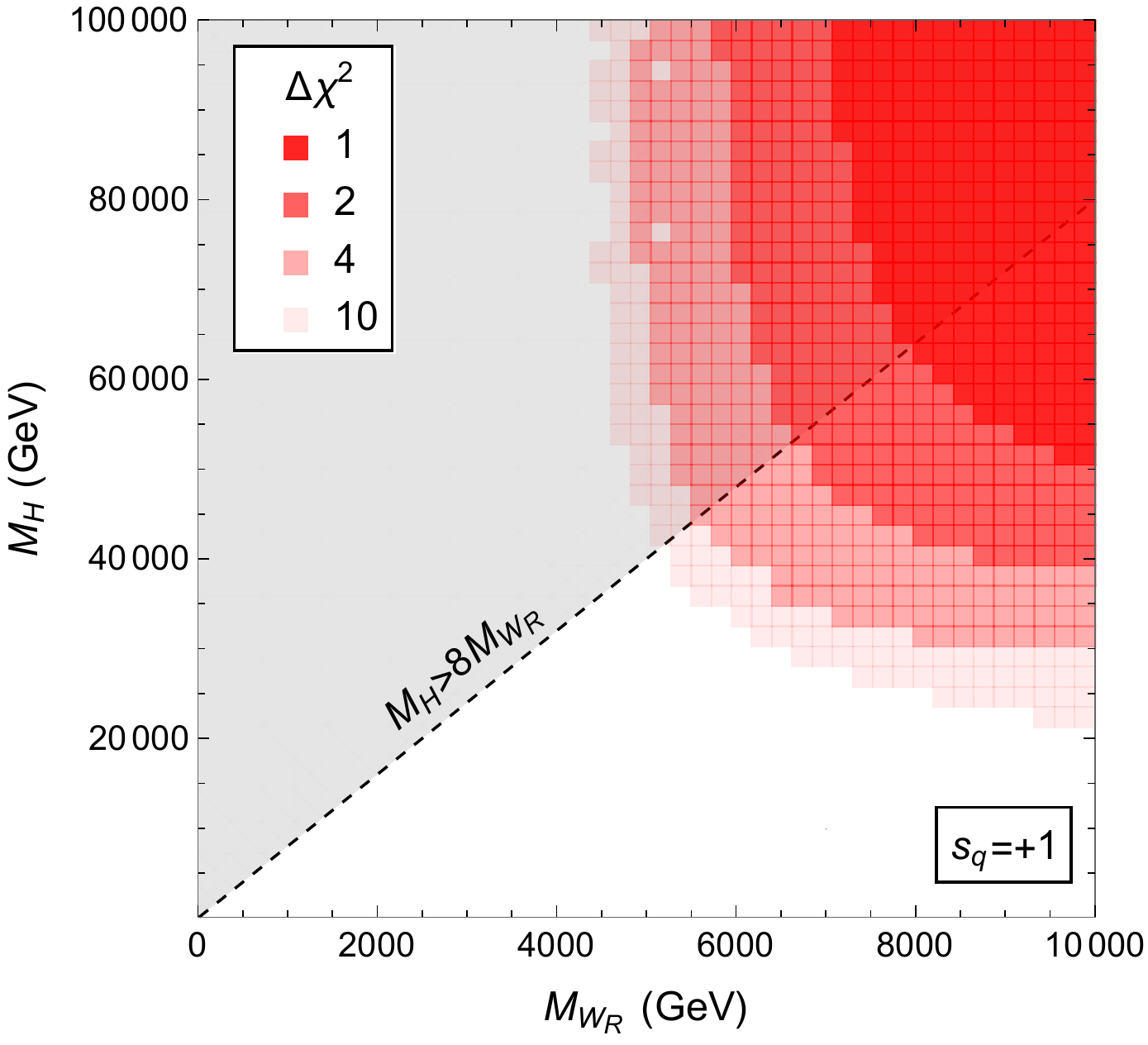}\hfill
\includegraphics[width=0.49\textwidth]{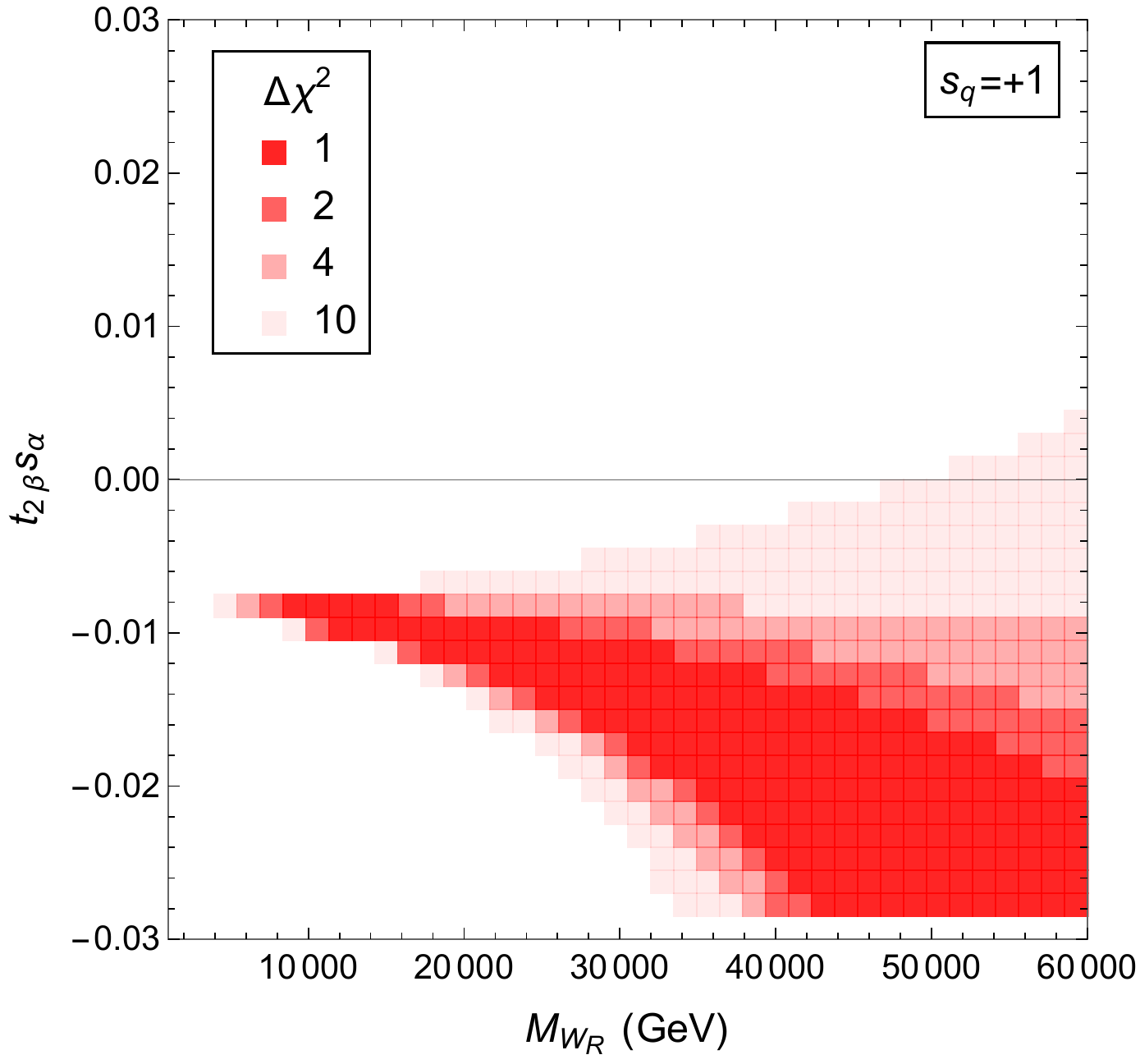}
\caption{The left panel depicts the $\Delta \chi^2 = \{1,2,4,10\}$ regions in the  $M_{W_R}$-$M_H$ plane, after marginalizing over the other LR and CKM parameters. A Peccei-Quinn mechanism is applied. The gray line shows the $M_H>8 M_{W_R}$ region where couplings in the Higgs potential become non-perturbatively large \cite{Basecq:1985cr}.  The right panel shows the allowed parameter space in the $M_{W_R}$-$ t_{2\beta}s_\alpha$ plane for fixed $M_H=6 M_{W_R}$, while marginalizing with respect to the remaining parameters. Both panels correspond to the choice $s_q=+1$. }
\label{fig:PQ}
\end{figure}

\subsection{Analysis with  a Peccei-Quinn mechanism}

We now consider the parity-conserving mLRSM in presence of a PQ mechanism. The strong CP problem is now resolved in the infrared and although EDMs still lead to significant constraints, they no longer effectively force $\alpha \simeq 0$. We start  our analysis by setting all signs to $s_q=+1$. This leads to the plots in Fig.~\ref{fig:PQ}. The left panel shows $\Delta \chi^2 = \{1,2,4,10\}$ contours in the  $M_{W_R}$-$M_H$ plane, after marginalizing with respect to the other parameters. We thus obtain a lower bound of $M_{W_R} \gtrsim 5.5$ TeV at $95\%$ C.L., in the parameter space where $M_H < 8 M_{W_R}$. This limit is significantly weaker than obtained in the no-PQ scenario, where a lower bound of $M_{W_R} \gtrsim 38$ TeV was obtained for the same choice of discrete signs (weakened to $\sim 17$ TeV for the most favorable sign combination). 

The weaker limit on $M_{W_R}$ compared to the scenario without a PQ mechanism is driven by the relaxed constraint on $\alpha$ and allows for a significant $t_{2\beta}s_\alpha\neq0$. As  $\varepsilon_K$ obtains contributions from both the CKM phase and the spontaneous phase $\alpha$ cancellations between the two terms now become possible \cite{Maiezza:2014ala,Bertolini:2019out}. 
This is depicted in the right panel of Fig.~\ref{fig:PQ} where small values of $M_{W_R}$ clearly require a nonzero value of $t_{2\beta}s_\alpha$. This rather specific value of  $t_{2\beta}s_\alpha$, illustrated by the funnel in the right panel leads to the mentioned cancellation which allows for much smaller values of $M_{W_R}$.

\begin{figure}[t!]\center
\includegraphics[width=0.49\textwidth]{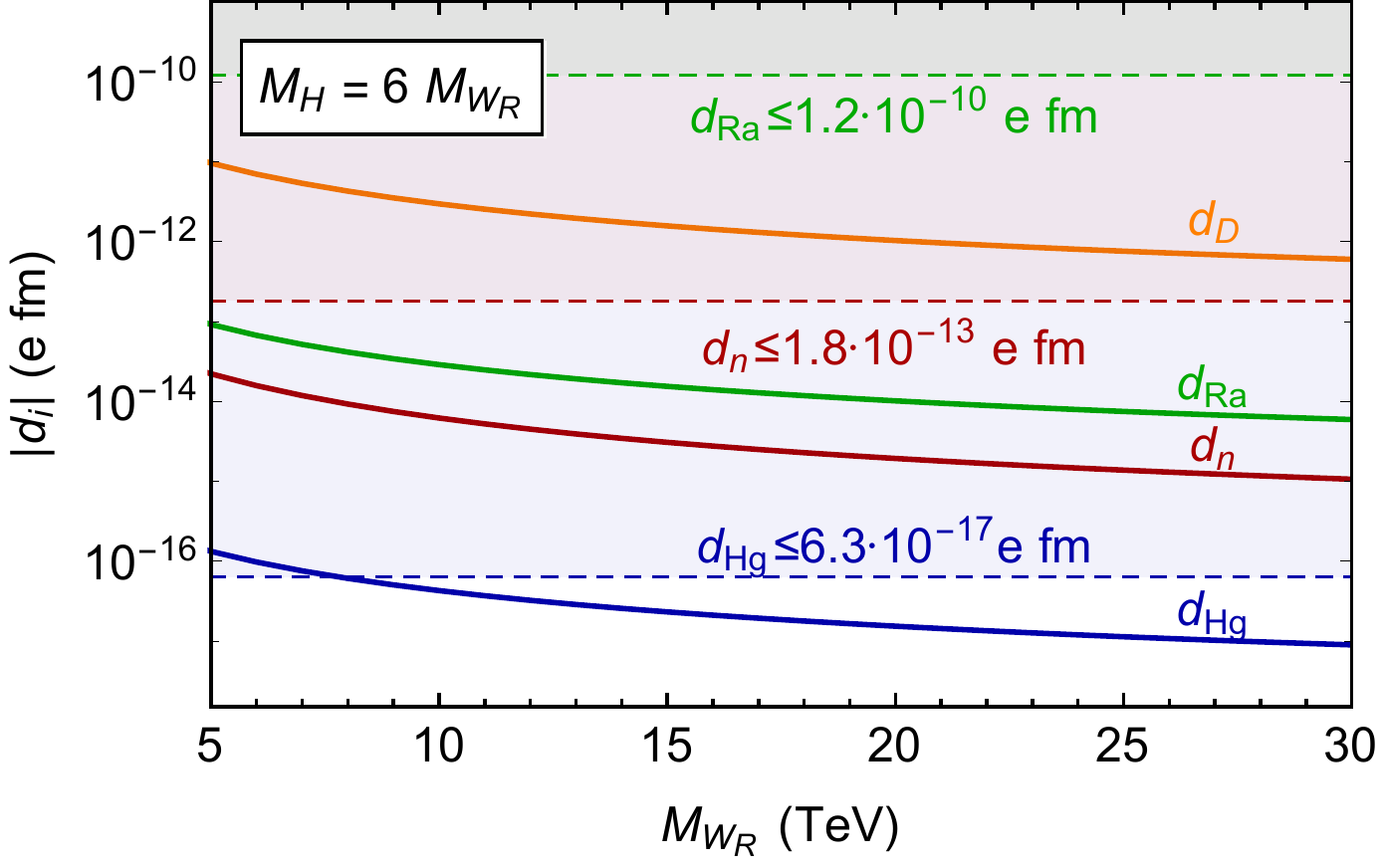}\hfill
\includegraphics[width=0.49\textwidth]{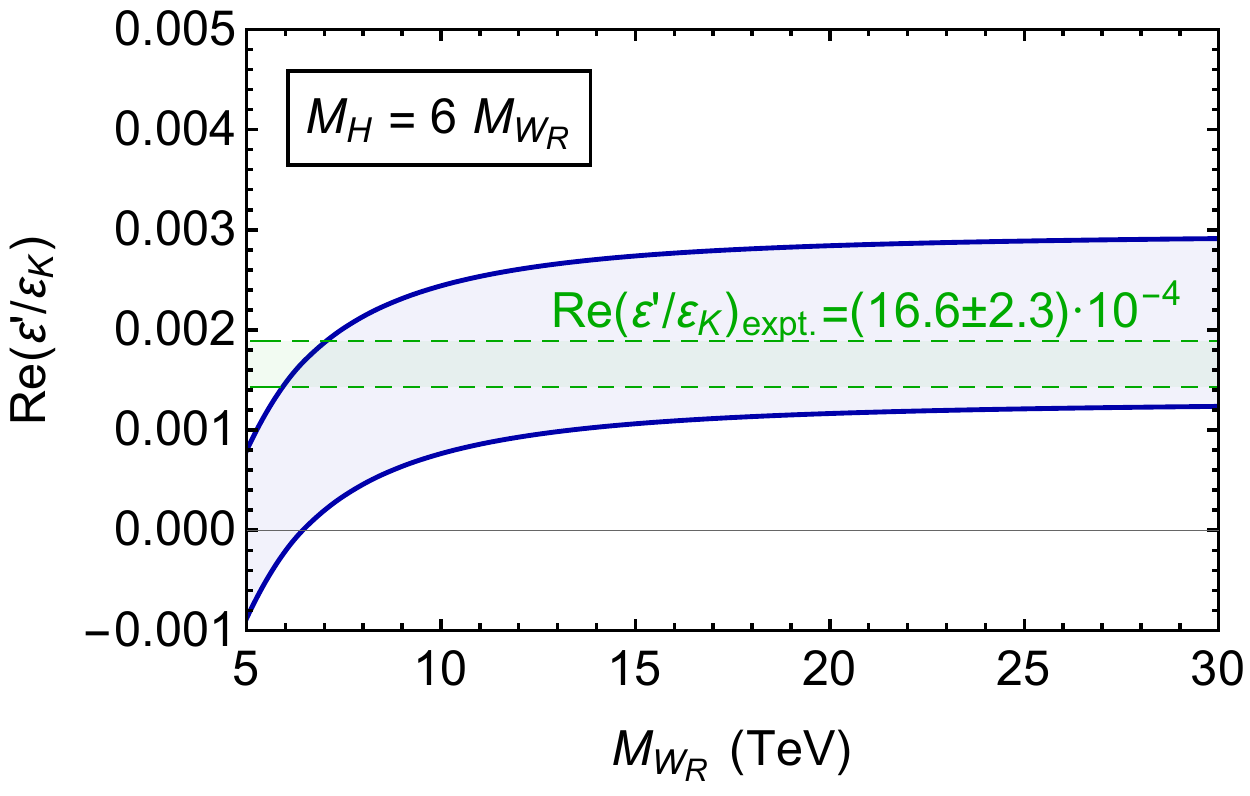}
\caption{The left panel shows the values of various EDMs as a function of $M_{W_R}$ inside the 'funnel' region where $t_{2\beta}s_\alpha\simeq -0.01$. The dashed lines indicate current limits. The right panel does the same for $\varepsilon'/\varepsilon_K$, where the width of the blue band indicates the uncertainty of the SM prediction. }
\label{fig:PQ_EDMs}
\end{figure}

The lowering of the limit on $M_{W_R}$ only goes so far. 
For small $M_{W_R}$ other CP-violating observables like $\varepsilon'/\varepsilon_K$ and EDMs become large, as these observables are induced by the CP-odd combination $t_{2\beta}s_\alpha$ which is forced to be sizable by $\varepsilon_K$. 
We illustrate this in Fig.~\ref{fig:PQ_EDMs}. Here we focus on the parameter space with $M_H = 6M_{W_R}$  and $M_{W_R}<30$ TeV as a representative example. 
The remaining parameters are set to the values preferred by the fit as a function of $M_{W_R}$.
In this region, the value of $t_{2\beta}s_\alpha$ then ranges between $-0.009$ and $-0.014$ with $t_\bt\simeq -0.05$ remaining constant~\footnote{The values of the SM CKM parameters preferred by the fit also remain roughly constant in this region, with $\la \simeq 0.226$, $A\simeq 0.79$, $\bar\rho \simeq 0.18$, and $\bar\eta \simeq 0.34$.}, corresponding to part of the funnel region in the right panel of Fig.\ \ref{fig:PQ}. We then plot values of the various EDMs as a function of $M_{W_R}$. The effect of the Schiff screening that affects the mercury EDM can clearly be seen from the relative sizes of $d_n$ and $d_{\rm Hg}$, while the relatively large values of $d_{\rm Ra}$ are due to the octupole enhancement discussed in Sect.\ \ref{sec:EDMs}.
The largest EDM is found to be that of the deuteron, which does not suffer from the suppression due to Schiff screening and is rather sensitive to the $\pi N$ couplings which receive large contributions in the mLRSM. 

We observe that several EDMs are predicted to lie only one or two orders of magnitude below the present limits. That is, next-generation EDM experiments can test the funnel region corresponding to low values of $M_{W_R}$. 
For instance, a ${}^{225}$Ra EDM measurement at the $10^{-14}e$ fm level might be possible \cite{Bishof:2016uqx} and would already go a long way in excluding small values of $M_{W_R}$. Similarly, a small improvement on $d_{\rm Hg}$ would have a big impact on the funnel region. Possible storage-ring experiment of $d_D\leq 10^{-16}e$ fm could have an even larger impact.  
We stress that a lower limit on $M_{W_R}$, assuming improved EDM measurements, cannot easily be deduced from the figure as it assumes values of $t_{2\beta}s_\alpha$ which resulted from a fit with current experimental input. Obtaining a new lower limit on $M_{W_R}$ would require one to perform a new global fit once improved EDM measurements are available. The right panel of Fig.~\ref{fig:PQ_EDMs} shows that future improvements in the theoretical prediction of $\varepsilon'/\varepsilon_K$, which would shrink the width of the blue band, are also excellent probes of the low $M_{W_R}$ regime. Apart from EDMs, there are several CP-even observables, particularly the $B$ and $K$ mass differences, which obtain significant corrections for $M_{W_R}$ in the TeV range.

\begin{figure}[t!]\center
\includegraphics[width=0.49\textwidth]{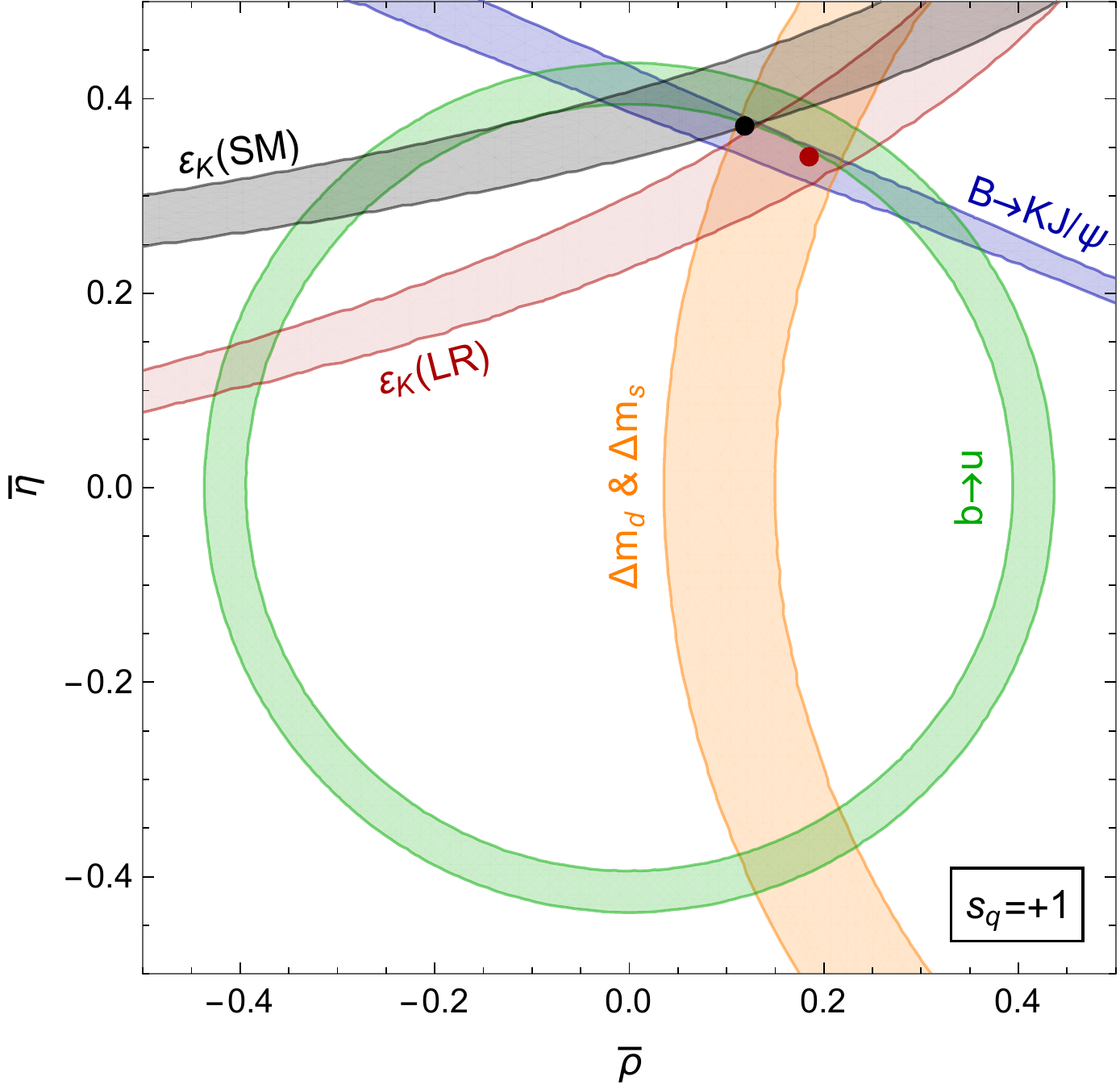}
\caption{
$68\%$ C.L. contours from various flavor observables in the $\bar\rho-\bar\eta$ plane for two scenarios, namely the SM, $M_{H,W_R}\to \infty$, and the case with the best fit values for the LR parameters, $\{M_{H},M_{W_R},t_{2\bt}s_\al, \al\} \simeq \{200\,{\rm TeV},21\,{\rm TeV}, -0.01,3.04\}$. The difference is only noticeable in the case of $\varepsilon_K$ for which the SM and LRM bands are shown in black and red, respectively. Each band was obtained including $s\to u$ and $b\to c$ observables in order to marginalize with respect to $A$ and $\la$. The best fit points in the SM and LRM are shown as black and red points, respectively.
 }
\label{fig:CKMplot}
\end{figure}

Finally, we note that the fit has a slight preference for finite values of $M_{W_R}$ and $M_H$ over the SM point, $M_{H,W_R}\to \infty$. This is due to a mild tension in the SM fit of the CKM parameters, which can be alleviated somewhat by LR contributions to $\varepsilon_K$, lowering the minimum $\chi^2$ by roughly $5$. To illustrate the impact of the LRM we show the different experimental constraints in the $\bar\rho-\bar\eta$ plane in Fig.\ \ref{fig:CKMplot}, both for the SM case ($M_{H,W_R}\to \infty$) and when using the best fit values for the LR parameters ($\{M_{H},M_{W_R},t_{2\bt}s_\al, \al\} \simeq \{200\,{\rm TeV},21\,{\rm TeV}, -0.01,3.04\}$). The figure shows the $68\%$ C.L.\ (for two parameters, $\Dt\chi^2=2.3$) bands for several flavor observables described in the previous sections. Each band was obtained by taking into account the $s\to u$ and $b\to c$ transitions, see Sect.\ \ref{sec:treeDecays}  and App.\ \ref{app:observables}, and marginalizing over $A$ and $\lambda$.  $\varepsilon_K$ is the only observable for which the change from the SM limit, shown in black, to the best fit point, shown in red, is noticeable. The shifted $\varepsilon_K$ band allows for better overlap with the preferred regions of the other observables, leading to a somewhat improved $\chi^2$. This change also leads to a noticeable shift in the best fit point in the $\bar\rho-\bar\eta$ plane, changing from $\{\bar\rho,\bar\eta\} =\{0.12,0.37\} $ in the SM to $\{\bar\rho,\bar\eta\} =\{0.19,0.34\} $ at the best fit point in the LRM, shown by the black and red points, respectively. Although the tension in the SM may not be very severe, the sizable shifts in the determinations of the CKM parameters due to the LRM do imply that the impact of fitting the CKM and LR parameters simultaneously can be significant.

Moving on to other possible sign choices, we find very similar allowed regions for the four cases with $s_d s_s = s_cs_t = s_us_t=+1$, while other combinations of the signs lead to more stringent constraints and require $M_{W_R}\gtrsim10$ TeV at $95\%$ C.L.  
All sign combinations now allow for a lower $\chi^2\vert_{\rm min}$ compared to the SM, though the corresponding best fit values for the LR parameters vary.
As the limits in the remaining cases are significantly tighter than those shown in Fig.~\ref{fig:PQ} we do not further pursue the other sign choices.

\subsection{$V_{ud}$, $V_{us}$, and CKM unitarity}\label{sec:CKM}
\begin{figure}[t!]\center
\includegraphics[width=0.49\textwidth]{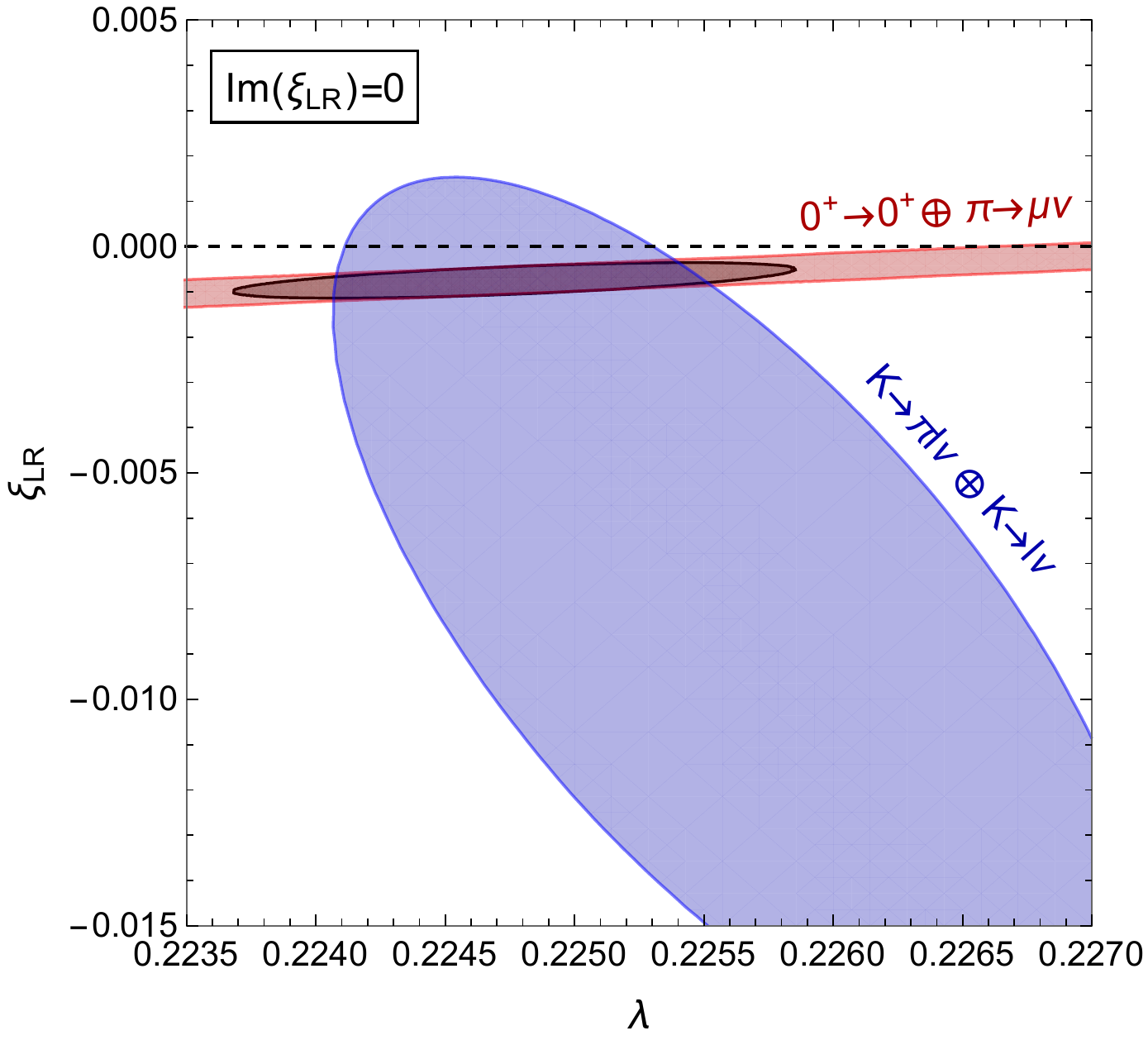}\hfill
\includegraphics[width=0.46\textwidth]{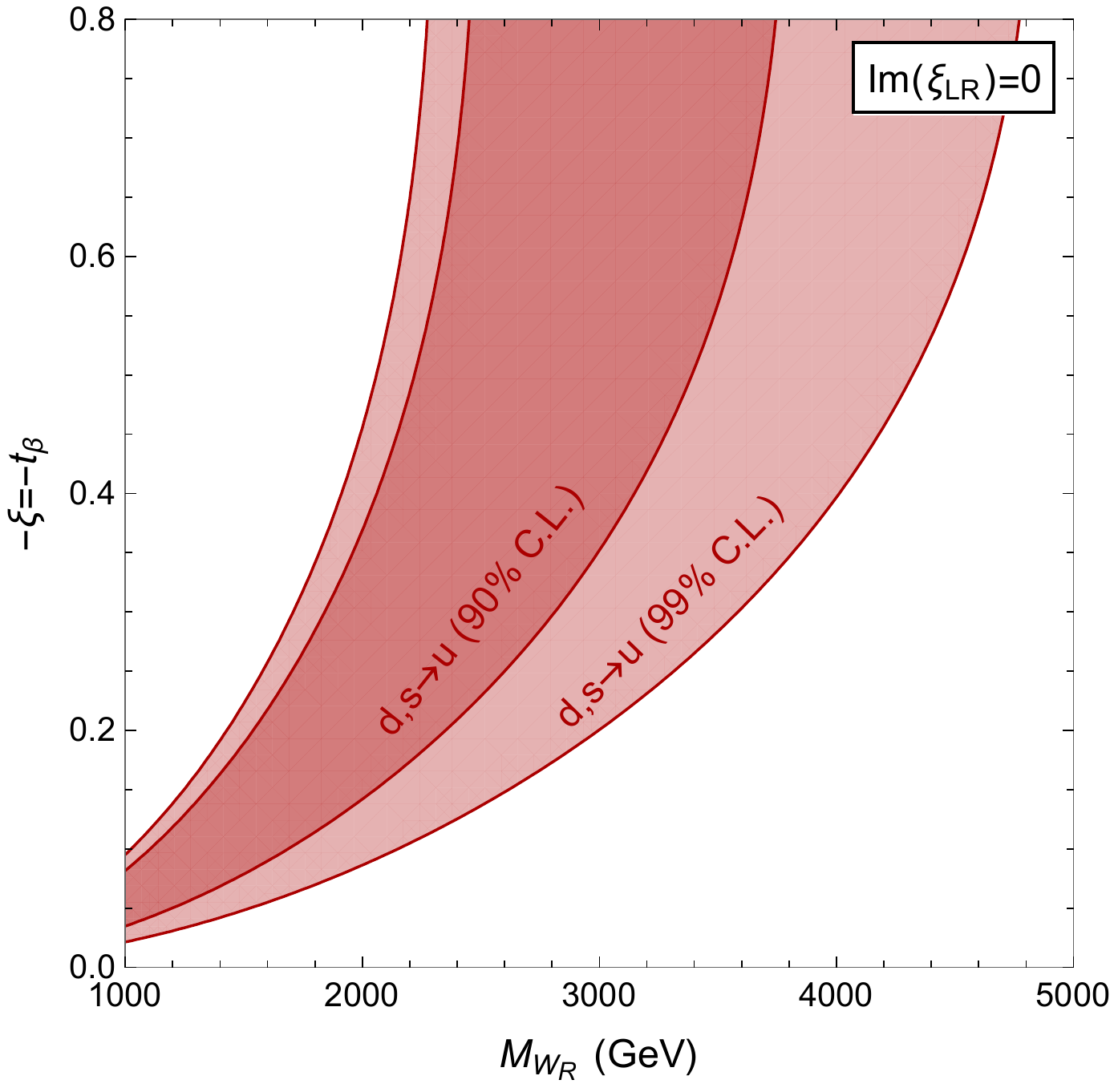}
\caption{The left panel depicts the  $\lambda - \xi_{LR}$ plane, with constraints at  $90\%$ C.L.\ ($\Dt\chi^2 = 4.6$)  from kaon decays in blue, those from $0^+\to 0^+$ and pion decays in red, and the combination in black. The right panel shows the preferred region at $90\%$ and $99\%$ C.L.\, projected onto the $M_{W_R}-\xi$ plane, while allowing the SM CKM parameter $\la$ to vary. Both panels assume Im$(\xi_{LR})=0$.}
\label{fig:VudVus}
\end{figure}
Before concluding we briefly discuss the discrepancy between the determinations of $V_{ud}$ and $V_{us}$, from $0^+\to 0^+$ and kaon decays, 
which recently sparked interest in possible BSM explanations \cite{Crivellin:2020lzu,Cheung:2020vqm,Belfatto:2019swo}. 
The inclusion of the SM CKM parameters within our analysis enables us to consider this anomaly in a consistent manner within the mLRSM and allows one to answer whether the tension is improved by LR interactions. Before embarking on a global analysis we first consider a simpler analysis in which we focus on the observables driving the discrepancy.

The discrepancy arises from a measured value of  $|V_{L\,ud}|^2+|V_{L\,us}|^2 \neq 1$, which implies a violation of unitarity (here $V_{L\,ub}$ is negligible with current sensitivities). Equivalently, using unitarity, one can obtain $V_{L\,ud}$ from the kaon decays of Eq.\ \eqref{eq:Vus}, which give  $V_{L\,ud} = [0.9743,\, 0.9746]$ at $1\sigma$. This result is in tension with the $0^+\to 0^+$ determination, which, in the SM, gives $\left| V_{L\,ud} \right| = 0.97370 \pm 0.00014$ \cite{Seng:2018yzq,Seng:2018qru,Czarnecki:2019mwq}. Note that this discrepancy worsens if we would use the $N_f = 2+1+1$ lattice results \cite{Aoki:2019cca} for the form factors in Eq.\ \eqref{eq:Vus} instead of the $2+1$ numbers used here.

It is interesting to see whether this  discrepancy can be resolved in the mLRSM. Taking $V_R = S_u V_L S_d$, which holds to good approximation, the above mentioned observables only involve two combinations of parameters, namely, $\lambda$ and 
\begin{equation}
\xi_{LR} \equiv \frac{m_W^2}{M_{W_R}^2}\frac{2\xi}{1+\xi^2} e^{i\al}\,.
\end{equation}
As any imaginary part of $\xi_{LR}$ is stringently constrained by EDMs as well as $\varepsilon'$, we will focus on the case where $\xi_{LR}$ is real in what follows~\footnote{In addition, allowing for an imaginary part does not significantly lower the minimal $\chi^2$.}.
The resulting constraints from kaon decays and $0^+\to0^+$ transitions are shown in the left panel of Fig.\ \ref{fig:VudVus} in blue and red, respectively. The SM prediction is depicted by the black dashed line and it does not fit the two types of decays very well since it intersects the red and blue regions at different points. Allowing for a non-zero $\xi_{LR}$ improves the fit significantly, as the minimum $\chi^2$ decreases from $19$ in the SM to around $3$. The improvement is most significant for the sign combinations with $s_d = s_s$, as both the kaon decays and $0^+\to 0^+$ prefer $\xi_{LR}V_R^{ud,us}\leq 0$~\footnote{The options with $s_d=-s_s$ lead to $\chi^2\vert_{\rm min}\simeq 5$.}. 
The preferred region in the $M_{W_R}-\xi$ plane due to the combination of $d\to u$ and $s\to u$ transitions is shown in the right panel of  Fig.\ \ref{fig:VudVus}, which also shows the preference for finite $M_{W_R}$ and $\xi$. 

Thus, the mLRSM can improve the discrepancy. However, although the kaon and $0^+\to0^+$ determinations are consistent at $90\%$ C.L.\ as can be seen from Fig.\ \ref{fig:VudVus}, the two contours do not overlap at $1\sigma$. The preferred size of ${\rm Re}\,\xi_{LR}$ is around $[-11,\,-4.5]\cdot 10^{-4}$ at $90\%$ C.L., which implies an upper limit on $M_{W_R}$ of $ M_{W_R}\lesssim 4$ TeV, as can be seen from the right panel of Fig.\ \ref{fig:VudVus}. This value lies below the bound $M_{W_R}\geq 5.5$ TeV even in the presence of a PQ mechanism. Indeed, once we include other observables discussed in Sect.~\ref{sec:obs} we find that while this region does improve the contributions from $0^+\to0^+$ and kaon decay to the total $\chi^2$, this improvement is offset completely by the increase due to other observables, mainly $\varepsilon_K$, which prefer larger values of $M_{W_R}$. Thus, a solution to the tension in CKM unitarity can be excluded within the $P$-symmetric mLRSM considered here. It would be interesting to see whether other variants, such as the $C$-symmetric mLRSM, can explain the discrepancy. 

\section{Conclusion}\label{sec:conclusion}
Left-right symmetric models are promising candidates for beyond-the-SM theories that provide an origin for $P$ violation, neutrino masses, and potentially the strong CP problem. They also lead to a very rich phenomenology. In this work, we perform a comprehensive study of the low-energy signatures of the $P$-symmetric mLRSM. We consider the case where the model itself accounts for the smallness of $\bar \theta$ by requiring small spontaneous CPV phases (the no-PQ case) as well as the scenario with a Peccei-Quinn mechanism (the PQ case). The most stringent constraints on the model arise from low-energy $\beta$-decay observables, flavor observables, and EDMs. These, with the exception of EDMs, also play a large role in determining the CKM parameters so that we are forced to perform a combined fit of CKM and mLRSM parameters. We do so by including a large number of different processes for which both accurate predictions as well as measurements exist. An important role is played by low-energy probes of CP violation. We have used updated SM predictions for $\varepsilon_K$ and $\varepsilon'$, using both chiral perturbation theory and lattice QCD calculations to determine mLRSM contributions. We have performed a comprehensive analysis of EDMs in the mLRSM including not just the neutron EDM, but also more complicated (and more sensitive) nuclear and atomic systems. 

We note that the mLRSM does not follow the flavor structure of minimal flavor violation (MFV) \cite{DAmbrosio:2002vsn}. 
 MFV requires invariance of the Lagrangian under $SU(3)_{Q_L}\times SU(3)_{u}\times SU(3)_{d}$, after treating the up- and down-type Yukawa couplings as spurions transforming as $Y_{u,d}\to U_{Q_L} Y_{u,d}U_{u,d}$.
Instead, the mLRSM becomes invariant under a smaller symmetry group, $SU(3)_{Q_L}\times SU(3)_{Q_R}$, if one treats the Yukawa couplings as spurions that transform as $\Gamma\to U_{Q_L}\Gamma U_{Q_R}^\dagger$ and $\tilde\Gamma\to U_{Q_L}\tilde\Gamma U_{Q_R}^\dagger$. This group is less restrictive and allows for additional interactions to arise unsuppressed by small Yukawa couplings. For example, $C_{Hud}$ is induced proportional to the right-handed CKM matrix, $\sim V_R$, while MFV would dictate $C_{Hud}\sim Y_{u}^\dagger Y_d$. Thus, assuming MFV would lead one to expect this operator to be negligibly small, while it is actually sizable in the mLRSM and leads to important effects in a number of observable such as EDMs and $\varepsilon'$.
This implies that although the mLRSM is well suited to an EFT approach, thanks to the large hierarchy in scales $M_{W_R}\gg m_W$, it does not follow the flavor  assumptions that are often employed in global SMEFT analyses. Due to the large number of operators appearing in the SMEFT, such works often take MFV as a working assumption and/or focus on high-energy collider observables \cite{Bruggisser:2021duo,Grojean:2018dqj,Aoude:2020dwv}. The mLRSM is a clear example of a scenario where such an approach does not apply as it does not follow MFV, making low-energy measurements very competitive compared to direct searches for signatures of left-right models, even in a global setting.

Our main findings are summarized in Figs.~\ref{fig:noPQ} and \ref{fig:PQ} where we show constraints in the $M_{W_R}$-$M_H$ plane in the no-PQ and PQ case respectively. In the no-PQ case, one obtains a calculable $\bar \theta$ that contributes significantly to $d_n$ and $d_{\rm Hg}$ forcing $\alpha \ll 1$,  leading to a lower bound $M_{W_R} \gtrsim 17$ TeV at $95\%$ C.L. driven by  $\varepsilon_K$. 
It will be hard to improve this bound with low-energy measurements unless theoretical predictions of $\varepsilon_K$ can significantly be improved. In the PQ case, there is no large contribution to EDMs from $\bar\theta$, allowing for a sizable $\al$. This makes it possible for contributions to $\varepsilon_K$ induced by $\sim\sin\al$ to cancel terms proportional to the phase in the SM CKM matrix. 
These cancellations weaken the constraints and we obtain $M_{W_R} \gtrsim 5.5$ TeV at $95\%$ C.L., not much higher than direct limits from colliders \cite{CMS:2018mgb,Aaboud:2017yvp,CMS:2021mux}. This bound can be tightened significantly with next-generation EDM measurements which would essentially limit the precision with which the different contributions to $\varepsilon_K$ can cancel each other. 

We also investigated whether the $P$-symmetric mLRSM can help resolve the CKM anomaly, finding that a relatively light $M_{W_R}\simeq 4$ TeV can in principle improve the tension found in the SM. Unfortunately, this region of parameter space is already excluded within a global analysis. 

This work focused on low-energy observables. It would be interesting to combine the global analysis with high-energy searches. Depending on the masses of new fields this can be done either in the SMEFT framework or has to be done in the full model. In addition, we have not considered the leptonic sector of the mLRSM. The mLRSM leads to a rich phenomenology of (semi-)leptonic observables such as the electron EDM \cite{Valle:1983nx,Nieves:1986uk,Nemevsek:2012iq}, charged-lepton flavor violation \cite{Bajc:2009ft,Cirigliano:2004mv,Lee:2013htl}, and neutrinoless double beta decay \cite{Tello:2010am,Nemevsek:2011aa,Li:2020flq} that can be included in a future analysis. 

In conclusion, we performed a systematic and global analysis of low-energy constraints on the parity-symmetric minimal left-right symmetric model. We find no significant evidence that this model is preferred over the Standard Model and set lower bounds on the masses of right-handed gauge bosons and scalar bosons that are more stringent that direct limits. 

\section*{Acknowledgments}

We thank Albert Young, Leendert Hayen, and Vincenzo Cirigliano for stimulating conversations.
L.~A is supported by the US Department of Energy under contract DE-SC0021027. E.~M. is supported  by the US Department of Energy through  
the Office of Nuclear Physics  and  the  
LDRD program at Los Alamos National Laboratory. Los Alamos National Laboratory is operated by Triad National Security, LLC, for the National Nuclear Security Administration of U.S.\ Department of Energy (Contract No. 89233218CNA000001). F.~O. is supported by the Dutch Organization for Scientific Research (NWO) under program 156.

\newpage

\appendix
\section{Solution of $V_R$ in the $P$-symmetric mLRSM}\label{app:VR}
In the $P$-symmetric limit a solution for $V_R$ can be derived  from the expressions of the mass matrices in Eq.\ \eqref{masses} \cite{Senjanovic:2014pva,Senjanovic:2015yea},
\bea\label{app:masses}
M_u = \sqrt{1/2}\ka( \Gamma +\xi e^{-i\al}\TG)\,,\qquad M_d = \sqrt{1/2}\ka (\xi e^{i\al} \Gamma +\TG)\,.\label{eq:Appmass}
\eea
Both mass matrices can generally be diagonalized using two unitary matrices, $L_q$ and $R_q$, so that $M_q = L_q m_q R_q^\dagger$, where $m_q$ are real and diagonal, and the CKM matrices become $V_L = L_u^\dagger L_d$ and  $V_R = R_u^\dagger R_d$.  If $L_q$ and $R_q$ diagonalize the mass matrices, then the same will be true for $ L_q S_q$ and $R_q S_q $, where $S_{u,d}$ are diagonal matrices of signs, meaning there will be $2^5$ distinct solutions for $V_R$.

To determine the number of physical parameters we can note that $P$ symmetry ensures that the Yukawa matrices, $\Gamma$ and $\tilde{\Gamma}$, are hermitian, each having $9$ parameters. This allows us to use a transformation of the form, $Q_{L,R}\to VQ_{L,R}$, so that $\Gamma\to V^\dagger \Gamma V$ becomes real and diagonal, leaving $V_{L,R}$ unchanged  \cite{Maiezza:2010ic} ~\footnote{This transformation affects the matrices needed to diagonalize the mass matrices as $L_q\to V^\dagger L_q$ and $R_q\to V^\dagger R_q$, while leaving the combinations  $V_L = L_u^\dagger L_d$ and  $V_R = R_u^\dagger R_d$ invariant.}. This rotation can be written as $V=V' S$, where $V'$ belongs to $SU(3)$ and $S$ is a diagonal matrix of phases. Since $S$ is not determined by the demand that $V^\dagger \Gamma V=V^{\prime\,\dagger} \Gamma V'$ is diagonal, we have the freedom to use the phases in $S$ to eliminate two of the off-diagonal phases in $\tilde\Gamma$. 
Since the mass matrices determine the CKM matrices and the quark masses, this implies that $m_q$ and $V_{L,R}$ are a function of $\xi$, $\al$, the three parameters in $V^\dagger \Gamma V$, and the seven remaining parameters in $V^\dagger \tilde\Gamma V$. Conversely, this means that $V_R$ and the $10$ parameters in $V^\dagger \Gamma V$ and $V^\dagger \tilde\Gamma V$ can be solved in terms of $\xi$, $\al$, the $6$ quark masses, and the $4$ SM CKM parameters in $V_L$.

The above was used in Refs.\ \cite{Senjanovic:2014pva,Senjanovic:2015yea} to obtain a solution for $V_R$ in terms of $\xi$, $\al$, $V_L$, and $m_q$. These references also obtained analytical approximations in terms of an expansion in $x\equiv \tan 2\bt \sin\al$. Here we follow a similar approach as Refs.\ \cite{Senjanovic:2014pva,Senjanovic:2015yea} and use the hermiticity of $\Gamma$ and $\tilde \Gamma$ to rewrite Eq.\ \eqref{app:masses} as,
\bea\label{app:VRsol}
U_u m_u U_u-m_u &=& -ix\left[\xi e^{i\al}m_u-V_L m_d V_R^\dagger\right]\,,\nn\\
U_d m_d U_d-m_d &=& ix\left[\xi e^{-i\al}m_d-V_L^\dagger m_u V_R\right]\,,\nn\\
V_R&=&U_u V_L U_d\,,\qquad  U_q = L^\dagger_q R_q\,.
\eea
These equations are useful as they allow one to obtain $U_q$ order by order after expanding both sides in terms of $x$,
\bea
V_R =  \sum_n x^n V_R^{(n)}\,, \qquad U_q =  \sum_n  x^n U_q^{(n)}\,,
\eea
in addition, we write $\xi e^{i\al} = \xi\cos\al +i \frac{1-\xi^2}{2}x$. Collecting terms at each order in $x$ one can then obtain $U_q^{(n)}$ from the first two lines in Eq.\ \eqref{app:VRsol}, which now only depend on the lower order terms, $V_R^{(m)}$ and $U_q^{(m)}$, with $m<n$. The third equation in Eq.\ \eqref{app:VRsol} then allows one to solve the $n$-th order in $V_R$ in terms of $V_R^{(m)}$ and $U_q^{(m)}$.
Thus, starting with the $x^0$ solution, $V_R^{(0)} = S_u V_L S_d$ and $U_q^{(0)}=S_q$, any higher order can be obtained iteratively. This procedure reproduces the analytical approximations of Refs.\ \cite{Senjanovic:2014pva,Senjanovic:2015yea}. In our analysis we use expressions for $V_R$ obtained in this way and take into account terms up to and including $x^4$. 

\subsection{Region of validity}\label{app:validity}
Eq.\ \eqref{eq:Appmass} does not allow for a solution for all values of $\xi=t_\bt$ and $\alpha$. A necessary condition was derived in Refs.\  \cite{Maiezza:2010ic,Senjanovic:2014pva,Senjanovic:2015yea}, and can roughly be stated as $|x|\lesssim 2m_b/m_t$. This condition can be obtained by considering the largest diagonal elements of the mass matrices, which we will take to be the $33$ entry, 
\bea
\bigg|\left(M_u M_u^\dagger\right)_{33}\bigg|&\geq &m_t^2-\bigg|\left(M_u M_u^\dagger\right)_{31}+\left(M_u M_u^\dagger\right)_{32}\bigg|\gtrsim m_t^2-2m_b m_t\,,\nn\\
\left[\left(M_d-M_d^\dagger\right) M_u\right]_{33}&\lesssim &2m_bm_t\,,
\eea
where the first inequality in the first line follows from eigenvalue equation for $M_u M_u^\dagger$. The second inequality can be derived by using that, for $i\neq j$, the matrix $\left(M_uM_u^\dagger\right)_{ij}$ can be expressed in terms of $M_dM_d^\dagger$, $M_uM_d^\dagger$, and $M_dM_u^\dagger$, and the fact that $|\left(M_{u,d}\right)_{ij}|\leq \sum_k m_{u_k,d_k}\lesssim m_{t,b}$. Using the above, one can derive the following inequality in the basis where $\Gamma$ is diagonal
\bea
\left|\frac{\left[\left(M_d-M_d^\dagger\right) M_u\right]_{33}}{\left[M_uM_u^\dagger-t_\bt^2 M_dM_d^\dagger\right]_{33}}\right| =2 \left|\frac{t_\bt \sin\al(1+t_\bt z e^{-i\al})}{1+2 z t_\bt(1-t^2_\bt)\cos\al-t_\bt^4}\right|\lesssim 2\frac{m_b}{m_t}\,,
\eea
where $z\equiv \Gamma_{33}/\tilde \Gamma_{33}$. Varying over the parameter $z$, gives a constraint that is very similar to the one discussed in Ref.\ \cite{Maiezza:2010ic} and numerically close to $|x|\lesssim 2m_b/m_t\simeq 0.036$. In practice, we consider the range $|x|\leq 0.03$ within which our approximate solutions of $V_R$ agrees with higher order solution to within $\lesssim 10\%$.

Finally, we can see that for values of $\xi=t_\bt\to 1$ the Yukawa matrices have to become large in order to explain the hierarchy between the up-type and down-type masses. In particular
\bea
\frac{1}{v^2}{\rm Tr} \left(M_uM_u^\dagger-M_dM_d^\dagger\right) =\frac{1}{2}\left( c_\bt^2-s_\bt^2\right){\rm Tr} \left(\Gamma^2-\tilde\Gamma^2\right)  \simeq \frac{m_t^2-m_b^2}{v^2}\,,
\eea
which implies that $\Gamma$ and/or $\tilde \Gamma$ have to become large in the $t_\bt\to 1$ limit. 
To avoid such large couplings we follow Ref.\ \cite{Maiezza:2010ic} and restrict  $|t_\bt|<0.8$ in our fits.
\section{Mass eigenstates of the Higgs fields}\label{app:higgses}
The spontaneous breaking of $SU(2)_{L,R}$ implies that the scalar fields, $\Dt_{L,R}$ and $\phi$,
should involve two neutral and two singly-charged would-be-Goldstone bosons. The remaining components are physical and make up six neutral, two singly-charged, and two doubly charged fields. 
The masses of these fields generally have lengthy expressions, we therefore only give approximate expressions for the $P$-symmetric case (setting some parameters in the Higgs potential to zero, $\bt_i=v_L=0$) and keep linear terms in $\ka/v_R$ and $\xi\equiv \ka'/\ka$.
With these approximations the  would-be-Goldstone bosons, that are absorbed by the $W_{L,R}$ and $Z_{L,R}$ bosons, can be written as 
\begin{align}
G_L^+ =&\, \phi^+_1 - \xi e^{-i\al}\phi_2^+\,, & G_R^\pm = &\,\dt_R^+-\frac{\ka}{\sqrt{2}v_R}\phi^+_2\, ,\nn\\
G_Z^0=& \,\sqrt{2}\im (\phi^{0*}_1 + \xi e^{-i\al}\phi_2^0)\,,& G_{Z'}^0 =&\, \sqrt{2}\im\dt_R^0 \,.
\end{align}
The masses of the remaining (physical) states are shown in Table \ref{Tab:HiggsMasses}, where the conventions for the parameters in the Higgs potential can be found in Ref.\ \cite{Zhang:2007da}. 

We finally discuss the masses and mixings of the $\phi$ fields in more detail, as they play a role in Sect.\ \ref{sec:matching}.
Writing the bidoublet in terms of two $SU(2)_L$ doublets, $\phi = (\phi_1,\, \phi_2)$, the breaking of $SU(2)_R$, gives rise to the following mass terms,
\bea
\vL \supset -(\tilde \phi_1^\dagger,\, \phi_2^\dagger)
\bma
v_R\sq \frac{\al_1}{2}-\mu_1\sq & 2\mu_2\sq e^{-i\dt_{\mu}}-\al_2 v_R\sq e^{i\dt_2}\\
2\mu_2\sq e^{i\dt_{\mu}}-\al_2 v_R\sq e^{-i\dt_2} & \frac{\al_1+\al_3}{2} v_R\sq-\mu_1\sq\ema \bma \tilde \phi_1\\\phi_2 \ema \,.
\eea
Here $\al_i$, $\dt_i$, and $\mu_i$ are parameters of the Higgs potential, with the notation as in Ref.\ \cite{Dekens:2014ina}. The above terms are the $\Or(v_R\sq)$ terms for the general potential in LR models, which include the $C$ and $P$ symmetric cases (the latter has $\dt_\mu=0$). In principle the above mass matrix has 2 nonzero eigenvalues, meaning that both doublets would obtain $\Or(v_R)$ masses. However, demanding that the Higgs potential resides in a minimum, $\partial V_H/\partial \{v_R,\,\kappa^{(\prime)},\,\al\} =0$, give rises to,
\bea
\mu_1\sq -\frac{v_R\sq}{2}\al_1 \approx  -\frac{v_R\sq}{2}\frac{\xi\sq}{1-\xi\sq}\al_3\,,\qquad
2\mu_2^2e^{-i\mu_2}-\al_2 v_R^2 e^{i\dt_2} = \frac{1}{2}\frac{\xi \al_3 v_R^2}{1-\xi^2} e^{-i\al}\,,
\eea
so that the mass terms become
\bea
\vL \supset -(\tilde \phi_1^\dagger,\, \phi_2^\dagger)
\, \frac{\al_3  v_R\sq}{2(1-\xi\sq)}\bma
\xi\sq & \xi e^{-i\al}\\
\xi e^{i\al} & 1\ema \bma \tilde \phi_1\\\phi_2 \ema \,,
\eea
which can be diagonalized as in Eq.\ \eqref{eq:HiggsRot}
\bea\label{HiggsRot}
\bma \tilde \phi_1\\\phi_2 \ema=\bma -c_\bt & s_\bt e^{-i\al}\\ s_\bt e^{i\al} & c_\bt
\ema
\bma \vp_{}\\ \vp_H\ema \,,
\eea
where $t_\bt =s_\bt/c_\bt= \xi$ and the signs are chosen such that $\langle \vp_{}\rangle=+\sqrt{\ka\sq+\ka^{\prime\,2 }}/\sqrt{2}= +v/\sqrt{2}$. The mass eigenstates then have the following eigenvalues, $m\sq_{\vp_{SM}}=0$ and $M\sq_{H}=\frac{\al_3  v_R\sq}{2}\frac{1+\xi\sq}{1-\xi\sq}$, which implies that the SM doublet only acquires an $\Or(\kappa\sq)$ mass after EWSB.

In the Higgs mass basis the Yukawa interactions of Eq.\ \eqref{Lagrangian} then take the following ($SU(2)_L$-invariant) form,
\bea\label{Yukawas}
-\vL_Y &=& \frac{\sqrt{2}}{v}\bar  Q_L\left[ \tilde \vp_{} M_u+\frac{1}{1-\xi\sq}\tilde \vp_{H} \left(M_d(1+\xi\sq)-2\xi e^{i\al}M_u\right)\right]U_R \nn\\
&&+ \frac{\sqrt{2}}{v}\bar  Q_L\left[ \vp_{} M_d+\frac{1}{1-\xi\sq} \vp_{H} \left(M_u(1+\xi\sq)-2\xi e^{-i\al}M_d\right)\right]D_R +{\rm h.c.}
\eea
In the mass basis for the quarks the neutral currents  become (up to $\Or(\xi\sq)$ terms) \cite{Zhang:2007da}
\bea
\vL_N &=& \bar U_L\bigg[Y_u(h^0-iG^0_Z)+(H_1^0-i A_1^0)(V_L Y_d V_R^\dagger-2\xi Y_u e^{i\al})\bigg] U_R\nn\\
&&+\bar D_L\bigg[Y_d(h^0+iG^0_Z)+(H_1^0+i A_1^0)(V_L^\dagger Y_u V_R-2\xi Y_d e^{-i\al})\bigg]D_R+{\rm h.c.}\,,
\eea
whereas the charged scalars give rise to the following interactions,
\bea
\vL_C &=& \sqrt{2}\bar U\bigg[(Y_u V_R-2\xi e^{-i\al}V_LY_d)P_R H_2^+-(V_RY_d -2\xi e^{-i\al}Y_uV_L)P_L H_2^+\nn\\
&&+(Y_uV_LP_L-V_LY_dP_R)G_L^+\bigg]D+{\rm h.c.}\,,
\eea
where $Y_{u,d}$ are diagonal matrices of Yukawas, $\left(Y_q\right)_{ii} = m_{q_i}/v$.

\renewcommand{\arraystretch}{1.3}
\begin{table}[t]
\begin{center}\small
\begin{tabular}{l|l}
\textbf{Mass eigenstate}& \textbf{Mass squared}\\
\hline\hline
Neutral scalars \\
\hline
$h^0 = \sqrt{2}\re (\phi_1^{0*}+\xi e^{-i\al}\phi_2^0)$ & $\frac{1}{2}\al_3v_R\sq \xi\sq +(2\la_1-\frac{1}{2}\al_1\sq/\rho_1)\ka\sq$ \\
$H_1^0=\sqrt{2} \re (\phi_2^0-\xi e^{i\al}\phi_1^{0*})$ & $\frac{1}{2}\al_3v_R\sq$\\
$A_1^0=\sqrt{2}\im (\phi_2^0-\xi e^{i\al}\phi_1^{0*})$ & $\frac{1}{2}\al_3 v_R\sq $\\
$\sqrt{2} \re \dt_R^0$& $2\rho_1v_R\sq$\\
$\sqrt{2}\re \dt_L^0$ & $\frac{1}{2}(\rho_3-2\rho_1)v_R\sq$\\
$\sqrt{2}\im\dt_L^0$ & $\frac{1}{2}(\rho_3-2\rho_1)v_R\sq$\\
\hline
Singly-charged scalars \\
\hline
$H_2^+=\phi_2^{+}+\xi e^{i\al}\phi_1^+ +\frac{\ka}{\sqrt{2}v_R} \dt_R^+$ & $\frac{1}{2}\al_3 v_R\sq$\\
$\dt_L^+ $ & $\frac{1}{2}(\rho_3-2\rho_1)v_R\sq +\frac{1}{4}\al_3 \ka\sq$\\
\hline
Doubly-charged scalars \\
\hline
$\dt_R^{++}$ & $2\rho_2 v_R\sq $\\
$\dt^{++}_L $& $\frac{1}{2}(\rho_3-2\rho_1)v_R\sq +\frac{1}{2}\al_3\ka\sq $
\end{tabular}
\end{center}
\caption{The physical  Higgs mass eigenstates and their masses for the $P$-symmetric potential, restricted to the $\bt_i=v_L=0$ case. Only linear terms in $\ka/v_R$ and $\xi\equiv \ka'/\ka $ have been kept \cite{Duka:1999uc,Kiers:2005gh,Zhang:2007da}. The definitions of the parameters from the Higgs potential can be found in \cite{Zhang:2007da}.}\label{Tab:HiggsMasses}
\end{table}
\renewcommand{\arraystretch}{1}

\section{Matching to the SMEFT in the Warsaw basis}\label{app:warsaw}

In Section \ref{sec:matching} we matched the mLRSM onto the SMEFT in a basis that is convenient for the discussion of low-energy observables. Here, we report the conversion between our basis and the standard ``Warsaw basis'' of Ref.\ \cite{Grzadkowski:2010es}.  We first note that Ref.\ \cite{Grzadkowski:2010es} as well as \cite{Jenkins:2013zja,Jenkins:2013wua,Alonso:2013hga} use a different sign convention for the gauge couplings, $g_{1,2,3}$ in their notation, and the Levi-Civita tensor. Explicitly,
\bea
g' = -g_1\,,\qquad g = -g_2\,,\qquad g_s = -g_3\,, \qquad \epsilon^{\al\bt\mu\nu}\vert_{\rm Here} = \epsilon^{\al\bt\mu\nu}\vert_\text{\cite{Grzadkowski:2010es}}\,.
\eea
With these identifications, our definition of the right-handed current operator $C_{Hud}$ agrees with that of Ref.\ \cite{Grzadkowski:2010es}. For the four-quark vector and scalar operators, we find
\begin{eqnarray}
\left[ C^{(1)}_{ud}\right]_{prst} &=& - \left[ C_{2\, RR} + \frac{1}{N_c} C_{1\, RR}\right]_{s r p t}\,, \nn\\
\left[ C^{(8)}_{ud}\right]_{prst} &=& - 2 \left[ C_{1\, RR} \right]_{s r p t}\,, \nn\\
\left[ C^{(1)}_{qu}\right]_{prst} &=&  \left[ C_{1\, q u} + \frac{1}{N_c} C_{2, qu} \right]_{p r s t}\,, \nn\\
\left[ C^{(8)}_{qu}\right]_{prst} &=&  2 \left[ C_{2, qu} \right]_{p r s t}\,, \nn\\
\left[ C^{(1)}_{qd}\right]_{prst} &=&  \left[ C_{1\, q d} + \frac{1}{N_c} C_{2, qd} \right]_{p r s t}\,, \nn\\
\left[ C^{(8)}_{qd}\right]_{prst} &=&  2 \left[ C_{2, qd} \right]_{p r s t}\,, \nn\\
\left[ C^{(1)}_{quqd}\right]_{prst} &=&  \left[ C_{1\, q u q d} + \frac{1}{N_c} C_{2, quqd} \right]_{p r s t}\,, \nn\\
\left[ C^{(8)}_{quqd}\right]_{prst} &=&  2 \left[ C_{2, quqd} \right]_{p r s t}\,. 
\end{eqnarray}
Note that a Fierz relation involving Dirac matrices was used to obtain the first two identities, so that they strictly speaking only hold at tree-level. For $d\neq 4$ the left- and right-hand sides will differ by evanescent operators, which can impact the finite parts of loop-level expressions. In practice we used the $O_{i\,RR}$ operators when computing the matching contributions described in Sect.\ \ref{sec:matching}, which may differ from the matching one would obtain using the SMEFT basis.

The dipole operators in Eq.\ \eqref{dipoles} agree with the definitions of Ref.\ \cite{Grzadkowski:2010es}, modulo factors of the gauge couplings,
\begin{eqnarray}
 & &C_{uW} = - \frac{g}{\sqrt{2}} \Gamma^{u}_W\,, \qquad C_{uB} = - \frac{g^\prime}{\sqrt{2}} \Gamma^{u}_B\,, \qquad C_{uG} = - \frac{g_s}{\sqrt{2}} \Gamma^{u}_g\,, \nn\\
 & &C_{dW} = - \frac{g}{\sqrt{2}} \Gamma^{d}_W\,, \qquad C_{dB} = - \frac{g^\prime}{\sqrt{2}} \Gamma^{d}_B\,, \qquad C_{dG} = - \frac{g_s}{\sqrt{2}} \Gamma^{d}_g\,. \end{eqnarray}

\subsection{Matching to the LEFT in the basis of Ref.\ \cite{Jenkins:2017jig}}
Similarly, below the electroweak scale, we matched onto bases that are traditionally used in the discussion of various observables, such as meson-antimeson oscillations or $B \rightarrow X_s \gamma$. A complete basis for the description of low-energy observables was established in Ref.\ \cite{Jenkins:2017jig}. Here we give the conversion between the operators introduced in Section \ref{sec:LEFTmatching} and Ref.\ \cite{Jenkins:2017jig,Jenkins:2017dyc}.
For the gauge couplings and epsilon tensor we now have,
\bea
g_s = -g\vert_\text{\cite{Jenkins:2017jig}}\,,\qquad e\vert_{\rm Here} = e\vert_\text{\cite{Jenkins:2017jig}}\,, \qquad \epsilon^{\al\bt\mu\nu}\vert_{\rm Here} = -\epsilon^{\al\bt\mu\nu}\vert_\text{\cite{Jenkins:2017jig}}\,.
\eea

For the four-quark operators, we find
\begin{eqnarray}
 \left[L^{V1\, LL}_{ud} \right]_{pr st} &=&   - \left[ C_{2\, LL} + \frac{1}{N_c} C_{1\, LL}\right]_{s r p t},\nn \\
 \left[L^{V8\, LL}_{ud} \right]_{pr st} &=&   - 2\left[ C_{1\, LL}\right]_{s r p t},\nn\\ 
 \left[L^{V1\, RR}_{ud} \right]_{pr st} &=&   - \left[ C_{2\, RR} + \frac{1}{N_c} C_{1\, RR}\right]_{s r p t}, \nn\\
 \left[L^{V8\, RR}_{ud} \right]_{pr st} &=&   - 2\left[ C_{1\, RR}\right]_{s r p t}, \nn\\
  \left[L^{V1\, LR}_{uddu} \right]_{pr st} &=&   - \left[ C^*_{1\, LR} + \frac{1}{N_c} C^*_{2\, LR}\right]_{r p t s}, \nn\\
 \left[L^{V8\, LR}_{uddu} \right]_{pr st} &=&   - 2\left[ C^*_{2\, LR}\right]_{r p  t s}, \nn\\
  \left[L^{V1\, LR}_{dd} \right]_{pr st} &=&    \left[ C_{4} + \frac{1}{N_c} C_{5}\right]_{p r s t}, \nn\\
 \left[L^{V8\, LR}_{dd} \right]_{pr st} &=&    2\left[ C_{5}\right]_{p r s t}, \nn\\
  \left[L^{S1\, RR}_{ud} \right]_{pr st} &=&    \left[ C_{1, qu qd} + \frac{1}{N_c} C_{2, qu qd}\right]_{p r v t} \left[V_L^*\right]_{vs}, \nn\\
 \left[L^{S8\, RR}_{ud} \right]_{pr st} &=&    2 \left[ C_{2, qu qd}\right]_{p r s t}\left[V_L^*\right]_{vs}, \nn\\
  \left[L^{S1\, RR}_{uddu} \right]_{pr st} &=&   - \left[ C_{1, qu qd} + \frac{1}{N_c} C_{2, qu qd}\right]_{v t p r}\left[V_L^*\right]_{vs},\nn \\
 \left[L^{S8\, RR}_{uddu} \right]_{pr st} &=&    -2 \left[ C_{2, qu qd}\right]_{v t p r}\left[V_L^*\right]_{vs},
 \end{eqnarray}
while, for the dipole operators, 
\begin{eqnarray}
& & \left[ L_{u\gamma}\right]_{pr} = - e \frac{Q_u}{2} m_{u_r} C^{pr}_{\gamma u}\,, \qquad 
  \left[ L_{uG}\right]_{pr} = -  \frac{g_s}{2} m_{u_r} C^{pr}_{g u}\,, \nn\\
& &  \left[ L_{d\gamma}\right]_{pr} = - e \frac{Q_d}{2} m_{d_r} C^{pr}_{\gamma d}\,, \qquad 
  \left[ L_{dG}\right]_{pr} = -  \frac{g_s}{2} m_{d_r} C^{pr}_{g d}\,.
\end{eqnarray}
Finally, the Weinberg operator in LEFT is given in terms of the coefficient in Eq.\ \eqref{eq:LowLag} by
\bea
L_{\tilde G} = -\frac{g_s}{3}C_{\tilde G}\,.
\eea

\section{Observables}\label{app:observables}

In this Appendix we give the expressions for the observables that are included in our $\chi^2$ function, but were not discussed in the main text.

\subsection{Leptonic and semileptonic decays}\label{app:treeDecays}

\begin{table}[t]
\center
\begin{tabular}{||c|c|| c |c ||}
\hline
               &Decay constant &   & Form Factor \\
\hline
$f_\pi$  		       &  $130.2 \pm 0.8$ MeV & $f^{K\pi}_+(0)$ 	  	& $0.9677 \pm 0.0027$\\ 
$f_{K}/f_{\pi} $     &  $1.1917 \pm 0.0037$   &  		  	& \\
$f_D  $ 		       &  $209.0 \pm 2.4$ MeV & $f_{+}^{D\pi}(0)$ 	& $0.666 \pm 0.029$ \\
$f_{D_s}$ 		       &  $248.0 \pm 1.6$ MeV & $f_+^{D K}(0)$    	& $0.747 \pm 0.019$ \\ 
$f_B $ 	 		       &  $192.0 \pm 4.3$ MeV &	$\mathcal F_{D}(1)$ 	& $1.035 \pm 0.040$			\\
$f_{B_s} $ 	 	       &  $228.4 \pm 3.7$ MeV & $\mathcal F_{D^*}(1)$ 	& $0.906 \pm 0.004 \pm 0.012$ 			\\
\hline
\end{tabular}
\caption{Pseudoscalar meson decay constants and form factors as determined from lattice QCD calculations. Here we use the FLAG lattice averages with $n_f = 2 + 1$\cite{Aoki:2019cca}. }
\label{LQCDinput}
\end{table}

\paragraph{$u\rightarrow d$ and $u \rightarrow s$ transitions}\label{app:ud}
In addition to the lifetime of superallowed $\beta$ emitters, 
the  $\pi\to \mu\nu_\mu$, $K \rightarrow \mu \nu_\mu$ 
and $K \rightarrow \pi l \nu_l$ branching ratios, which were discussed in Section \ref{sec:treeDecays},
we use the triple correlation $\langle \vec J\, \rangle\cdot ( \vec p_e \times \vec p_\nu)$,
where $\vec J$ is the neutron or $\Sigma$ baryon polarization, which is sensitive to time-reversal violation. The mLRSM contributions to this correlation in neutron decay and $\Sigma^- \rightarrow n e^- \bar{\nu}$ can be written as \cite{Jackson:1957zz},
 \begin{eqnarray}
D_n      &=& \frac{4 g_A}{1+3g_A\sq}\, \mathrm{Im}\,\frac{v^2C_{Hud}^{ud}}{2V_{L\,ud}}\simeq 0.87\, \mathrm{Im}\,\frac{v^2C_{Hud}^{ud}}{2V_{L\,ud}}\,,  \nn\\
D_\Sigma &=& \frac{4 g_{A\, \Sigma n}}{1+3g_{A\,  \Sigma n}\sq}\, \mathrm{Im}\,\frac{v^2C_{Hud}^{us}}{2V_{L\, us}}\simeq 1.01 \, \mathrm{Im}\,\frac{v^2C_{Hud}^{us}}{2V_{L\, us}}\,,  
\end{eqnarray}
where  $g_A = 1.27$, and $g_{A\, \Sigma n}=0.340 \pm 0.017$ \cite{Zyla:2020zbs} are the axial coupling of the nucleon and that of the $\Sigma$ to the neutron.
The SM contribution, as well as contamination from fake $T$-odd signals from final-state interactions,  are negligible with current experimental accuracy (see Ref.~\cite{Vos:2015eba} for a more detailed discussion). 
Current measurements give \cite{Mumm:2011nd,Hsueh:1988ar}
\bea
D_n = (-0.96\pm 1.89\pm 1.01)\cdot 10^{-4}\,,\qquad D_{\Sigma} = 0.11 \pm 0.10\,. 
\eea

\paragraph{$c\rightarrow d$ transitions}\label{app:cd}
Here we use we the leptonic and semileptonic decays of the $D$  mesons,
$D^+ \rightarrow \mu^+ \nu_\mu$ and  $D \rightarrow \pi l \nu_l$, to constrain the axial and vector couplings, respectively. The experimental input is \cite{Zyla:2020zbs,Amhis:2019ckw}
\begin{align}
D \rightarrow \pi l \nu_l:&\qquad &f_+^{D\pi}(0) \left|V_{L\,cd} + \frac{v^2}{2}C_{Hud}^{cd}\right| &= 0.1426 \pm 0.0019 \,  , \nn\\
D^+ \rightarrow \mu^+ \nu_\mu,\, \tau^+ \nu_\tau:&\qquad &f_D \left|V_{L\,cd} - \frac{v^2}{2}C_{Hud}^{cd}\right| &= 45.91 \pm 1.05 \, \textrm{MeV}\, .
\end{align}
\paragraph{$c\rightarrow s$ transitions}\label{app:cs}
Analogously to the $c\to d$ case, the leptonic $D_s$ decay and semileptonic decay of the $D$ to kaons can be used to constrain $c\to s$ transitions.
We use \cite{Zyla:2020zbs,Amhis:2019ckw}
\begin{align}
D \rightarrow K l \nu_l:&\qquad &f_+^{DK}(0) \left|V_{L\,cs}  +\frac{v^2}{2}C_{Hud}^{cs}\right| &= 0.7226 \pm 0.0034\,   ,\nn\\
D_s^+ \rightarrow \mu^+ \nu_\mu,\,  \tau^+ \nu_\tau:&\qquad & f_{D_s} \left|V_{L\,cs} -  \frac{v^2}{2}C_{Hud}^{cs}\right| &= 250.9 \pm 4.0 \, \textrm{MeV}\, .
\end{align}

\paragraph{$b \rightarrow c$ transitions}\label{app:cb}
The vector component of the charged $Wcb$ current is constrained by the semileptonic decay $B \rightarrow D l \nu_l$. 
For the axial component, the purely leptonic decay of the $B_c$ meson has not yet been observed, while the decay $B \rightarrow D^* l \nu_l$ depends on both  the vector and axial current. In the zero-recoil limit,   when $w = v \cdot v^\prime = 1$, where $v$ and $v^\prime$ are the $B$ and $D$ mesons four-velocities, only the axial contribution survives \cite{Manohar:2000dt}.  
Using the HFLAV averages \cite{Amhis:2019ckw}, we can write 
\begin{align}
B \rightarrow D l \nu_l:&\qquad &\eta_{EW}\mathcal F_{D}(1) |V_{L\,cb} +  \frac{v^2}{2}C_{Hud}^{cb}| =   \left( 42.00 \pm 0.45 \pm 0.89 \right) \cdot 10^{-3} \, , \nonumber \\
B \rightarrow D^* l \nu_l:&\qquad &\eta_{EW}'\mathcal F_{D^*}(1) |V_{L\,cb} - \frac{v^2}{2}C_{Hud}^{cb}| =  \left( 35.27 \pm 0.11 \pm 0.36 \right) \cdot 10^{-3}\, ,\label{eq:Vcbex}
\end{align}
	where $\eta_{EW}=1.012\pm0.005$ and $\eta_{EW}^{\prime}=1.0066\pm0.0050$ \cite{Sirlin:1981ie,Bailey:2014tva,Lattice:2015rga,Zyla:2020zbs} are electroweak corrections and
$\mathcal F_{D}(1)$ and $\mathcal F_{D^*}(1)$ denote the form factors, evaluated at $w=1$,  which are given in Table~\ref{LQCDinput}.

Apart from these exclusive decays, $V_{L\, cb}$ and $C_{Hud}^{cb}$ can also be constrained through the inclusive decays $\bar{B} \rightarrow X_c l \bar\nu_l$. Neglecting power corrections of order $\mathcal O(\Lambda_{\textrm{QCD}}/m_b)$, the inclusive semileptonic width into charmed final states is given by
\begin{eqnarray}
\Gamma(B \rightarrow X_c l \nu) &=& \frac{G_F^2 m_b^5 |V_{L\, cb}|^2}{192 \pi^3} \left[ \left( 1 + \left|\frac{v^2C_{Hud}^{cb}}{2V_{L\, cb}}\right|^2\right) \left( 1 - 8 \rho + 8 \rho^3 - \rho^4 - 12 \rho^2 \log\rho \right)  \right. \nn \\
& & \left. - 4 \frac{m_c}{m_b}  \textrm{Re} \left( \frac{v^2C_{Hud}^{cb}}{2V_{L\, cb}} \right)\,  \left(  1 + 9 \rho - 9 \rho^2 - \rho^3  + 6 \rho (1+\rho) \log\rho \right)
\right]\, ,\label{VcbWidth}
\end{eqnarray}
where $\rho = m_c^2/m_b^2$. We then set constraints by using the PDG average  \cite{Zyla:2020zbs},
\bea
B\to X_c l\nu:\qquad|V_{cb}^{\rm eff}| = (42.2 \pm 0.8)\Ex{-3}  \, , \label{VcbIncl}
\eea
where $|V_{cb}^{\rm eff}|\sq = |V_{L\, cb}|^2\, \Gamma(B \rightarrow X_c l \nu) /\Gamma^{{\rm SM}}(B \rightarrow X_c l \nu) $. 

The limits obtained from these inclusive decays and $B \rightarrow D^* l \nu_l$ should be interpreted as an order-of-magnitude constraint only. The reason is that Eq.\ \eqref{VcbWidth} does not include power corrections \cite{Bauer:2002sh,Gambino:2011cq,Gambino:2013rza,Alberti:2014yda}, while both Eqs.\ \eqref{eq:Vcbex} and \eqref{VcbWidth} rely on SM fits to the leptonic and hadronic moments of the decay distributions that do not include modifications due to $C^{cb}_{Hud}$. For a recent discussion in the case of $B \rightarrow D^* l \nu_l$, see Ref.\ \cite{Huang:2021fuc}.
A complete analysis that properly takes these issues into account is beyond the scope of the current work and we will use Eq.\ \eqref{eq:Vcbex} and \eqref{VcbWidth} to estimate the limits from the exclusive and inclusive measurements, while referring to Refs.\ \cite{Dassinger:2008as,Feger:2010qc} for a more detailed discussion.

\paragraph{$b \rightarrow u$ transitions}\label{app:ub}
In the case of $b\to u$ transitions, the leptonic channel $B^+ \rightarrow \tau^+ \nu_\tau$ constrains the axial current, while the vector current is probed by $B \rightarrow \pi l \nu_l$.
In what follows we  will use the HFLAV average of the BaBar and Belle results, $\textrm{Br} (B^+ \rightarrow \tau \nu) = (1.06 \pm 0.19 ) \cdot 10^{-4}$  \cite{Amhis:2019ckw}, and we employ the FLAG extraction for the semileptonic case \cite{Aoki:2019cca},
\begin{align}
B \rightarrow \pi l \nu_l:&\qquad &|V_{L\, ub} +  \frac{v^2}{2}C_{Hud}^{ub}| & = (3.74 \pm0.14 ) \cdot 10^{-3}\, ,\nn \\
B^+ \rightarrow \tau^+ \nu_\tau:&\qquad &f_B|V_{L\, ub} -  \frac{v^2}{2}C_{Hud}^{ub}|  &= ( 0.77 \pm 0.12)\,{\rm MeV} ,
\end{align}
where the decay constant, $f_B$, is given in Table \ref{LQCDinput}.

In addition, inclusive decays lead to the following constraint  \cite{Zyla:2020zbs},
\bea\label{eq:InclVub}
B\to X_u l\nu:\qquad \sqrt{|V_{L\, ub}|\sq +| \frac{v^2}{2}C_{Hud}^{ub}|\sq}= (4.25 \pm 0.12^{+0.15}_{-0.14}\pm0.23)\Ex{-3}\, .
\eea
These inclusive  decays suffer from similar problems as those in the $b\to c$ transitions; ideally, power corrections should be included \cite{Bauer:2001rc,Lange:2005yw} and the leptonic spectrum should be refitted to take into account $C_{Hud}^{ub}$ contributions. However, such an analysis is beyond the scope of the current work, and we estimate constraints from inclusive decays by using Eq.\ \eqref{eq:InclVub}.
\newline

Finally, the measurements of $\Lambda_b$ baryon decays, in particular the ratio $\textrm{Br}(\Lambda^0_b \rightarrow p \mu^- \bar{\nu})_{q\sq > 15\, {\rm GeV}}/\allowbreak\textrm{Br}(\Lambda^0_b \rightarrow \Lambda_c^+ \mu^- \bar{\nu})_{q\sq > 7\, {\rm GeV}}$, are sensitive to both the $b\to u$ and $b\to c$ charged currents. Here we use the form factors from the lattice QCD calculation of Ref.\ \cite{Detmold:2015aaa} and  obtain  the following partially integrated decay widths,
\bea \label{eq:LambdaB}
\Gamma(\Lambda_b^0\to p\mu^- \bar \nu)_{q\sq > 15\, {\rm GeV}} &=&4.17\, {\rm ps}^{-1} \,|V_{L\,ub}+ \frac{v^2}{2}C_{Hud}^{ub}|\sq +8.17\, {\rm ps}^{-1} \,|V_{L\,ub}- \frac{v^2}{2}C_{Hud}^{ub}|\sq\nn\\
\pm \sigma_{\rm stat}^{(p)}\pm \sigma_{\rm syst}^{(p)}\,,\nn\\
\Gamma(\Lambda_b^0\to \Lambda_c^+\mu^- \bar \nu)_{q\sq > 7\, {\rm GeV}} &=&1.41\, {\rm ps}^{-1} \,|V_{L\, cb}+ \frac{v^2}{2}C_{Hud}^{cb}|\sq +6.99\, {\rm ps}^{-1} \,|V_{L\,cb}- \frac{v^2}{2}C_{Hud}^{cb}|\sq\nn\\
\pm \sigma_{\rm stat}^{(\Lambda_c^+)}\pm \sigma_{\rm syst}^{(\Lambda_c^+)}\,,
\eea
where the lattice uncertainties are given by
\bea
(\sigma_{\rm stat}^{(p)}\, {\rm ps}) \sq &=&  0.10\, \left|V_{L\,ub}+ \frac{v^2}{2}C_{Hud}^{ub}\right|^4+0.33\, \left|V_{L\,ub}- \frac{v^2}{2}C_{Hud}^{ub}\right|^4 +0.16\,  \left|V_{L\,ub}\sq-\left( \frac{v^2}{2}C_{Hud}^{ub}\right)\sq \right|   \sq\,,\nn\\
(\sigma_{\rm syst}^{(p)}\, {\rm ps}) \sq &=&  0.10\, \left|V_{L\,ub}+ \frac{v^2}{2}C_{Hud}^{ub}\right|^4+0.44\, \left|V_{L\,ub}- \frac{v^2}{2}C_{Hud}^{ub}\right|^4 +0.050\,  \left|V_{L\,ub}\sq-\left( \frac{v^2}{2}C_{Hud}^{ub}\right)\sq\right|\sq \,,\nn\\
(\sigma_{\rm stat}^{(\Lambda_c^+)}\, {\rm ps}) \sq &=& 0.0023\, \left|V_{L\,cb}+ \frac{v^2}{2}C_{Hud}^{cb}\right|^4+0.017\, \left|V_{L\,cb}-\frac{v^2}{2}C_{Hud}^{cb}\right|^4 +0.0052\,  \left|V_{L\,cb}\sq-\left(\frac{v^2}{2}C_{Hud}^{cb}\right)\sq\right|\sq\,,\nn\\
(\sigma_{\rm syst}^{(\Lambda_c^+)}\, {\rm ps}) \sq &=& 0.0053\, \left|V_{L\,cb}+\frac{v^2}{2}C_{Hud}^{cb}\right|^4 +0.11\, \left|V_{L\,cb}-\frac{v^2}{2}C_{Hud}^{cb}\right|^4+0.0027\,  \left|V_{L\,cb}\sq-\left(\frac{v^2}{2}C_{Hud}^{cb}\right)\sq\right| \sq\,.
\eea
We then set constraints by combining this theory prediction with the experimental determination \cite{Aaij:2015bfa,Zyla:2020zbs},
\bea
\frac{\textrm{Br}(\Lambda^0_b \rightarrow p \mu^- \bar{\nu})_{q\sq > 15\, {\rm GeV}}}{\textrm{Br}(\Lambda^0_b \rightarrow \Lambda_c^+ \mu^- \bar{\nu})_{q\sq > 7\, {\rm GeV}} }= \left( 0.92\pm 0.04\pm0.07\right)\Ex{-2}\,.
\eea

\subsection{$\Delta B =1$  and $\Delta S =1$ processes}\label{app:DeltaB1}

Here we consider two types of processes, namely decays induced at tree level through charged currents, and loop-induced flavor-changing neutral currents. The decay $B\to J/\psi K_S$ is in the former category and is important in the determination of the SM CKM elements. In particular, it allows for a precise determination of the phase $\beta\simeq {\rm Arg }\,\left(-V_{L\, td}\right)$, while it is not expected to be very sensitive to mLRSM contributions. 

Instead, $\Delta B = 1$ and $\Delta S = 1$  FCNC processes such as $B \rightarrow X_{s,d}\, \gamma$ and $K_L \rightarrow \pi^0 e^+ e^-$
lead to stringent constraints on the elements of $C_{Hud}$ involving the top quark, as they benefit from an enhancement factor of $m_t/m_b$ compared to the SM contributions. 

The theoretical expressions for $\Delta B =1$ FCNC observables are usually written in terms of the $C_{7,8}^{(\prime)}$ coefficients, see e.g.\ Refs.\ \cite{Altmannshofer:2012az,Altmannshofer:2011gn}, which are related to the couplings of the dipole operators in Eq.\ \eqref{eq:Lagdip} as follows 
\begin{eqnarray}\label{eq:C7}
C_7(m_W) &=&-\frac{4\pi\sq Q_d }{V_{L\, tb}V_{L\, tq}^*}\, v\sq C_{\g d}^{qb}\,, \qquad C^\prime_7(m_W) = -\frac{4\pi\sq Q_d }{V_{L\, tb}V_{L\, tq}^*}\frac{m_q}{m_b}\, \big(v\sq C_{\g d}^{bq}\big)^* \, ,  \nn \\
C_8(m_W)& =&  \frac{4\pi\sq  }{V_{L\, tb}V_{L\, tq}^*}\, v\sq C_{g d}^{qb} \,,  \qquad  C^\prime_8(m_W) = \frac{4\pi\sq  }{V_{L\, tb}V_{L\, tq}^*}\frac{m_q}{m_b}\, \big(v\sq C_{g d}^{bq}\big)^*\, .
\end{eqnarray}
Below we closely follow the analysis of Ref.\ \cite{Alioli:2017ces} and focus on the $B \rightarrow X_{s,d} \gamma$ branching ratios, the CP asymmetries in inclusive  $B\to X_{d,s}\g$ decays, and in the exclusive channel $B\to K^{*0}\g$. We summarize the relevant experimental results \cite{Zyla:2020zbs,Amhis:2019ckw}  in Table \ref{DB1exp}.

\subsubsection{$B\to J/\psi K_S$}\label{app:Beta}
In the SM, the time-dependent CP asymmetry in $B \rightarrow J/\psi K_S$ is sensitive to the angle $\beta = {\rm Arg} \big(-\frac{V_{L\,cd}V_{L\,cb}^*}{V_{L\,td}V_{L\,tb}^*}\big)$. The CP asymmetry is defined as
\bea
\frac{\Gamma(\bar B\to J/\psi K_S)-\Gamma(B\to J/\psi K_S)}{\Gamma(\bar B\to J/\psi K_S)+\Gamma(B\to J/\psi K_S)} = S_{J/\psi K_S}\sin (\Delta m_d t)+C_{J/\psi K_S}\cos (\Delta m_d t)\, .
\eea
Here 
\bea\label{eq:BJpsi1}
S_{J/\psi K_S}=\frac{2{\rm Im}\lambda_{J/\psi K_S}}{1+|\lambda_{J/\psi K_S}|\sq}\,,\qquad \lambda_{J/\psi K_S} = \left(\frac{q}{p}\right)_{B_d}\frac{\bar A_{J/\psi K}}{A_{J/\psi K}}\,,
\eea
where $(q/p)_{B_d}$ is related to the mixing parameters in  $B^0_d- \bar{B}^0_d$ oscillations, and the ratio of amplitudes is given by
\begin{equation}\label{eq:BJpsi2}
\frac{\bar A_{J/\psi K}}{A_{J/\psi K}} = \left(\frac{p}{q} \right)_K \frac{\langle J/\psi \bar K_0 | \, \mathcal H_w | \bar B^0_d \rangle}{ \langle J/\psi  K_0 | \, \mathcal H_w |  B^0_d \rangle}\,.
\end{equation}
In both the  $K - \bar K$ and $B - \bar B$ systems,  the ratio $|q/p|$ can be shown to be very close to 1 without the need for additional theoretical assumptions, so that we have  \cite{Branco:1999fs}
\begin{equation}
\left(\frac{q}{p}\right)_{B_d} = \exp( i \arg(M_{12}^*)_{B_d})\,, \qquad \left(\frac{q}{p}\right)_{K} = \exp( i \arg(M_{12}^*)_{K})\,,
\end{equation}
up to very small corrections.
In the SM, these phases can be expressed in terms of ratios of CKM elements, while the corrections to $\left(M_{12}\right)_{B_d,K}$ within the mLRSM are discussed in Sects.\ \ref{sec:BBbar} and \ref{sec:S=2}.

In addition, there are corrections to the ratio of the $ r_{J/\psi K} = \frac{\langle J/\psi \bar K_0 | \, \mathcal H_w | \bar B^0_d \rangle}{ \langle J/\psi  K_0 | \, \mathcal H_w |  B^0_d \rangle}$. Within the SM, these transitions are mediated by the tree-level charged-current operators, $C_{i\, LL}$. In this case, the non-perturbative matrix elements drop out in the ratio leaving only CKM elements. Within the mLRSM there are additional contributions from the $C_{i\, RR}$ and $C_{i\, LR}$ operators. Expanding the ratio to first order in $1/M_{W_R}^2$ we have,
\bea
r_{J/\psi K} &=& - \frac{V_{L\,cb} V_{L\,cs}^*}{V_{L\,cb}^* V_{L\,cs}} \Bigg[1-2i {\rm Im}\,\Bigg(\frac{C_{1\, LR}^{bccs}+C_{1\, LR}^{csbc}+r_{LR}(C_{2\, LR}^{bccs}+C_{2\, LR}^{csbc})+C_{1\, RR}^{bccs}+r_{LL } C_{2\, RR}^{bccs}}{C_{1\, LL}^{bccs}+r_{LL } C_{2\, LL}^{bccs}}\Bigg)\Bigg]\,,\nn\\
r_{LL} &=&\frac{\langle J/\psi \bar K_0 | \bar s_L^\al \gamma^\mu c_L^\bt \, \bar c_L ^\bt \gamma^\mu b_L^\al | \bar B^0_d \rangle}{ \langle J/\psi \bar K_0 | \bar s_L \gamma^\mu c_L \, \bar c_L \gamma^\mu b_L | \bar B^0_d \rangle}\,,\qquad
r_{LR} =  \frac{\langle J/\psi \bar K_0 | \bar s_L^\al \gamma^\mu c_L^\bt \, \bar c_R ^\bt \gamma^\mu b_R^\al | \bar B^0_d \rangle}{ \langle J/\psi \bar K_0 | \bar s_L \gamma^\mu c_L \, \bar c_R \gamma^\mu b_R | \bar B^0_d \rangle}\,,
\eea
where the matrix elements and the Wilson coefficients are to be evaluated at the same scale. As the ratios of matrix elements, $r_{LL,LR}$, are currently unknown, the non-standard contributions to $r_{J/\psi K}$ are hard to estimate. However, these terms do not come with any enhancement factors. In addition, within the $P$-symmetric scenario, the phases of $C_{i\, LR,RR}$ are expected to be closely aligned to those of $C_{i\,LL}$ due to the relation between $V_L$ and $V_R$, Eq.\ \eqref{CKMP}, and the fact that $\al$ is stringently constrained by CP-violating $\Dt F=0$ observables. We therefore expect these contributions to be below the experimental sensitivity for $M_{W_R}\gtrsim 1$ TeV and neglect them in our analysis. We thus use $ r_{J/\psi K} =-\frac{V_{L\,cb} V_{L\,cs}^*}{V_{L\,cb}^* V_{L\,cs}} $ in combination with Eqs.\ \eqref{eq:BJpsi1} and \eqref{eq:BJpsi2}, which we compare with the experimental value \cite{Amhis:2019ckw}
\bea
S_{J/\psi K_S} = 0.695\pm0.019\, .
\eea

\subsubsection{The $B\to X_{d,s}\g$  branching ratio}\label{BRbsg}

For the $B \rightarrow X_{d,s}\gamma$ branching ratios, we employ the expressions derived in Ref.\ \cite{Hurth:2003dk} rescaled by the SM predictions of Refs.\ \cite{Misiak:2006zs,Misiak:2015xwa,Czakon:2015exa},
\bea
 \text{BR}\,(B\to X_q\g) &=& r_q\frac{\mathcal N}{100} \frac{\left|V_{L\, tq}^* V_{L\, tb}\right|^2}{|V_{L\, cb}|\sq+|\frac{v^2}{2}{C_{Hud}^{cb}|\sq}}\bigg[ a+a_{77}(|R_7|\sq+|R_7'|\sq)+a_7^r \,{\rm Re}\, R_7+a_7^i \,{\rm Im}\, R_7\nn\\&&+
a_{88}(|R_8|\sq+|R_8'|\sq)+a_8^r \,{\rm Re}\, R_8+a_8^i \,{\rm Im}\, R_8 +a_{\epsilon\epsilon}|\epsilon_q|\sq +a_\epsilon^r \,{\rm Re}\, \epsilon_q\nn\\
&&+a_\epsilon^i \,{\rm Im}\, \epsilon_q + a_{87}^r\,{\rm Re}\,(R_8 R_7^*+R_8' R_7^{\prime\,*})+a_{87}^i\,{\rm Im}\,(R_8 R_7^*+R_8' R_7^{\prime\,*})\nn\\
&&+a_{7\epsilon}^r\,{\rm Re}\,(R_7 \epsilon_q^*)+a_{7\epsilon}^i\,{\rm Im}\,(R_7 \epsilon_q^*)+a_{8\epsilon}^r\,{\rm Re}\,(R_8 \epsilon_q^*)+a_{8\epsilon}^i\,{\rm Im}\,(R_8 \epsilon_q^*)\bigg]\, ,
\label{BR}\eea
where $R_{7,8} = \frac{C_{7,8}(m_t)}{C_{7,8}^{\rm SM}(m_t)}$, $R_{7,8}' = \frac{C'_{7,8}(m_t)}{C_{7,8}^{\rm SM}(m_t)}$, $C_7^{\rm SM}(m_t) = -0.189$, and $C_8^{\rm SM}(m_t) = -0.095$ and we neglect the SM contributions to $C_{7,8}'$ which are suppressed by $m_q/m_b$. In addition, $\mathcal N =  2.567(1\pm 0.064)\cdot 10^{-3}$, while $r_q$ are  factors that rescale the above expression to the SM predictions of Refs.\ \cite{Misiak:2006zs,Misiak:2015xwa,Czakon:2015exa} for which we use $r_s = \frac{3.36}{3.55}$ and $r_d = \frac{1.73}{1.47}$. Finally, $\epsilon_q = \frac{V_{L\, uq}^*V_{L\, ub}}{V_{L\, tq}^*V_{L\, tb}}$ and  the coefficients  $a_{ij}$ can be found in Ref.\ \cite{Hurth:2003dk}. We applied the expressions valid for a cut on the photon energy of $E_\gamma > 1.6 $ GeV, which, for $B \rightarrow X_d\, \gamma$,  requires extrapolating the branching ratio quoted in Ref.\ \cite{Amhis:2019ckw}, as discussed in Ref. \cite{Misiak:2015xwa}.

To set constraints we compare the branching ratios in Eq. \eqref{BR}  with the current experimental world averages  \cite{Zyla:2020zbs,Amhis:2019ckw},  shown in Table \ref{DB1exp}. To take into account theoretical uncertainties, we follow Refs.\ \cite{Altmannshofer:2012az,Altmannshofer:2011gn}  and use the following theory errors  $\sigma_d = \frac{0.22}{1.73}{\rm BR}(B\to X_d\g) $ and $\sigma_s = \frac{0.23}{3.36}{\rm BR}(B\to X_s\g) $, which are added in quadrature to the experimental ones.

\subsubsection{The $B\to X_{d,s}\g$  CP asymmetry}\label{ACPinc}
The $B\to X_s\g$ CP asymmetry provides a probe of the phase of the $tb$ element of $C_{Hud}$. We employ the expression derived in Ref.~\cite{Benzke:2010tq},
\bea\label{eq:BsgammaCP}
\frac{A_{CP}(B\to s\g)}{\pi}&\equiv & \frac{1}{\pi}\frac{\Gamma(\bar B\to X_s\g)-\Gamma(B\to X_{\bar s}\g)}{\Gamma(\bar B\to X_s\g)+\Gamma(B\to X_{\bar s}\g)} \nn\\&\approx & \bigg[\bigg(\frac{40}{81}-\frac{40}{9}\frac{\Lambda_c}{m_b}\bigg)\frac{\al_s}{\pi}+\frac{\Lambda_{17}^c}{m_b}\bigg]\text{Im}\,\frac{C_2}{C_7}-\bigg(\frac{4\al_s}{9\pi}+4\pi\al_s\frac{\Lambda_{78}}{3m_b}\bigg)\text{Im}\,\frac{C_8}{C_7}\nn\\
&&-\bigg(\frac{\Lambda_{17}^u-\Lambda_{17}^c}{m_b}+\frac{40}{9}\frac{\Lambda_c}{m_b}\frac{\al_s}{\pi}\bigg)\text{Im}\,\bigg( \ep_s \frac{C_2}{C_7}\bigg)\ ,
\eea 
where the Wilson coefficients should be evaluated at the factorization scale $\mu_b\simeq 2$ GeV and $C_2$ denotes the coefficient of the SM charged-current operator $\mathcal O^{sc\, cb}_{1\, LL}$, $C_2 = C^{sc\, cb}_{1\, LL}/(2\sqrt{2}G_FV_{L\, cb} V^*_{L\, cs})$. We use the following values for the SM parts of these coefficients  \cite{Benzke:2010tq},
\bea
C_2^{\rm SM}(2\, {\rm GeV}) = 1.204\, ,\qquad C_7^{\rm SM}(2\, {\rm GeV}) = -0.381\, ,\qquad C_8^{\rm SM}(2\, {\rm GeV}) = -0.175\, .\label{BcoeffSM}
\eea
Furthermore, $\Lambda_c\simeq 0.38\, \text{GeV}$, while the three hadronic parameters, $\Lambda_{17}^{u,c}$ and $\Lambda_{78}$,  are estimated to lie in the following ranges \cite{Benzke:2010tq},
\bea
\Lambda_{17}^u\in [-0.33,\, 0.525]\, \text{GeV},\qquad \Lambda_{17}^c\in [-0.009,\, 0.011]\, \text{GeV},\qquad \Lambda_{78}\in [0.017,\, 0.19]\, \text{GeV}\,.\label{lambdas}
\eea
We compare the above expressions with the experimental result in Table \ref{DB1exp}.

\subsubsection{The $B\to K^{*0}\g$  CP asymmetry}\label{ACPex}
In addition we consider the time-dependent CP asymmetry in  $B\to K^{*0}\g$ decays
\bea
\frac{\Gamma(\bar B\to \bar K^{*0}\g)-\Gamma(B\to K^{*0}\g)}{\Gamma(\bar B\to \bar K^{*0}\g)+\Gamma(B\to K^{*0}\g)}
=S_{K^*\g}  \cos(\Delta m_d t)+C_{K^*\g}  \sin(\Delta m_d t)\, ,\eea
where we focus on $S_{K^*\g}$, which can be expressed as
\bea
S_{K^*\g}  = 2\frac{{\rm Im}\, \lambda_{K^{*}\g}}{1+|\lambda_{K^{*}\g}|\sq}, \qquad \lambda_{K^{*}\g} = \frac{q}{p} \frac{A( \bar B\to \bar K^{*0}\g)}{A(B\to  K^{*0}\g)}\, ,
\eea
where the ratio  $\frac{q}{p} =\sqrt{\frac{M_{12}^*}{M_{12}}}$ arises from the phase of the $B_d-\bar B_d$ mixing amplitude $M_{12}$ discussed in Sect.\ \ref{sec:BBbar}. This asymmetry is generated by the electromagnetic dipole operators, $C_7$ and $C_7^\prime$, at leading order and vanishes as $C_7'\to 0$. The latter coefficient is suppressed by $m_s/m_b$ in the SM, while it is enhanced in the presence of $C_{Hud}$, making it a probe of right-handed currents. Using the fact that the largest BSM modifications will arise from the enhanced $C_7'$ contributions we can approximate the ratio $q/p$ by its SM value,   $q/p \simeq (V_{L\, tb}V_{L\, td}^*)/(V_{L\, tb}^*V_{L\, td})$.
The leading-order expression is then given by \cite{Altmannshofer:2011gn,Paul:2016urs},
\bea
S_{K^*\g} = \frac{2\, {\rm Im}\bigg(\frac{V_{L\, tb}V_{L\, td}^*}{V_{L\, tb}^*V_{L\, td}} \frac{V_{L\, tb}V_{L\, ts}^*}{V_{L\, tb}^*V_{L\, ts} }\,C_7 C_7'\bigg)}{|C_7|\sq+|C_7'|\sq}\, ,
\eea 
while the SM prediction is rather small \cite{Ball:2006cva,Ball:2006eu}
\bea
S_{K^*\g}^{\rm SM} = (-2.3\pm1.6)\Ex{-2}\, .
\eea
The  experimental value for $S_{K^*\g}$ is given in Table \ref{DB1exp}.

\subsubsection{Corrections to the B meson widths}\label{app:Bwidths}
The absorptive part of the box diagrams that induce $B - \bar B$ oscillations give rise to the $B_q$ meson widths.
The corrections due to $W_L-W_R$ mixing were computed in Ref.\ \cite{Alioli:2017ces}, and  are given by 
\bea
\Gamma_{12}^{(q)}(\xi) &=& -\frac{1}{2}\frac{G_F\sq m_b\sq  m_{B_q}f_{B_q}\sq}{\pi}\sqrt{z} \left(\lambda^{(q)\,2}_c\big(\sqrt{1-4z}-(1-z)\sq\big)-\lambda^{(q)}_c\lambda^{(q)}_t (1-z)\sq\right)\times\nn\\
&& \bigg[\left(\left[\frac{2}{3}B_1-\frac{5}{6}B_2R \right]\frac{\xi_{cb}}{V_{L\,cb}}+\frac{1}{3}B_5 \left(R + \frac{3}{2}\right)\frac{\xi_{cq}^*}{V_{L\,cq}^*}\right) \eta_{11LL}\eta_{11LR} \nn\\
&&+\left(\left[\frac{2}{3}B_1+\frac{1}{6}B_3 R\right]\frac{\xi_{cb}}{V_{L\,cb}}+B_4 \left( R + \frac{1}{6}\right)\frac{\xi_{cq}^*}{V_{L\,cq}^*}\right)\big(\eta_{11LL}\eta_{21LR}+\eta_{21LL}\eta_{11LR} + 3 \,\eta_{21LL}\eta_{21LR}\big)\bigg]\, , \nn\\
\eea
where $z\equiv m_c\sq/m_b\sq$, $\lambda_i^{(q)} = V_{L\, ib}V_{L\, iq}^*$, and $ \xi_{i j } \equiv \frac{v^2}{v_R^2} \frac{\xi e^{i \alpha}}{1 + \xi^2} V_{R\,i j}$.
The bag factors, $B_i$, are again given in Table \ref{TabBag}, where the $B_1$ factors are related to the RG-invariant definition in Table  \ref{TabBag} by an RG factor, $B_1(m_b) =  \hat B_{B_{d,s}}/1.517$ for the $B_{d,s}$ systems  \cite{Aoki:2019cca}. The $\eta$ factors describe the RGE evolution of the four-fermion operators between $m_W$ and $m_b$, through $C_{i\, LL(LR)}(m_b)=\eta_{ij LL(LR)}C_{j \,LL(LR)}(m_W)$. Explicitly we have
\bea
\eta_{11\,LL} &=& \frac{1}{2}\big(\eta^{6/23}+\eta^{-12/23}\big)\,,\qquad \eta_{11\,LR} = \eta^{3/23}\,,\nn\\
\eta_{21\,LL} &=& \frac{1}{2}\big(\eta^{6/23}-\eta^{-12/23}\big)\,,\qquad \eta_{21\,LR} = \frac{1}{3}\big(\eta^{-24/23}-\eta^{3/23}\big)\,,
\eea
where $\eta = \al_s(m_W)/\al_s(m_b)$.

Additional contributions arise from  diagrams involving $C_{i\, RR}$, due to $W_R$ exchange
\begin{eqnarray}
\Gamma^{(q)} (C_{i\, RR})&=& - \frac{1}{4}\frac{z}{2\pi} \sqrt{1-4 z} \, m_{B_q} f^2_{B_q} m_b^2  \left\{ 
B_4 \left( R(\mu) + \frac{1}{6} \right) C_{1 RR}^{q c c b} C_{1\, LL}^{q c c b}  \right. \nonumber \\ && \left.
+ \frac{1}{3} B_5 \left(R + \frac{3}{2}\right) \left( C_{2\, RR}^{q c c b} C^{q c c b}_{1\, LL} + C_{1\, RR}^{q c c b}C_{2\, LL}^{q c c b} + N_c C_{2\, RR}^{q c c b} C_{2\, LL}^{q c c b}
\right)
\right\}.
\end{eqnarray}

The real part of these contributions to $\Gamma_{12}$ can be constrained by the width difference between the mass eigenstates, whereas $a_{\rm fs}^q$ is sensitive to the imaginary part  \cite{Buras:1997fb},
\bea
\Delta \Gamma^{(q)}=4\frac{{\rm Re}\, \big(\Gamma_{12}^{(q)*} M^{(q)}_{12}\big)}{\Delta m_{q}}\, ,\qquad a_{\rm fs}^q=1-\bigg|\frac{q}{p}\bigg|\sq =- {\rm Im}\bigg(\frac{\Gamma_{12}^{(q)} }{M^{(q)}_{12}}\bigg)\, .
\eea
These expressions only depend on the ratio of $\Gamma_{12}^{(q)}/M_{12}^{(q)}$ and $\Dt m_q = 2\vert M_{12}^{(q)}\vert$, which we expand in terms of the BSM contributions as follows, 
\bea
\frac{\Gamma_{12}^{(q)}}{M_{12}^{(q)}}\simeq \frac{\Gamma_{12}^{(q)}({\rm SM})}{M_{12}^{(q)}({\rm SM})}\left(1-\frac{M_{12}^{(q)}({\rm LR})}{M_{12}^{(q)}({\rm SM})}\right) +\frac{\Gamma_{12}^{(q)}({\rm LR})}{M_{12}^{(q)}({\rm SM})}\,,
\eea
where $\Gamma_{12}^{(q)}({\rm LR}) = \Gamma_{12}^{(q)}(\xi)+\Gamma_{12}^{(q)}(C_{i\, RR})$, while $M_{12}^{(q)}({\rm LR})$ is given by Eq.\ \eqref{eq:BBbarM12}. We combine the mLRSM contribution with the SM prediction, which is given by \cite{Artuso:2015swg},
\bea
\frac{\Gamma_{12}^{(q)}}{M_{12}^{(q)}}\Bigg|_{\rm SM} = -10^{-4}\left[c^{(q)}+a^{(q)}\frac{\lambda_u^{(q)}}{\lambda_t^{(q)}} + b^{(q)}\left(\frac{\lambda_u^{(q)}}{\lambda_t^{(q)}}\right)^2\right]\,,
\eea
with
\bea
a^{(d)} &=& 11.7\pm 1.3\,,\qquad a^{(s)} = 12.3\pm 1.4\,, \nn\\
b^{(d)} &=& 0.24\pm0.06\,,\qquad b^{(s)} = 0.79\pm0.12\,, \nn\\
c^{(d)} &=& -49.5\pm8.5\,,\qquad c^{(s)} = -48.0\pm8.3\,. 
\eea
The experimental determinations are shown in Table \ref{DB1exp}.

\subsubsection{$K_L\to \pi^0 e^+e^-$}\label{app:Kpiee}
This decay is sensitive to the dipole operators $C^{ds}_{\gamma d}$ and $C^{sd}_{\gamma d}$. Due to the enhancement factors of $m_t/m_{s,d}$ and $m_c/m_{s,d}$ that appear in the matching of these Wilson coefficients, the LR model can give rise to large contributions to the branching fraction.
Within the SM, this decay is mediated by the semi-leptonic penguin operators $C_{7V} \bar s \gamma^\mu d \, \bar e \gamma_\mu e $ and $C_{7A} \bar s \gamma^\mu d \, \bar e \gamma_\mu \gamma_5 e $ \cite{Buchalla:1995vs}, that give rise to direct CP violation. In addition, there are long-distance and indirect CPV contributions that are harder to estimate. 

The above contributions involve the following vector and tensor form factors,
\begin{eqnarray}
\langle \pi^0 | \bar s \gamma^\mu d | K_L \rangle &=&  \frac{1}{\sqrt{2}} f^{K^0\pi^+}_+(q^2) (p^\mu_K + p^\mu_\pi)\, , \nn \\
\langle \pi^0 | \bar s \sigma^{\mu\nu} d | K_L \rangle &=& i f^{K\pi}_{T}(q^2) \frac{\sqrt{2}}{m_K + m_\pi} (p^\mu_\pi p^\nu_K - p_K^\mu p_\pi^\nu)\, ,
 \end{eqnarray}
where $f^{K\pi}_+$ (see Table~\ref{LQCDinput}) is related to the vector form factor in $K^+ \rightarrow \pi^0 e^+ \nu$, while $f^{K\pi}_T$ has been computed in Ref. \cite{Baum:2011rm}, $f_{T}^{K\pi}= 0.417 \pm 0.015$, at a renormalization scale $\mu = 2$ GeV.
This allows us to express the branching fraction as 
\begin{eqnarray}\label{eq:KLpiee}
\textrm{Br}(K_L \rightarrow \pi^0 e^+ e^-) = \kappa_e \left[ \left( \textrm{Im} \lambda_t\, \tilde y_{7V} +  \frac{2}{m_K + m_\pi}  \frac{f_T^{K\pi}(0)}{f^{K\pi}_+(0)} 
16 \pi^2 \textrm{Im} (v^2 C_T)  \right)^2 + \textrm{Im} \lambda_t^2\, \tilde{y}^2_{7A}\right] \, ,
\end{eqnarray}
where $\lambda_t = V^*_{L\, ts} V_{L\, td}$ and  $\kappa_e$ is introduced to cancel the SM dependence on the vector form factor $f^{K\pi}_+$ by normalizing to the $K^+ \rightarrow \pi^0 e^+ \nu$ decay rate. $\kappa_e$ is defined as
\begin{eqnarray}
\kappa_e &=& \frac{1}{|V_{L\,us} + \frac{v^2}{2}C_{Hud}^{us}|^2} \frac{\tau(K_L)}{\tau(K^+)} \left( \frac{\alpha_{\textrm{em}}}{2\pi} \right)^2 \textrm{Br} (K^+ \rightarrow \pi^0 e^+ \nu) \,,\nn\\
&\simeq& \left(\frac{0.225}{|V_{L\, us} + \frac{v^2}{2}C_{Hud}^{us}|} \right)^2\, 6 \cdot 10^{-6}\, ,
\end{eqnarray}
where we used the experimental values of Ref.\ \cite{Zyla:2020zbs}.
The BSM contributions in Eq.\ \eqref{eq:KLpiee} arise from $C_T$
\bea
C_T(\mu) = -\frac{Q_d}{4} \left( m_s C^{ds *}_{\gamma d}(\mu) + m_d C^{sd}_{\gamma d}(\mu) \right)\, ,
\eea
while the Wilson coefficients of the SM penguin operators are given by \cite{Buchalla:1995vs}
\begin{eqnarray}
\tilde y_{7V}(\mu) &=& P_0(\mu)  - 4 \left( C_0(x_t) + \frac{1}{4} D_0(x_t) \right)  + \frac{Y_0(x_t)}{s_w^2}, \qquad \tilde y_{7 A} = - \frac{Y_0(x_t) }{s_w^2}  \, ,
\end{eqnarray}
with 
\begin{eqnarray}
Y_0(x_t) &=& \frac{x_t}{8} \left( \frac{4-x_t}{1-x_t} + \frac{3 x_t}{(1-x_t)^2} \log x_t \right)\, , \nonumber \\
C_0(x_t) &=& \frac{x_t}{8} \left( \frac{x_t-6}{x_t-1} + \frac{3 x_t + 2}{(1-x_t)^2} \log x_t \right)\, , \nonumber \\
D_0(x_t) &=& -\frac{4}{9} \log x_t +  \frac{-19 x_t^3 + 25 x^2_t}{36 (x_t-1)^3} + \frac{x_t^2 (5 x_t^2 - 2 x_t - 6)}{18 (1-x_t)^4} \log x_t\, ,
\end{eqnarray}
where $x_t = m_t(m_W)^2/m_W^2$ and, neglecting resummation, $P_0 = -4/9 \log x_c$. The value of $P_0(\mu)$ at different scales can be found in Ref.\ \cite{Buchalla:1995vs}. 

In principle, there are additional BSM contributions to Eq.\ \eqref{eq:KLpiee} as the mLRSM can also induce the semi-leptonic penguin operators. However, these contributions are not enhanced by factors of $m_t/m_{s,d}$. In addition, the contributions from heavy Higgs exchange are suppressed by small Yukawa couplings while those from loops involving $W_R$ bosons have the same form as  the SM contributions with $m_W\to M_{W_R}$ and $ x_t\to m_t^2/M_{W_R}^2$ so that they are suppressed compared to the SM.
It should be noted that Eq.\ \eqref{eq:KLpiee} only contains the direct CPV contributions from the SM and we neglected CP-even terms and indirect contributions due to $K$-$\bar K$ mixing  \cite{Buchalla:1995vs}. 
We nevertheless use this expression to estimate the branching ratio as the experimental limit is currently sensitive to branching ratios roughly two orders of magnitude larger than the SM prediction \cite{Zyla:2020zbs},
\bea
{\rm BR}(K_L\to \pi^0e^+e^-) < 2.8 \Ex{-10} \quad (90\% \, {\rm C.L.)}\,\, .
\eea

\section{Renormalization group equations}\label{app:RGE}
In this appendix we give several semi-analytical results for the RGE effects of the four-fermion operators discussed in Sect.\ \ref{sec:matching}. As mentioned in Sect.\ \ref{sect:RGEsummary} the Wilson coefficients of these operators in general depend on the scale at which we integrate out the heavy LR fields. In our analysis we take this to be a single scale $\mu_0 = M_{W_R}$. The resulting $\mu_0$ dependence of the right-handed charged currents is then approximately given by
\bea
v_R^2C_{1,2RR}^{ijkl}(\mu_{\rm low}) &=&\left[0.40\eta^{2/7}\pm 0.79\eta^{-4/7}\right]\left(V_R\right)_{ji}^*\left(V_R\right)_{kl}\,,\nn
\eea
where $\eta = \frac{\al_s(\mu_0)}{\al_s(m_t)}$ and we set $\mu_{\rm low}=2$ GeV. Similar expressions can be derived for the $C_{i\, quqd}$ coefficients
\bea
C_{1,\, quqd}^{ijkl}(\mu_{\rm low})&=&\eta^{\frac{1+\sqrt{241}}{21}}\left[
0.0045-0.093\eta^{-6/7}+0.86\eta^{-2\sqrt{241}/21}+2.1\eta^{-\frac{18+2\sqrt{241}}{21}}
\right] \frac{Y_{dH}^{kl}Y_{uH}^{ij}}{M_H^2}\nn\\
&&+
\eta^{\frac{1+\sqrt{241}}{21}}\left[
-0.0045-0.093\eta^{-6/7}-0.86\eta^{-2\sqrt{241}/21}+2.1\eta^{-\frac{18+2\sqrt{241}}{21}}
\right] \frac{ Y_{dH}^{il} Y_{uH}^{kj}}{M_H^2}\,,\nn\\
C_{2\, quqd}^{ijkl}(\mu_{\rm low})&=&
\eta^{\frac{1+\sqrt{241}}{21}}\left[
0.017+0.17\eta^{-6/7}-0.056\eta^{-2\sqrt{241}/21}-0.57\eta^{-\frac{18+2\sqrt{241}}{21}}
\right]
 \frac{Y_{dH}^{kl}Y_{uH}^{ij}}{M_H^2}\nn\\
 &&\eta^{\frac{1+\sqrt{241}}{21}}\left[
-0.017+0.17\eta^{-6/7}+0.056\eta^{-2\sqrt{241}/21}-0.57\eta^{-\frac{18+2\sqrt{241}}{21}}
\right]\frac{ Y_{dH}^{il} Y_{uH}^{kj}}{M_H^2}\,,\nn
\eea
where $Y_{qH}$ are to be evaluated at $\mu=\mu_0$.
Finally, the Wilson coefficients for the $\Dt F=2$ operators can be written as
\bea
C_4^{ijkl}(\mu_{\rm low}) &=& \frac{g_R^2}{M_{W_R}^2}\sum_{a,b}a_{ab}^{(4)}\frac{m_{u_a}m_{u_b}}{m_t^2}V^*_{L\,ai}V_{L\,bj}\left(V_R\right)_{bk}^*\left(V_R\right)_{al}\,,\\
C_5^{ijkl} (\mu_{\rm low})&=& -1.26\eta^{-8/7}\frac{1}{M_H^2}\left( Y_{dH}\right)^*_{jk}Y_{dH}^{il}+\frac{g_R^2}{M_{W_R}^2}\sum_{a,b}a_{ab}^{(5)}\frac{m_{u_a}m_{u_b}}{m_t^2}V^*_{L\,ai}V_{L\,bj}\left(V_R\right)_{bk}^*\left(V_R\right)_{al}\,,\nn
\eea
with $Y_{qH}$ again evaluated at $\mu=\mu_0$, while the coefficients $a^{(4,5)}$ are now functions of $\mu_0$ and are given by,
\bea
a^{(4)} &=& -0.024\left[
a_1^{(4)}+a_2^{(4)} \eta^{-6/7}+a_3^{(4)} \ln\eta
\right]\eta^{2/7}\,,\nn\\
a^{(5)} &=& -0.024\left[
a_1^{(5)}\eta^{1/7}+a_2^{(5)} \eta^{-2/7}+a_3^{(5)}\eta+a_4^{(5)} \eta\ln\eta
\right]\eta^{-5/7}\,,
\eea
where the coefficients for $a^{(4)}$ are
\bea
a_1^{(4)} = \bma
1&1&0.53\\
1&0.97&0.55\\
0.53&0.55&0.50
\ema\,,\quad
a_2^{(4)} = -\bma
1.75&1.75&0.53\\
1.75&1.86&0.55\\
0.53&0.55&0.50
\ema\,,\quad
a_3^{(4)} = -0.42\bma
1&1&1\\
1&1&1\\
1&1&1
\ema\,,\nn
\eea
while those for $a^{(5)}$ are
\bea
a_1^{(5)} &=& \bma
1&1&-1.58\\
1&1.22&-1.53\\
-1.58&-1.53&-1.64
\ema\,,\qquad
a_2^{(5)} = 3.03\bma
1&1&1\\
1&1&1\\
1&1&1
\ema\,,\nn\\
a_3^{(5)} &=&\bma
0.90&0.90&-1.25\\
0.90&0.97&-1.19\\
-1.25&-1.19&-1.34
\ema\,,\qquad 
a_4^{(5)} = 0.14\bma
1&1&1\\
1&1&1\\
1&1&1
\ema\,.
\eea
The terms $\sim \log\eta$ arise from the fact that the anomalous dimension matrix in Eq.\ \eqref{eq:4fermiRGE} has degenerate eigenvalues at $n_f=6$, leading to contributions of the form $\sim \frac{\eta^{\epsilon}-1}{\epsilon}$ with $\epsilon\propto n_f-6$.

\bibliographystyle{h-physrev3} 
 \bibliography{bibliography}

\end{document}